\documentclass{aa}  
\usepackage{graphicx}
\usepackage{amsmath,amssymb}
\usepackage{txfonts}
\begin{document}

   \title{Physical properties of dense cores in Orion B9
   \thanks{This publication is based on data acquired with the Atacama 
Pathfinder Experiment (APEX) under programme 084.F-9312, and observations with 
the 100-m telescope of the Max-Planck-Institut f\"ur Radioastronomie (MPIfR) 
at Effelsberg. APEX is a collaboration between the MPIfR, the European Southern
Observatory, and the Onsala Space Observatory.}}

   \author{O. Miettinen\inst{1,2}, J. Harju\inst{2}, L.~K. Haikala\inst{2}, 
\and M. Juvela\inst{2}}

   \offprints{O. Miettinen}

   \institute{Finnish Centre for Astronomy with ESO (FINCA), University of 
Turku, V\"ais\"al\"antie 20, FI-21500 Piikki\"o, Finland\\ 
\email{oskari.miettinen@helsinki.fi} \and Department of Physics, P.O. Box 64, 
00014 University of Helsinki, Finland\\}

\date{Received ; accepted}

\authorrunning{Miettinen et al.}
\titlerunning{Dense cores in Orion B9}

  \abstract
   {}
   {We aim to determine the physical and chemical properties of 
the starless and protostellar cores in Orion B9, which represents a relatively 
quiescent star-forming region in Orion B.} 
   {We observed the NH$_3$ $(J,\,K)=(1,\,1)$ and $(2,\,2)$ inversion lines 
and the N$_2$H$^+(3-2)$ rotational lines, with the Effelsberg 100-m and 
APEX telescopes, respectively, towards the submillimetre peak positions in 
Orion B9. These data are used in conjunction with our APEX/LABOCA 870 $\mu$m 
dust continuum data of the region.}
   {The gas kinetic temperature in the cores derived from the NH$_3$ data
is between $\sim9.4-13.9$ K. The non-thermal velocity dispersion 
is subsonic in most of the cores. The non-thermal linewidth in protostellar 
cores appears to increase with increasing bolometric luminosity. 
The core masses, $\sim2-8$ M$_{\sun}$, are very 
likely drawn from the same parent distribution as the core masses in Orion B 
North. Based on the virial parameter analysis, starless 
cores in the region are likely to be gravitationally bound, and thus 
prestellar. Some of the cores have a lower radial velocity than the systemic 
velocity of the region, suggesting that they are members of the ``low-velocity 
part'' of Orion B. The observed core-separation distances deviate from the 
corresponding random-like model distributions. The distances between the 
nearest neighbours are comparable to the thermal Jeans length. The fractional 
abundances of NH$_3$ and N$_2$H$^+$ in the cores are $\sim1.5-9.8\times10^{-8}$ 
and $\sim0.2-5.9\times10^{-10}$, respectively. The NH$_3$ abundance appears to 
decrease with increasing H$_2$ column and number densities. 
The NH$_3$/N$_2$H$^+$ column density ratio is larger in starless cores than in 
cores with embedded protostars. } 
   {The core population in Orion B9 is comparable in physical properties to 
those in nearby low-mass star-forming regions. The Orion B9 cores also seem 
to resemble cores found in isolation rather than those associated with 
clusters. Moreover, because the cores may not be randomly distributed within 
the region (contrary to what was suggested in our Paper I), it is unclear 
whether the origin of cores could be explained by turbulent fragmentation. 
On the other hand, many of the core properties conform to the picture of 
dynamic core evolution. The Orion B9 region has probably been influenced by 
the feedback from the nearby Ori OB 1b group, and the 
fragmentation of the parental cloud into cores could be caused by 
gravitational instability.}

    \keywords{Stars: formation - ISM: clouds - ISM: individual objects: 
Orion B - ISM: molecules - ISM - Radio lines: ISM}

    \maketitle
%

\section{Introduction}

The Orion B molecular cloud (L1630) is a useful target for studying dense cores 
and the processes of star formation. The cloud complex lies at a relatively
close distance to the Sun ($\sim450$ pc), and it contains a wide range of 
star-forming environments, such as the high-mass star-forming region NGC 2024 
and several other regions of clustered star formation (which is the dominant 
mode of star formation in the Galaxy), but also more quiescent 
areas (see, e.g., \cite{bally2008}; \cite{ikeda2009}; \cite{buckle2010}).
We recently mapped region of the central part of Orion B of 
about $36\arcmin \times 27\arcmin$ (4.7 pc $\times$ 3.5 pc), called Orion B9 
($l=206\fdg1$, $b=-15\fdg8$), at 870 $\mu$m dust continuum using LABOCA on 
APEX, and discovered 12 dense submm cores (\cite{miettinen2009}; hereafter 
Paper I). The cores were classified into starless and protostellar ones by 
using the Spitzer data, yielding the result that half the cores have embedded 
protostar(s). In the present paper, we examine the physical 
characteristics of the Orion B9 cores further. To determine the gas kinetic 
temperature, kinematics, and the dynamical state of the cores, we performed 
NH$_3(1,\,1)$ and $(2,\,2)$, and N$_2$H$^+(3-2)$ observations towards 
the core submm peak positions with the Effelsberg 100-m and APEX telescopes, 
respectively. In this paper, the derived temperatures are also used to 
recalculate the temperature-dependent core parameters presented in Paper I.

This paper is organised as follows. The observations and data-reduction 
procedures are described in Sect.~2. The observational results are presented 
in Sect.~3. In Sect.~4, we describe the analysis of the physical and chemical 
properties of the cores. Discussion of the results is given in Sect.~5, 
and in Sect.~6, we summarise our main conclusions.

\section{Observations and data reduction}

\subsection{Effelsberg 100-m observations}

Pointed observations of the NH$_3(1,\,1)$ and $(2,\,2)$ inversion 
line emission towards the dense cores in Orion B9 were performed with the 
Effelsberg 100-m telescope of the MPIfR on 2009 November 23-25. 
The target positions listed in Table~\ref{table:cores} were drawn from the 
submm peak positions in the LABOCA 870 $\mu$m dust continuum map of
the region. In Table~\ref{table:cores}, we also show the source classification 
from Paper I. 

The 1.3~cm HEMT receiver was tuned to a frequency of
23\,708.564 MHz, lying midway between the rest frequencies of the
NH$_3(1,\,1)$ and $(2,\,2)$ lines, which are separated by about 28 MHz. For
the sky frequency conversion the local standard of rest (LSR) velocity was set
to 9.2 km~s$^{-1}$. The receiver measures orthogonal linear polarisations. 
The sum of the two channels was fed into the Fast Fourier Transfrom 
Spectrometer (FFTS) with a 100 MHz bandwidth. In this manner, the Stokes $I$ 
spectra of the two lines were measured simultaneously. The backend is a 
modified copy of the FFTS at the APEX telescope (see Sect.~2.2). 
The 100 MHz band was split into 16\,384 channels. This resulted
in a channel separation of 6.1 kHz which corresponds to about 77 m~s$^{-1}$
at 23.7 GHz. The spectral resolution (equivalent noise bandwidth) is
about 9.8 kHz (124 m~s$^{-1}$). One of the sources, SMM 6, was also observed
in NH$_3(1,\,1)$ (23\,694.4955 MHz) using a 20 MHz bandwidth to resolve the
hyperfine structure in more detail. In this configuration, the channel
separation is 1.22 kHz (15 m~s$^{-1}$).

The FWHM (full width at half maximum) beamsize at the observed 
frequencies is $40\arcsec$ (0.09 pc at 450 pc), and the main beam efficiency 
is $\eta_{\rm MB}=0.52$.
Observations were conducted in frequency switching mode with a frequency
throw of 6 MHz. The single-sideband (SSB) system temperatures were about 
200-310 K in the main-beam brightness temperature, $T_{\rm MB}$, scale during 
the observations. Typical integration time was $\sim120$ min per position, 
resulting in a $1\sigma$ rms noise of about 38-75 mK. For further information 
of the telescope and the receiving system, see 
\texttt{http://www.mpifr-bonn.mpg.de/radioteleskop/}.

The NH$_3(1,\,1)$ and $(2,\,2)$ emission was detected towards all submm cores
except IRAS 05412-0105 (rms$\sim$79 mK). This source was also very weak 
(0.17 Jy~beam$^{-1}$) in our LABOCA 870 $\mu$m map (Paper I; Fig.~1 therein).
We performed test measurements of the NH$_3(3,\,3)$ inversion line at
23\,870.1292 MHz towards two strong NH$_3(1,\,1)$ sources in our sample:
SMM 4 and IRAS 05405-0117. No lines were detected after 30 min integration
per position (rms $\sim200-300$ mK at the velocity resolution 15 m~s$^{-1}$).

Telescope pointing and focus were checked about every 1-1.5 h by 
continuum scans on radio quasars PKS 0420-014 and 3C147, and the radio galaxy 
3C213. The pointing was typically accurate to within $\sim7\arcsec$.
Absolute flux calibration was based upon continuum cross scans on the planetary
nebula NGC 7027, and the radio galaxies/quasars 3C123, 3C147, and 3C286, 
for which we adopted the flux densities $5.58\pm0.10$, 2.93, $1.91\pm0.11$, 
and $2.46\pm0.05$ Jy at 1.3 cm, respectively (\cite{ott1994}; 
\cite{peng2000}). We were not able to retrieve the zenith opacity, 
$\tau_{\rm z}$, by doing a sky-dipping measurement or by observing calibrators 
in good weather in a wide range of elevations (e.g., \cite{pillai2006}; 
\cite{frieswijk2007}). Instead, we estimated the average $\tau_{\rm z}$ at 1.3
cm by fitting all the calibration measurements with the function 
$G(\theta)\times{\rm e}^{-\tau_{\rm z}/\sin \theta}$, where $G(\theta)$ 
is the instrumental gain-elevation curve given on the telescope {\tt www} 
pages (see above). In this manner, we obtained an average $\tau_{\rm z}$ of 
0.15. This value is consistent with previous estimates at centrimetre 
wavelengths (\cite{frieswijk2007}; Appendix A therein).
The typical uncertainty in absolute flux calibration is $\sim15\%$ (excluding 
the systematic error in $\tau_{\rm z}$). 
Note that this uncertainty does not affect the parameters that depend on the 
line intensity ratios (e.g, optical thickness, kinetic temperature), 
or the kinematical parameters 
(centroid velocity and width of the line). The calibration uncertainty 
propagates to the values of the line excitation temperature, and NH$_3$
column densities and fractional abundances (see Sect.~4.1).

The spectra were reduced using the CLASS package of the IRAM's GILDAS 
software\footnote{{\tt http://www.iram.fr/IRAMFR/GILDAS}}.
For a given source, the spectra were averaged and folded. A first order 
polynomial, and in one case (SMM 4) a third order polynomial, was applied 
to correct the baseline in the NH$_3(1,\,1)$ lines. 
A polynomial baseline of order 3 was subtracted
from the NH$_3(2,\,2)$ lines. We fitted the hyperfine structure of the 
NH$_3(1,\,1)$ line using ``method NH3(1,\,1)'' of the CLASS package to 
derive the LSR velocity (${\rm v}_{\rm LSR}$) of the emission, FWHM linewidth 
($\Delta {\rm v}$), and the line optical thickness (see Sect.~3.1). 
The emission of the $(2,\,2)$ satellite lines was not detected. 
Nevertheless, the radial velocities and linewidths were determined using 
``method NH3(2,\,2)'' of CLASS.

\begin{table}
\caption{Submillimetre cores in Orion B9.}
\centering
\renewcommand{\footnoterule}{}
\label{table:cores}
\begin{tabular}{c c c c}
\hline\hline 
Source & \multicolumn{2}{c}{Peak position} & Class\\
name & $\alpha_{2000.0}$ [h:m:s] & $\delta_{2000.0}$ [$\degr$:$\arcmin$:$\arcsec$] & \\
\hline
IRAS 05399-0121 & 05 42 27.4 & -01 19 50 & 0/I\\
SMM 1 & 05 42 30.5 & -01 20 45 & prestellar\\
SMM 2 & 05 42 32.9 & -01 25 28 & prestellar\\
SMM 3 & 05 42 44.4 & -01 16 03 & 0\\
IRAS 05405-0117 & 05 43 02.7 & -01 16 21 & 0\\
SMM 4 & 05 43 03.9 & -01 15 44 & 0\\
SMM 5 & 05 43 04.5 & -01 17 06 & prestellar\\
SMM 6 & 05 43 05.1 & -01 18 38 & prestellar\\
Ori B9 N & 05 43 05.7 & -01 14 41 & prestellar\\
SMM 7 & 05 43 22.1 & -01 13 46 & prestellar\\
IRAS 05412-0105 & 05 43 46.4 & -01 04 30 & 0\\
IRAS 05413-0104 & 05 43 51.3 & -01 02 50 & 0\\
\hline 
\end{tabular} 
\end{table}

\subsection{APEX observation}

The N$_2$H$^+(3-2)$ observations towards nine cores in the central region of 
Orion B9 (i.e., IRAS 05399-0121 and 05405-0117, SMM 1, and 3--7, and Ori B9 N) 
were carried out on 2009 September 2 and 11 with the 
APEX telescope\footnote{Other spectral lines observed under this programme 
include C$^{17}$O$(2-1)$, N$_2$D$^+(3-2)$, DCO$^+(4-3)$, and 
H$^{13}$CO$^+(4-3)$. These observations will be presented in the forthcoming 
paper.}. The backend was the MPIfR FFTS (\cite{klein2006}) with a 1 GHz 
bandwidth divided into 8192 channels. The resulting channel width was 122 kHz 
which corresponds to 0.13 km~s$^{-1}$ at the observed frequency 279.5 GHz. 
As frontend, we used APEX-2 of the Swedish Heterodyne Facility Instrument 
(SHFI; \cite{vassilev2008}). At 279.5 GHz, the APEX beamsize is 
$\simeq 22\farcs3$, and the main beam efficiency is $\eta_{\rm MB}\simeq0.74$. 

The observations were performed in the wobbler-switching mode with a 
$100\arcsec$ throw (symmetric offsets) and a chopping rate of 0.5 Hz.
The typical total integration time was $\sim11$ min per position, and 
the SSB system temperature was between 210-360 K 
($T_{\rm MB}$ scale) during these observations.
The telescope pointing was checked using the planets Mars and 
Uranus, and the stars $\alpha$ Orionis and V1259 Ori, and was found to be 
accurate to $\sim3\arcsec$. The calibration was achieved
by the chopper-wheel method, and the intensity scale given by the system
is $T_{\rm A}^*$, the antenna temperature corrected for atmospheric
attenuation. The observed intensities were converted to the main-beam
brightness temperature scale by $T_{\rm MB}=T_{\rm A}^*/\eta_{\rm MB}$.

The spectra were reduced using the CLASS. The individual spectra were averaged,
and linear baselines were subtracted from the resulting sum spectra. 
The resulting $1\sigma$ rms noise values are 35--130 mK.

The $J=3-2$ transition of N$_2$H$^+$ contains 38 hyperfine components.
We fitted the hyperfine structure of the lines using ``method hfs'' of the 
CLASS. For the rest frequencies of the hyperfine components, we used the 
values from Pagani et al. (2009; Table 4 therein). The adopted central 
frequency, 279\,511.832 MHz, is that of the 
$J_{F_1 F} = 3_{45} \rightarrow 2_{34}$  hyperfine component which has 
a relative intensity of 17.46 \%

\section{Observational results}

\subsection{NH$_3$}

\begin{figure*}
\begin{center}
\includegraphics[width=3.2cm, angle=-90]{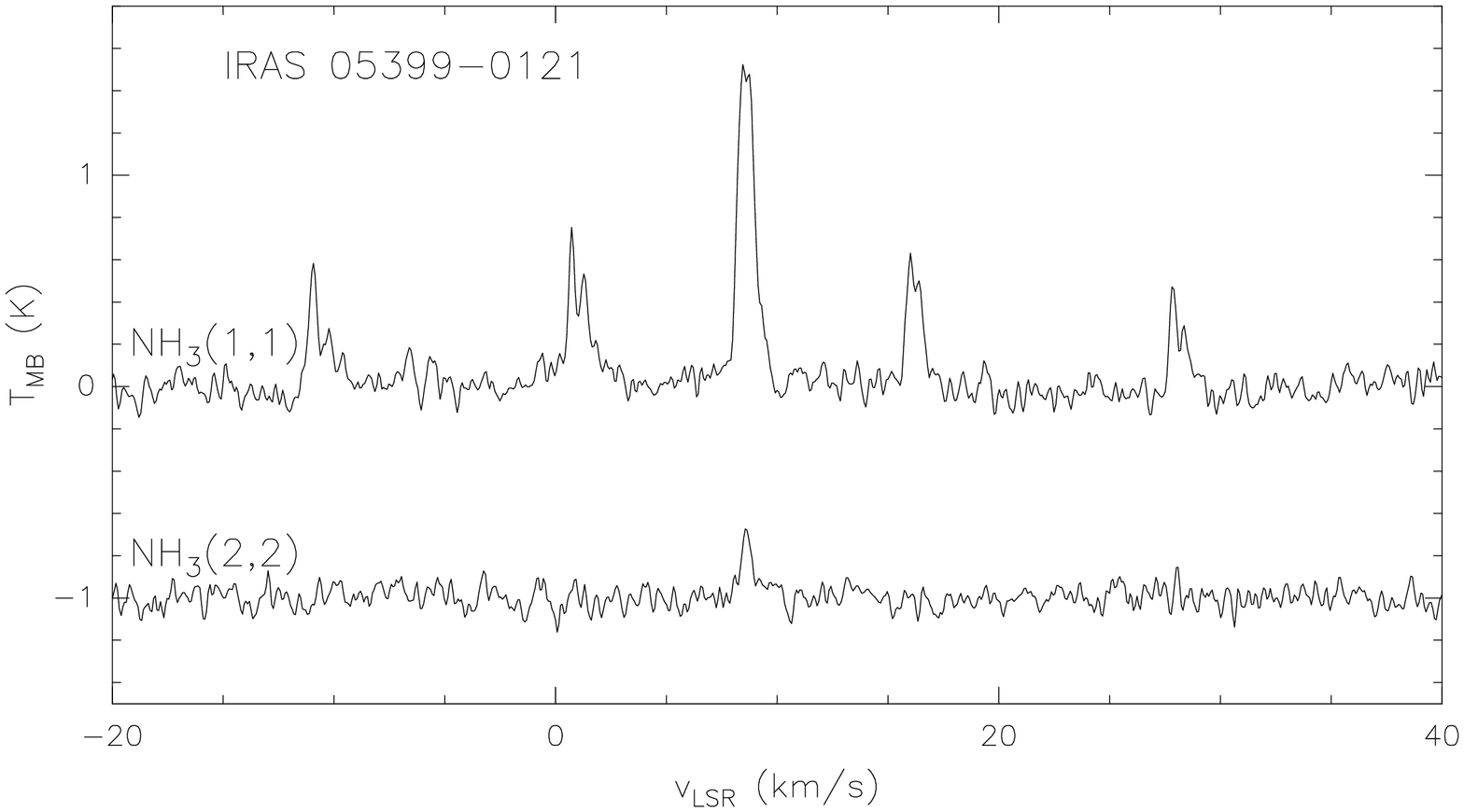}
\includegraphics[width=3.2cm, angle=-90]{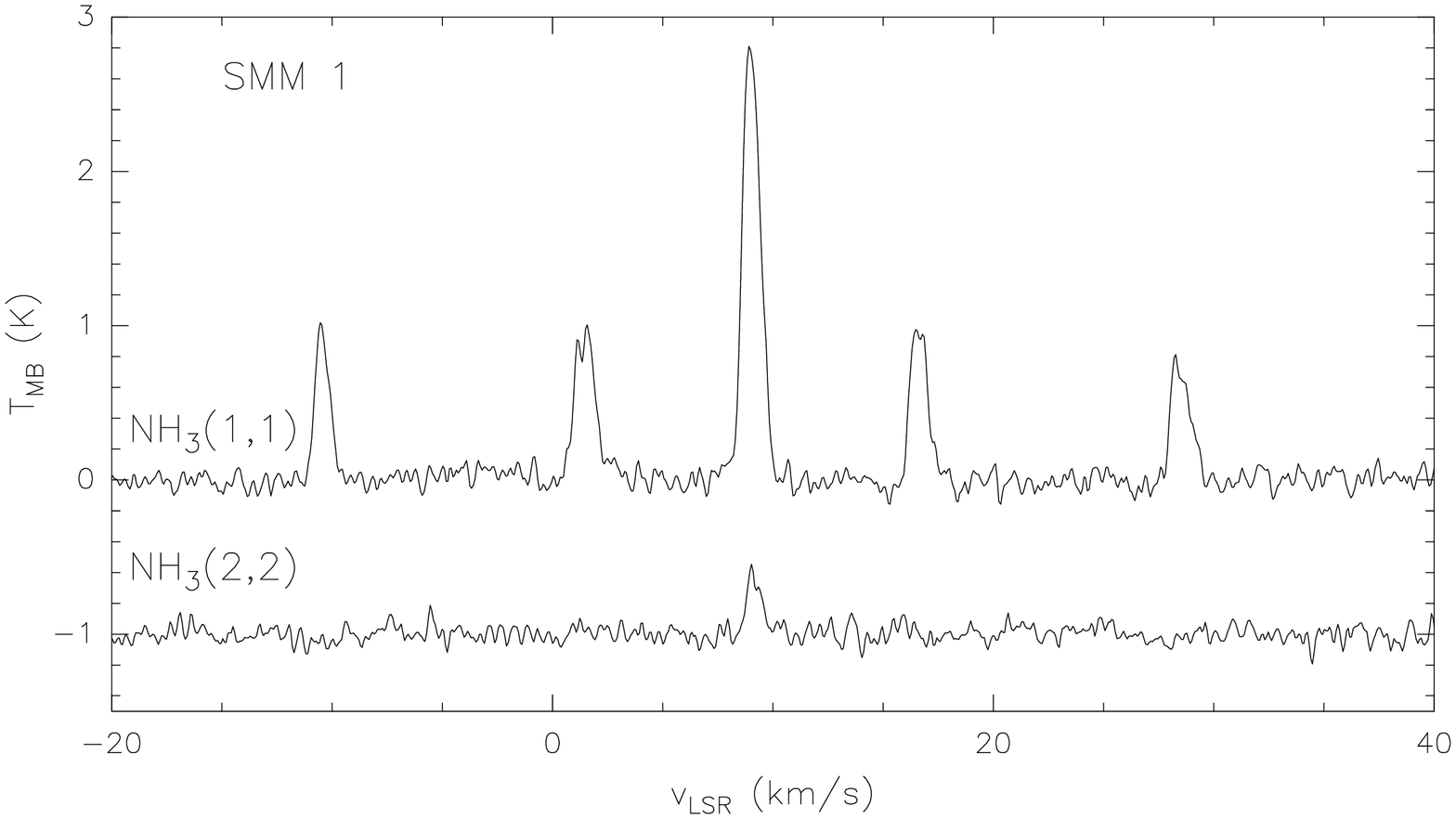}
\includegraphics[width=3.2cm, angle=-90]{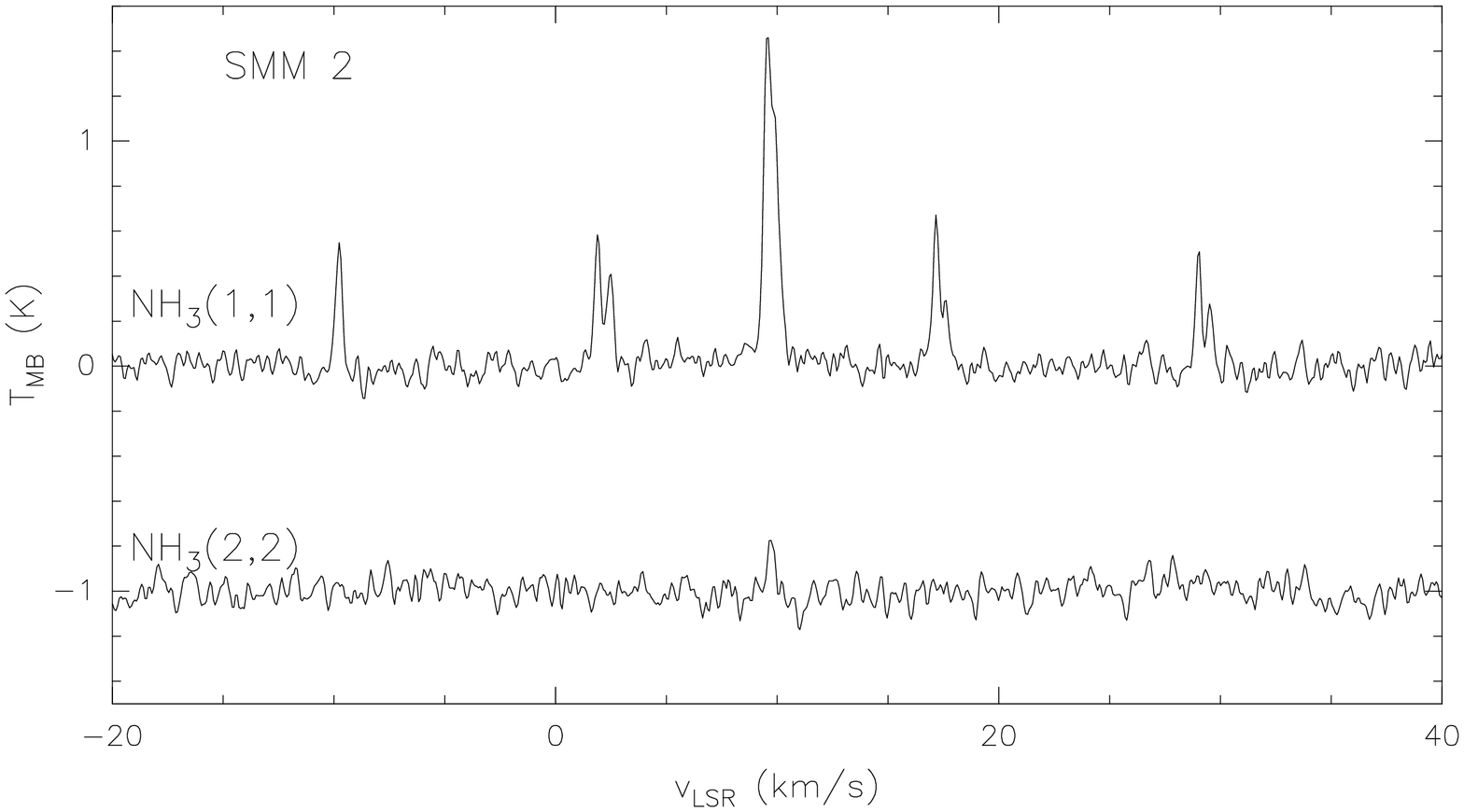}
\includegraphics[width=3.2cm, angle=-90]{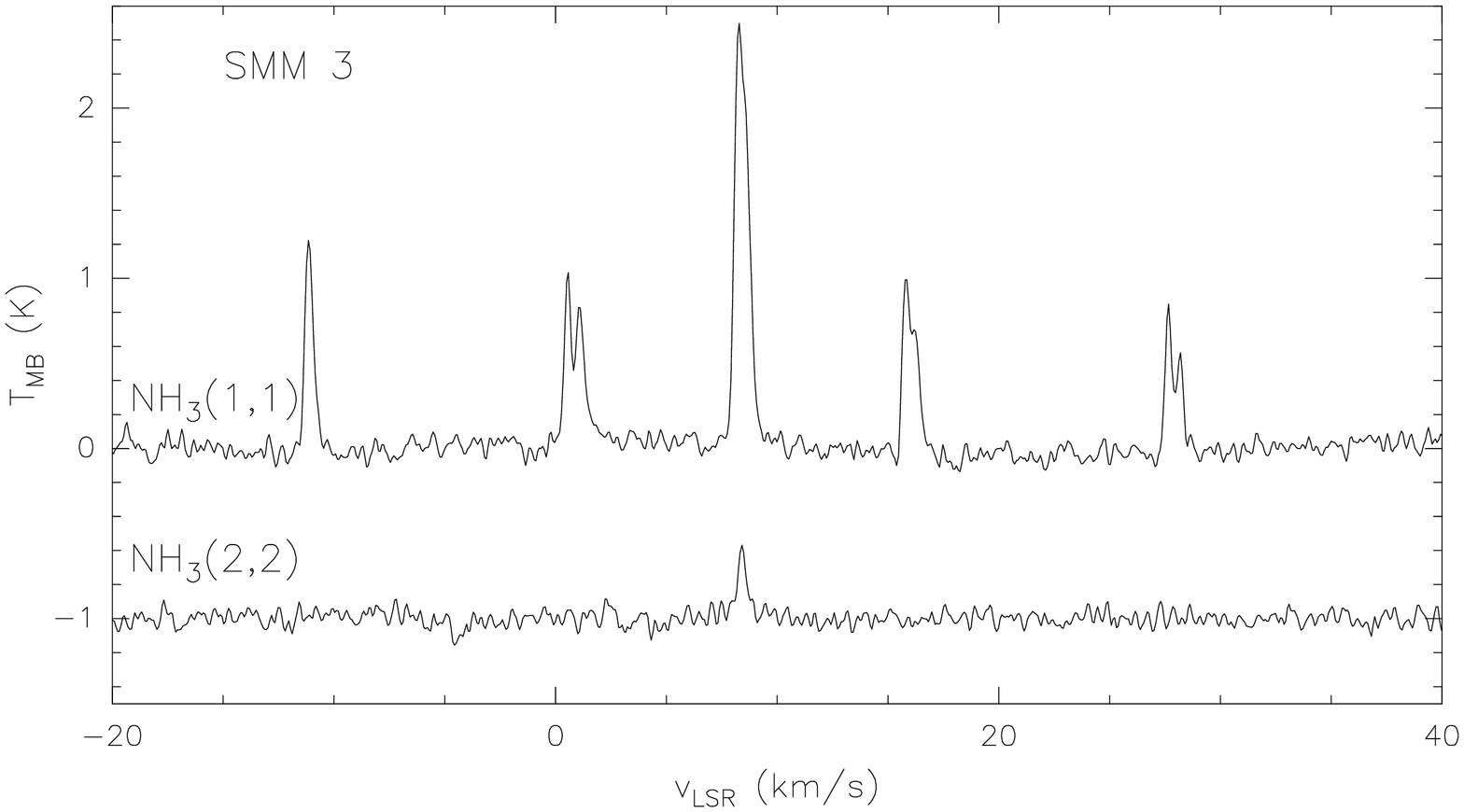}
\includegraphics[width=3.2cm, angle=-90]{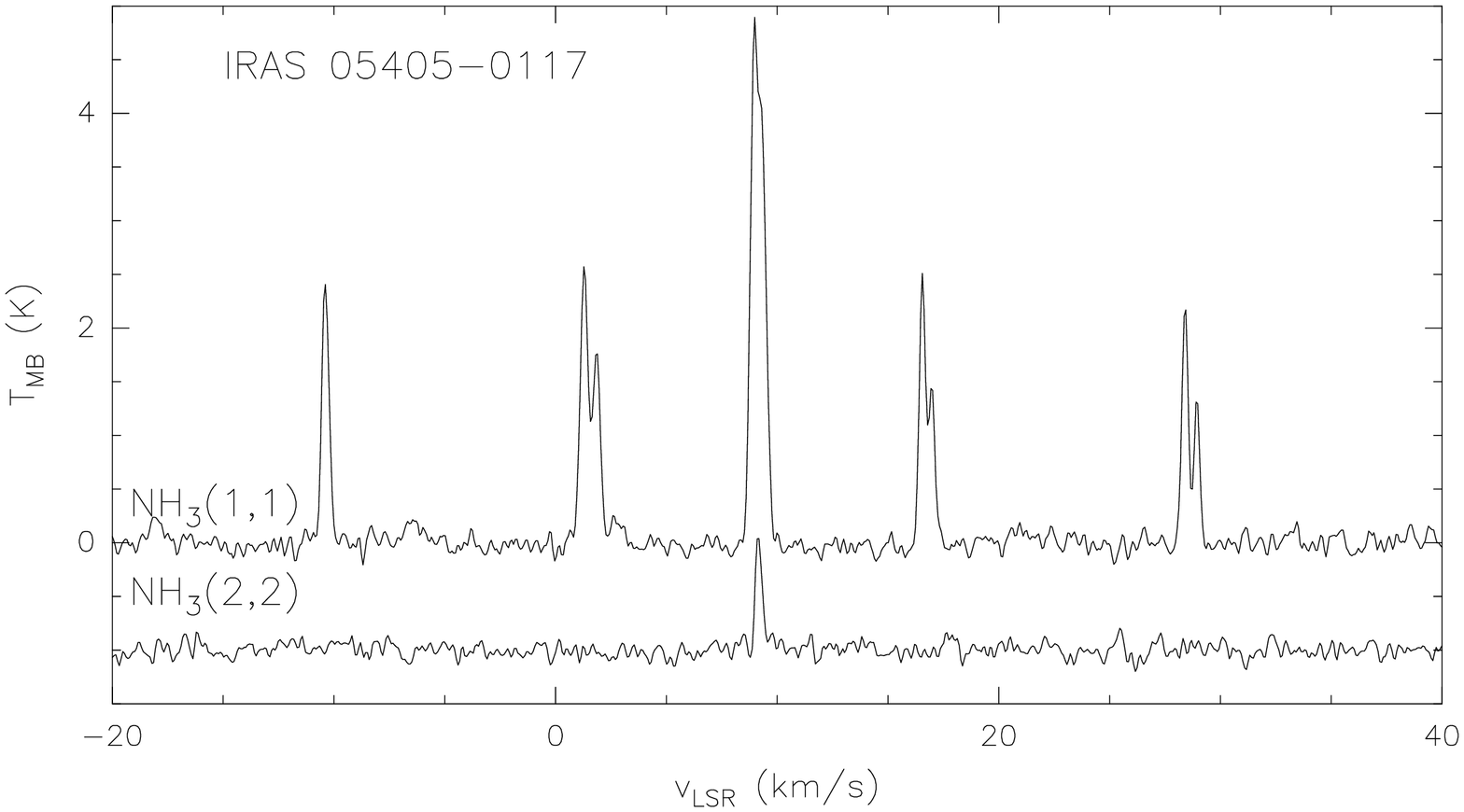}
\includegraphics[width=3.2cm, angle=-90]{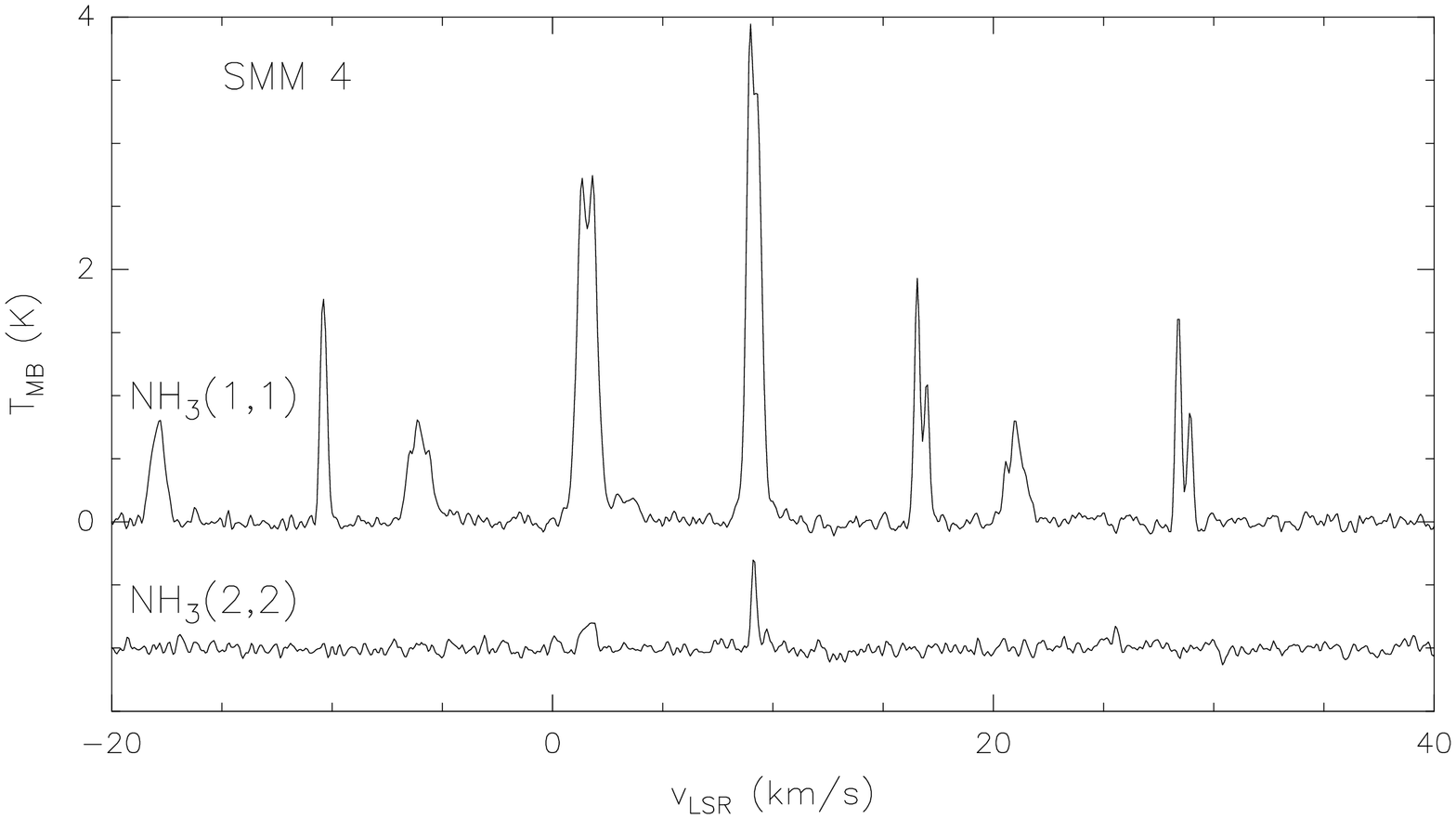}
\includegraphics[width=3.2cm, angle=-90]{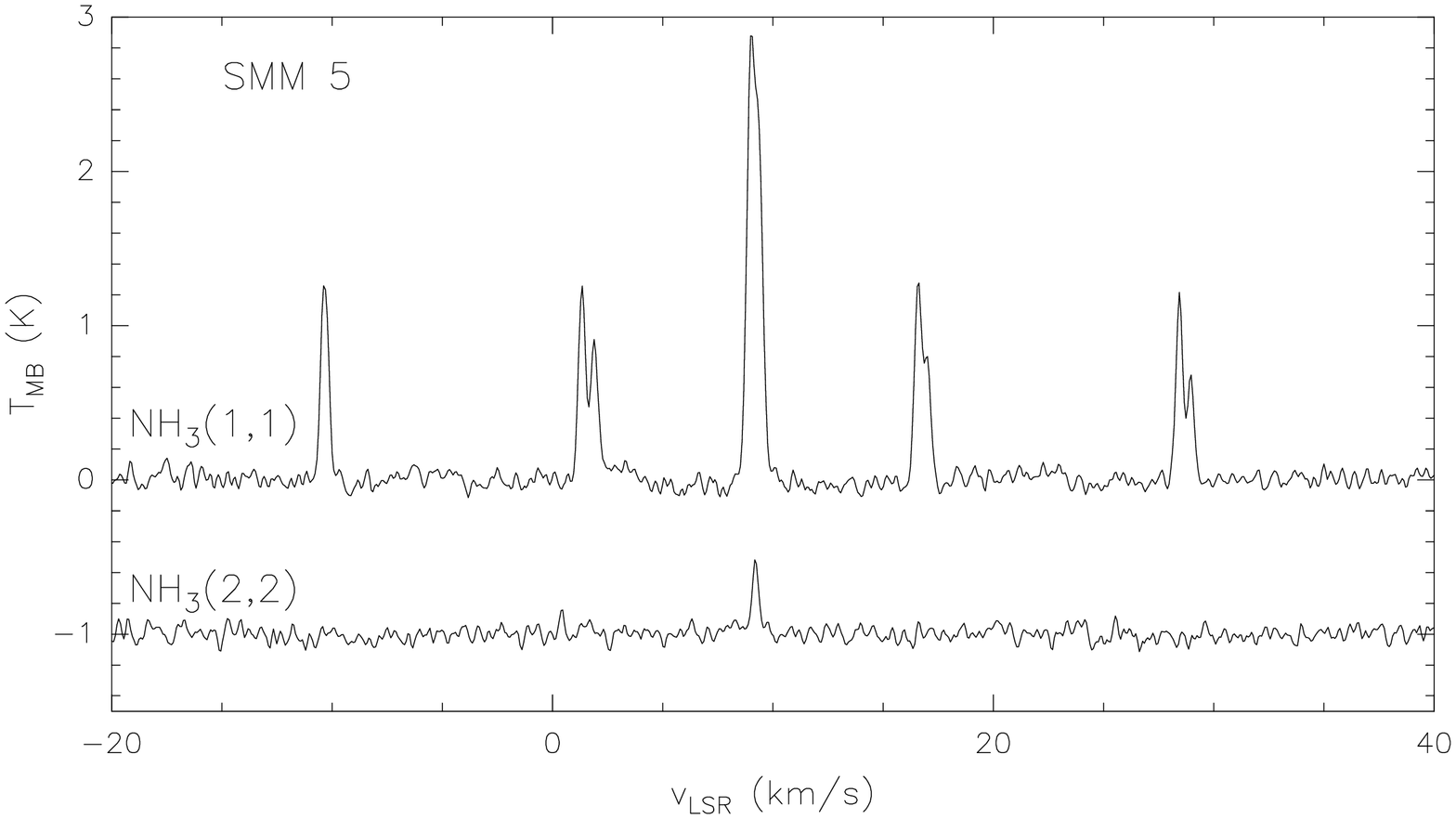}
\includegraphics[width=3.2cm, angle=-90]{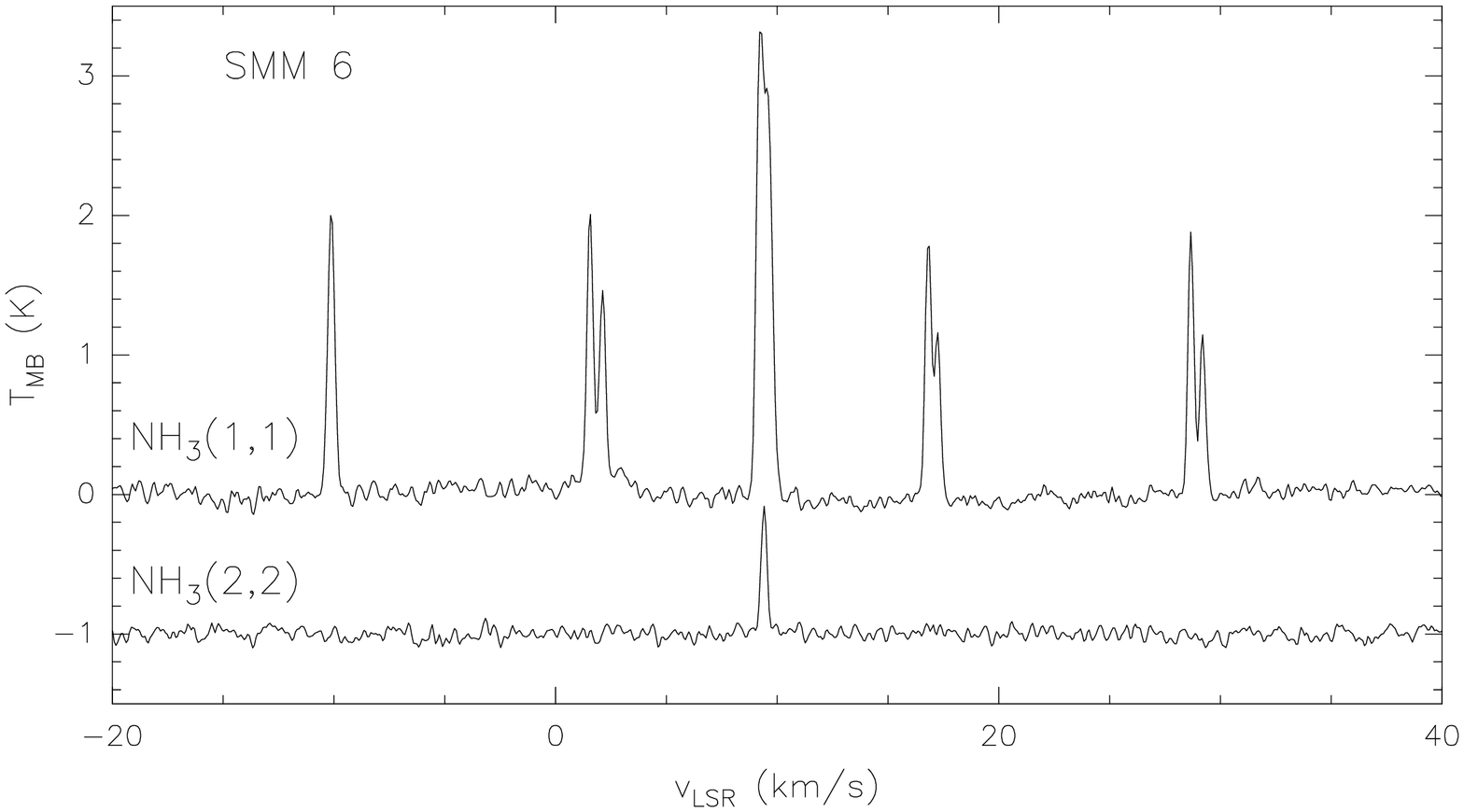}
\includegraphics[width=3.2cm, angle=-90]{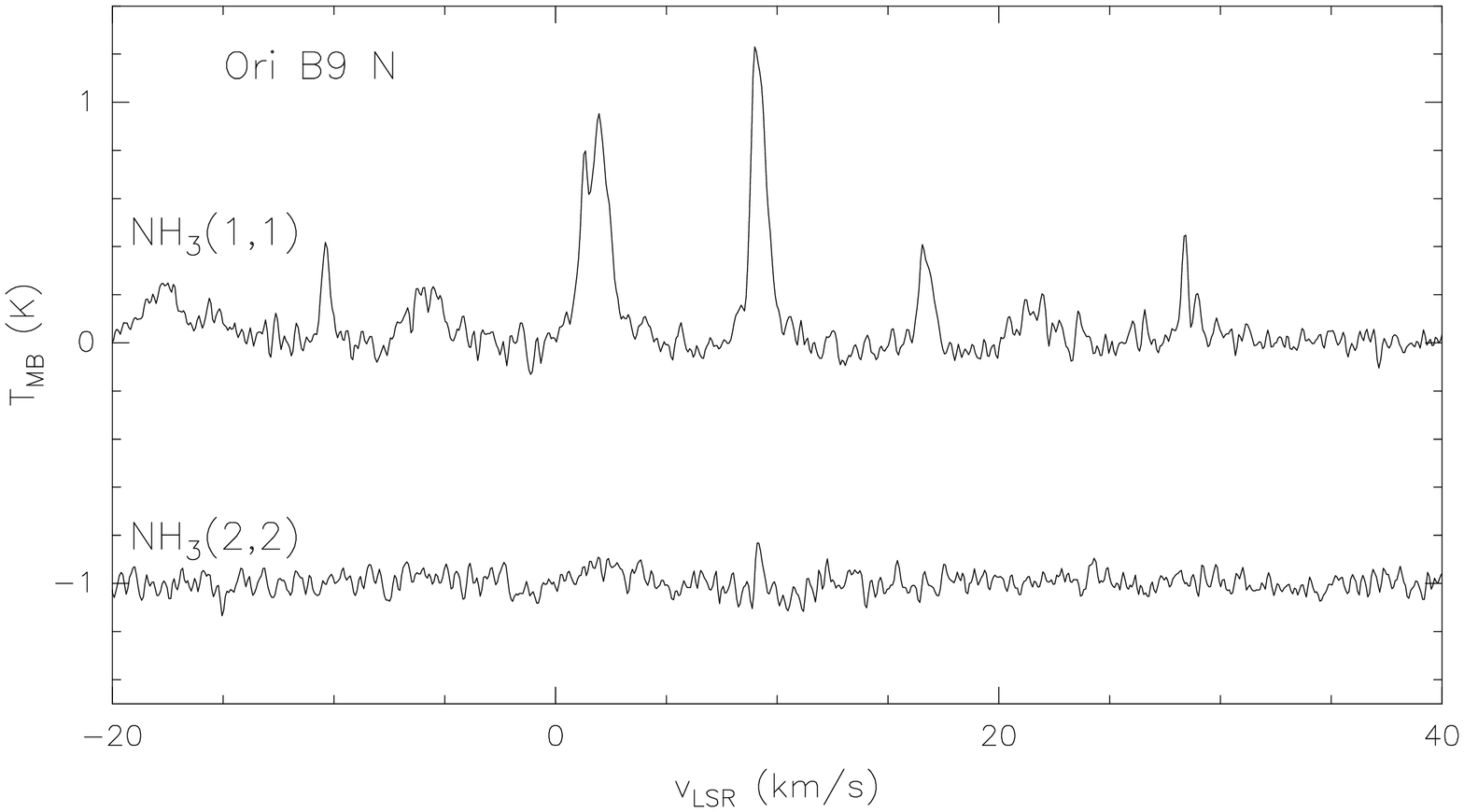}
\includegraphics[width=3.2cm, angle=-90]{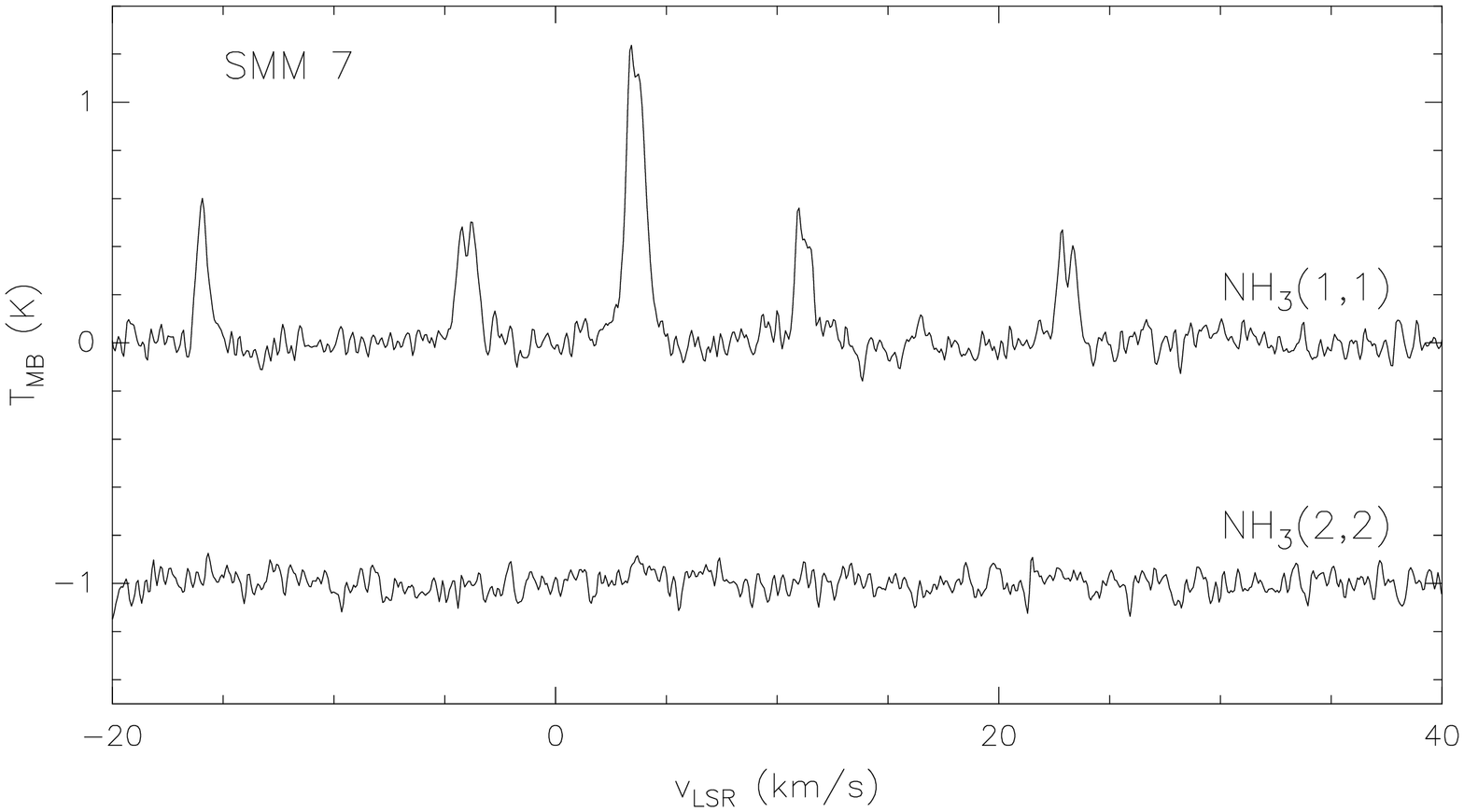}
\includegraphics[width=3.2cm, angle=-90]{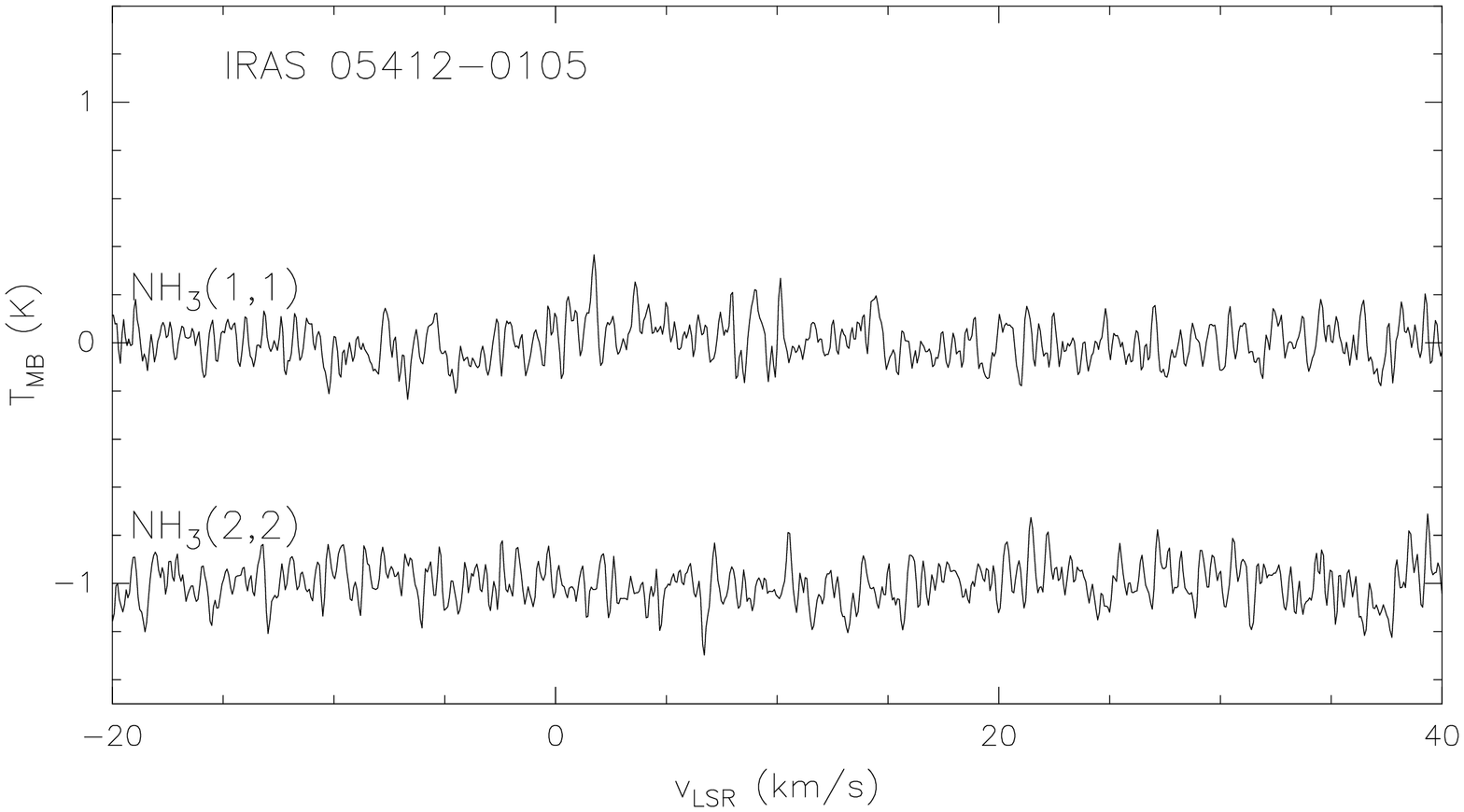}
\includegraphics[width=3.2cm, angle=-90]{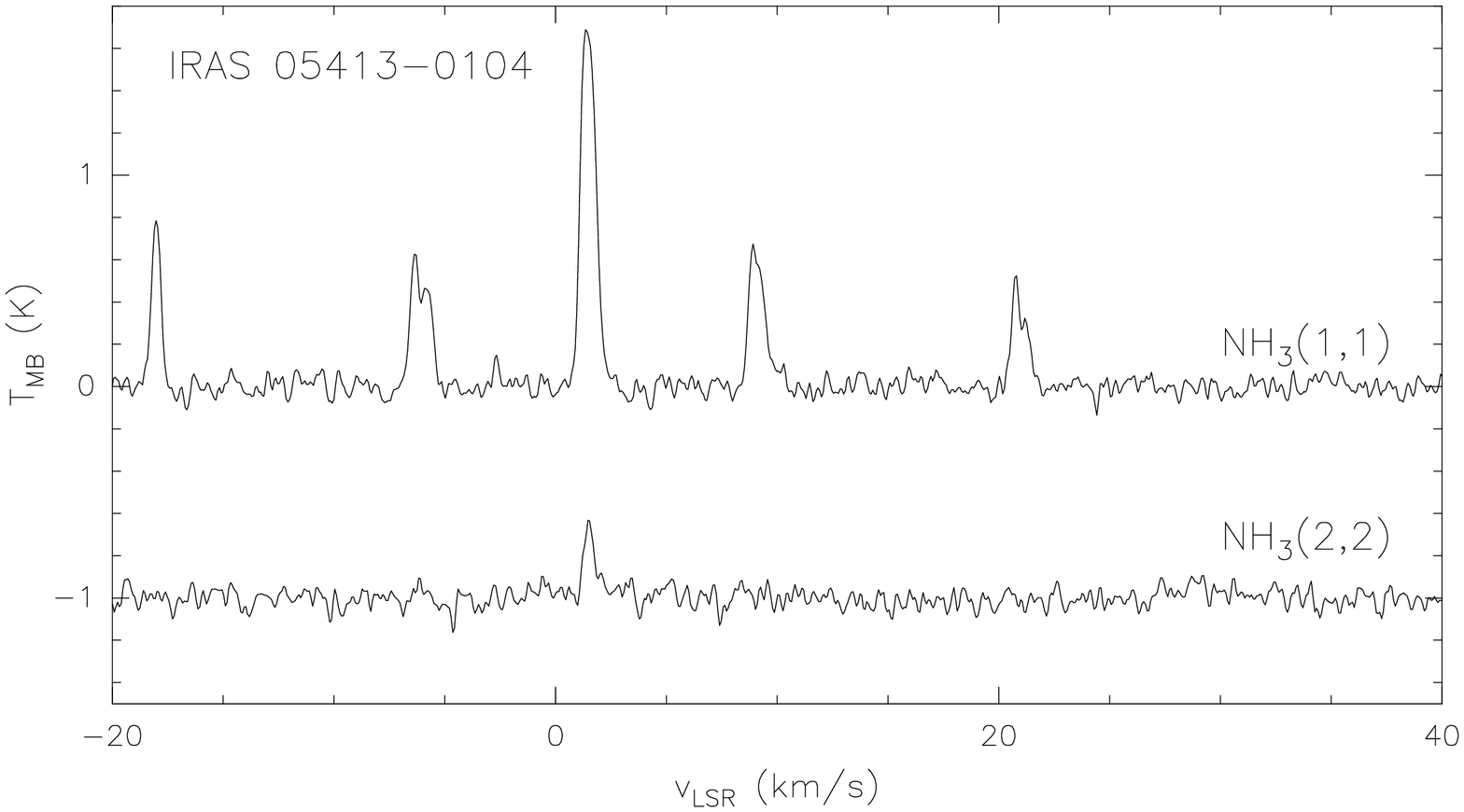}
\caption{NH$_3(1,\,1)$ and $(2,\,2)$ spectra measured from the pre- and 
protostellar cores in Orion B9. The temperature scale is in $T_{\rm MB}$. 
NH$_3(2,\,2)$ spectra are offset by -1 K from zero baseline for clarity. 
Note that there are two velocity components towards SMM 4 and Ori B9 N, and 
that no lines were detected towards IRAS 05412-0105.}
\label{figure:spectra}
\end{center}
\end{figure*}

\begin{figure}[!h]
\resizebox{\hsize}{!}{\includegraphics{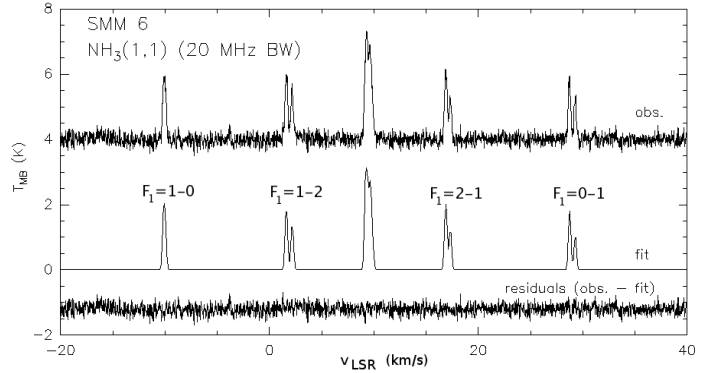}}
\caption{NH$_3(1,\,1)$ spectrum measured from SMM 6 using the FFTS with a 
20 MHz bandwidth. Below the spectrum are shown the model fit to the hyperfine 
structure and the residual spectrum. Satellite hyperfine components in the
$F_1^{'}\to F_1$ format are indicated.}
\label{figure:SMM6}
\end{figure}

The NH$_3(1,\,1)$ and $(2,\,2)$ spectra are shown in 
Fig.~\ref{figure:spectra}. The NH$_3(1,\,1)$ spectrum observed from SMM 6
with a 20 MHz bandwidth (see Sect.~2.1) is shown in Fig.~\ref{figure:SMM6}. 
In Table~\ref{table:parameters1}, we give the line 
parameters resulting from the hyperfine/Gaussian fits to the NH$_3$ lines. 
The so-called 'main group optical thickness', $\tau_{\rm m}$, given
in Table~\ref{table:parameters1} is the sum of peak optical thicknesses of the
eight hyperfine components included in the main group. For the $(1,\,1)$
transition this equals to the corresponding sum of the ten satellite
components, so the total of all peak optical thicknesses is
$2\times \tau_{\rm m}$. Assuming that the line profile (due to
velocity dispersion) is Gaussian, the integrated optical thickness
can be obtained from $\int \tau({\rm v}){\rm dv}=\sqrt{\pi/\ln 2}\tau_{\rm m}\Delta {\rm v}$, where $\Delta {\rm v}$ is the linewidth (FWHM) in velocity 
units. The uncertainties associated with ${\rm v}_{\rm LSR}$, 
$\Delta {\rm v}$, and $\tau_{\rm m}$ are the $1\sigma$ fitting errors. 
The uncertainty in $T_{\rm MB}$, $\sigma(T_{\rm MB})$, includes
the absolute calibration uncertainty of 15\%, and the $1\sigma$ rms noise in 
the spectrum, and was obtained by quadratic summing of these two errors as 
$\sigma(T_{\rm MB})=\sqrt{\sigma_{\rm cal}^2+\sigma_{\rm rms}^2}$. 
The determination of the line excitation temperature, $T_{\rm ex}$, listed 
in the last column of Table~\ref{table:parameters1}, is described in 
Sect.~4.1.1. 

The NH$_3$ spectra towards SMM 4 and Ori B9 N show evidence of
a second velocity component in addition to the strong line at about
9.1 km~s$^{-1}$ corresponding to the systemic velocity of Orion B. 
In the case of IRAS 05405-0117, there is also a hint of a second velocity 
component, probably resulting from the fact that the $40\arcsec$ beam 
slightly overlaps with the nearby core SMM 4. In the
direction of SMM 4, the second component has a velocity of about 
1.6 km~s$^{-1}$, whereas towards Ori B9 N the velocity is 1.9 km~s$^{-1}$. 
In both cases, the main hyperfine group of the second velocity component 
overlaps with the inner satellite line $F_1=1-2$ of the principal velocity
component. Consequently, the inner satellite line $F_1=2-1$ of the
second velocity component contaminates the main hyperfine group of the
principal velocity component. These spectra were analysed using a
two-component fit to the NH$_3(1,\,1)$ hyperfine structure. We note that
the secondary velocity components towards SMM 4 and Ori B9 N are close
to the radial velocities of IRAS 05413-0104 (1.5 km~s$^{-1}$) and SMM 7 
(3.6 km~s$^{-1}$). We also note that the second velocity components influence
the core parameters derived from the dust continuum emission (see Sect.~4.3). 
Part of the observed flux density may be due to the dust component associated 
with another source along the line of sight. However, because continuum 
observations do not provide velocity information, it is impossible to solve 
this ``overlap'' problem.

The NH$_3(1,\,1)$ line towards SMM 6 measured using a 20 MHz bandwidth is 30 
m~s$^{-1}$ narrower and has a 15\% larger optical thickness than that
measured using a 100 MHz bandwidth. The difference in the linewidths
corresponds to effect expected from instrumental broadening with the
two configurations. As is expected, the product 
$\Delta {\rm v}\tau_{\rm m}$ is roughly constant at the two spectral 
resolutions (see Eq.~(\ref{eq:N11})). As the high resolution spectrum 
is only available for one object, we use in the subsequent analysis the 
100 MHz spectra for all of them.

\begin{table*}
\caption{NH$_3(1,\,1)$ and $(2,\,2)$ line parameters.}
\begin{minipage}{2\columnwidth}
\centering
\renewcommand{\footnoterule}{}
\label{table:parameters1}
\begin{tabular}{c c c c c c c}
\hline\hline 
       & Transition & ${\rm v}_{\rm LSR}$ & $\Delta {\rm v}$ & $T_{\rm MB}$\footnote{The error in $T_{\rm MB}$ includes the $\sim15\%$ calibration uncertainty (see the text in Sect.~3.1).} & $\tau_{\rm m}$\footnote{$\tau_{\rm m}$ is the optical thickness of the $(1,\,1)$ main group. The total optical thickness, i.e., the sum over all satellites, is $\tau_{\rm tot}=2\times \tau_{\rm m}$.} & $T_{\rm ex}$\\
Source & $(J,\,K)$ & [km~s$^{-1}$] & [km~s$^{-1}$] & [K] & & [K]\\
\hline
IRAS 05399-0121 & $(1,\,1)$ & $8.61\pm0.01$ & $0.59\pm0.02$ & $1.6\pm0.2$ & $1.10\pm0.13$ & $5.4\pm0.5$\\
                & $(2,\,2)$ & $8.64\pm0.02$ & $0.42\pm0.08$ & $0.3\pm0.1$ &\\
SMM 1 & $(1,\,1)$ & $9.04\pm0.003$ & $0.67\pm0.01$ & $2.9\pm0.4$ & $1.23\pm0.06$& $7.2\pm0.7$\\
      & $(2,\,2)$ & $9.13\pm0.02$ & $0.53\pm0.05$ & $0.4\pm0.1$ &\\
SMM 2 & $(1,\,1)$ & $9.71\pm0.003$ & $0.34\pm0.01$ & $1.4\pm0.2$ & $1.38\pm0.16$& $5.3\pm0.4$\\
      & $(2,\,2)$ & $9.72\pm0.02$ & $0.26\pm0.04$ & $0.2\pm0.1$ &\\
SMM 3 & $(1,\,1)$ & $8.40\pm0.002$ & $0.38\pm0.01$ & $2.5\pm0.4$ & $2.14\pm0.10$& $6.1\pm0.5$\\
      & $(2,\,2)$ &  $8.42\pm0.01$ & $0.42\pm0.06$ & $0.4\pm0.1$ &\\
IRAS 05405-0117 & $(1,\,1)$ & $9.12\pm0.001$ & $0.32\pm0.003$ & $4.9\pm0.7$ & $3.36\pm0.09$ & $8.4\pm0.9$\\
                & $(2,\,2)$ & $9.17\pm0.01$ & $0.30\pm0.02$ & $1.1\pm0.2$ &\\
SMM 4 & $(1,\,1)$ & $9.12\pm0.001$ & $0.34\pm0.01$ & $3.2\pm0.5$\footnote{$T_{\rm MB}(1,\,1)$ was determined by substracting the second velocity component from the spectrum.} & $1.33\pm0.03$ & $8.6\pm0.9$\\
      & $(2,\,2)$ & $9.15\pm0.01$ & $0.23\pm0.02$ & $0.7\pm0.1$ &\\
SMM 4 (2nd v-comp.) &  $(1,\,1)$ & $1.62\pm0.003$ & $0.71\pm0.01$ & $2.0\pm0.2$ & $1.50\pm0.06$ & $5.1\pm0.5$ \\
                   & $(2,\,2)$ & $1.66\pm0.03$ & $0.51\pm0.07$ & $0.2\pm0.1$ &\\SMM 5 & $(1,\,1)$ & $9.16\pm0.002$ & $0.35\pm0.004$ & $3.0\pm0.5$ & $2.28\pm0.08$ & $6.6\pm0.6$\\
      & $(2,\,2)$ & $9.21\pm0.01$ & $0.31\pm0.03$ & $0.5\pm0.1$ &\\
SMM 6 & $(1,\,1)$ & $9.38\pm0.001$ & $0.29\pm0.002$ & $3.4\pm0.5$ & $4.66\pm0.11$ & $6.3\pm0.5$\\
      & $(2,\,2)$ & $9.41\pm0.004$ & $0.30\pm0.01$ & $0.9\pm0.1$ &\\
SMM 6 (20 MHz BW) & $(1,\,1)$ & $9.44\pm0.001$ & $0.26\pm0.002$ & $3.3\pm0.5$ & $5.38\pm0.13$ & $6.2\pm0.5$\\
Ori B9 N & $(1,\,1)$ & $9.14\pm0.004$ & $0.40\pm0.01$ & $1.1\pm0.2^c$ & $1.08\pm0.15$ & $5.0\pm0.4$\\
         & $(2,\,2)$ & $9.18\pm0.02$ & $0.21\pm0.04$ & $0.2\pm0.05$ &\\
Ori B9 N (2nd v-comp.) & $(1,\,1)$ & $1.93\pm0.01$ & $1.50\pm0.04$ & $0.7\pm0.1$& $0.37\pm0.09$ & $5.1\pm0.6$\\
                     & $(2,\,2)$ & $2.29\pm0.11$ & $2.53\pm0.28$ & $0.1\pm0.04$ &\\
SMM 7 & $(1,\,1)$ & $3.60\pm0.01$ & $0.60\pm0.02$ & $1.2\pm0.2$ & $2.14\pm0.14$ & $4.1\pm0.2$\\
      & $(2,\,2)$ & $3.89\pm0.10$ & $0.67\pm0.14$ & $0.1\pm0.05$ &\\
IRAS 05413-0104 & $(1,\,1)$ & $1.48\pm0.003$ & $0.46\pm0.01$ & $1.8\pm0.3$ & $1.68\pm0.09$ & $5.2\pm0.4$\\
                & $(2,\,2)$ & $1.52\pm0.02$ & $0.53\pm0.04$ & $0.4\pm0.1$ &\\
\hline 
\end{tabular} 
\end{minipage}
\end{table*}

\subsection{N$_2$H$^+$}

\begin{figure*}
\begin{center}
\includegraphics[width=3.2cm, angle=-90]{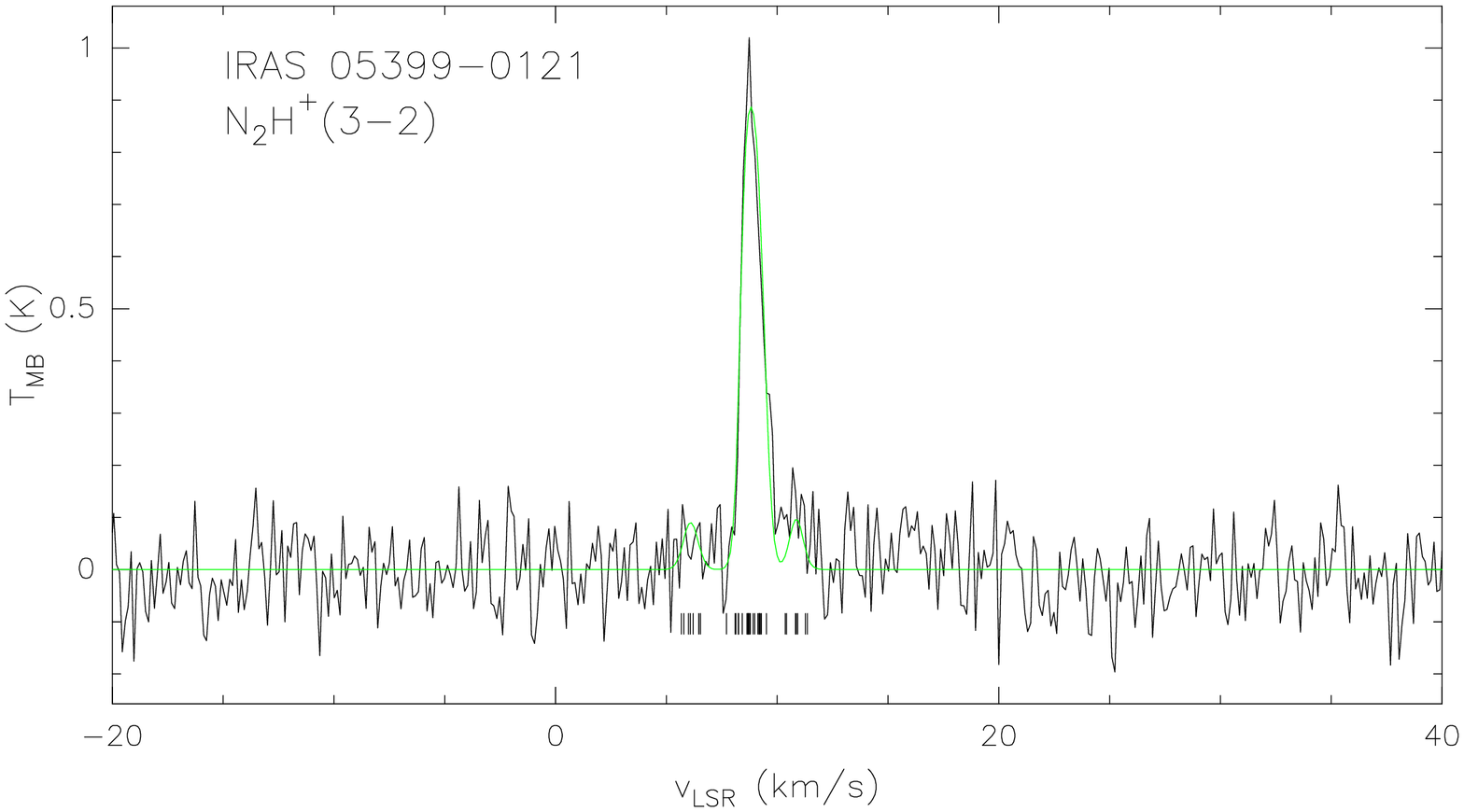}
\includegraphics[width=3.2cm, angle=-90]{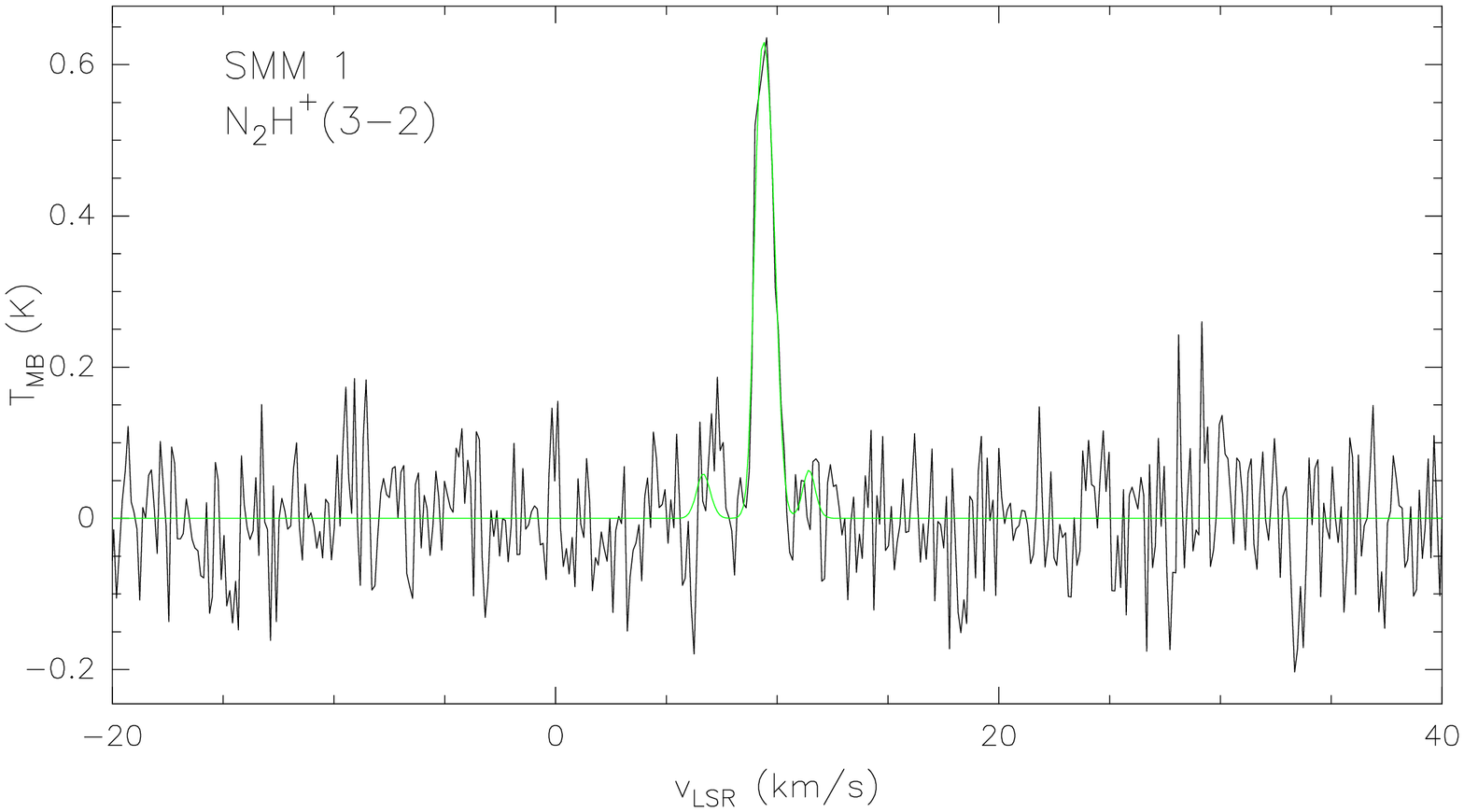}
\includegraphics[width=3.2cm, angle=-90]{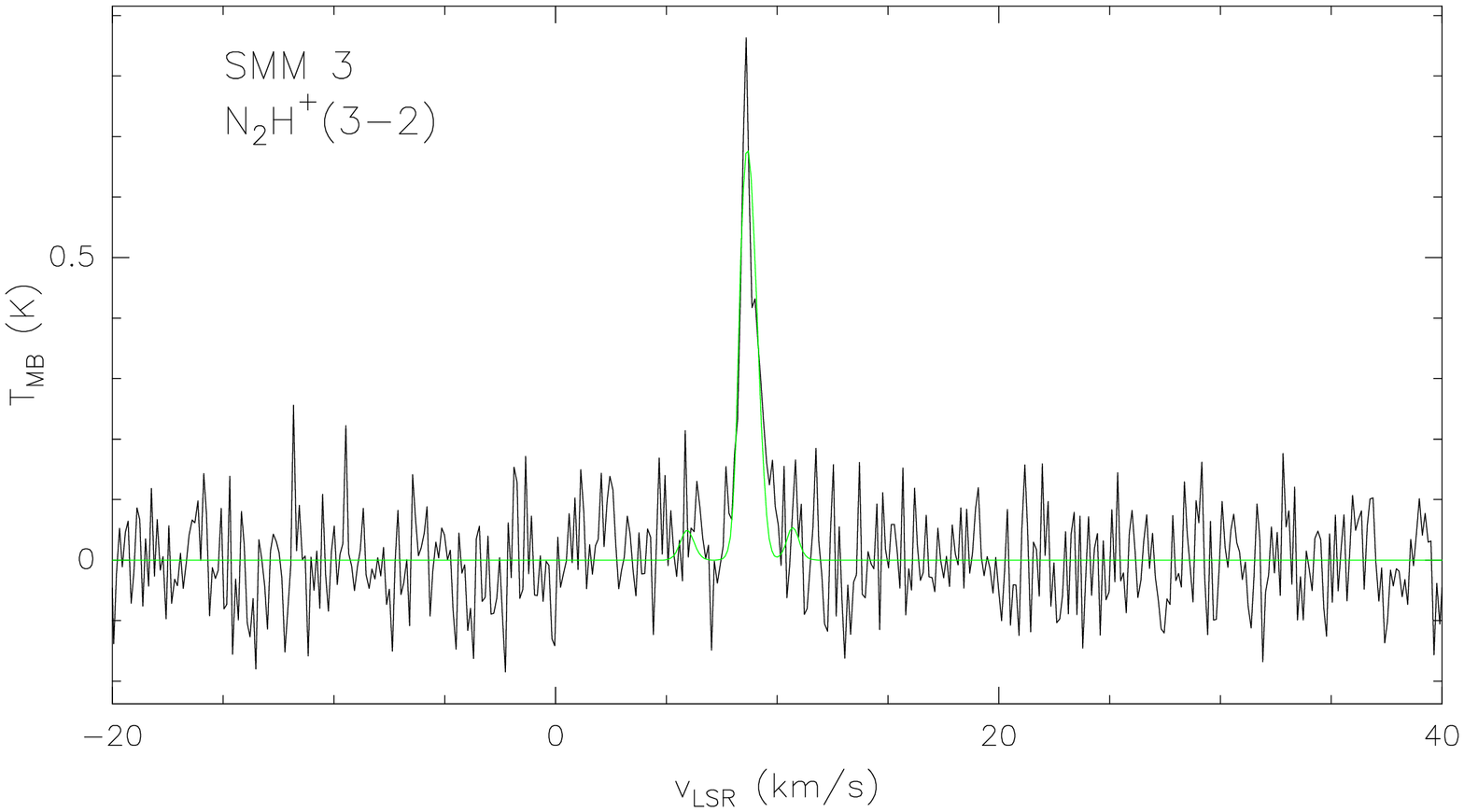}
\includegraphics[width=3.2cm, angle=-90]{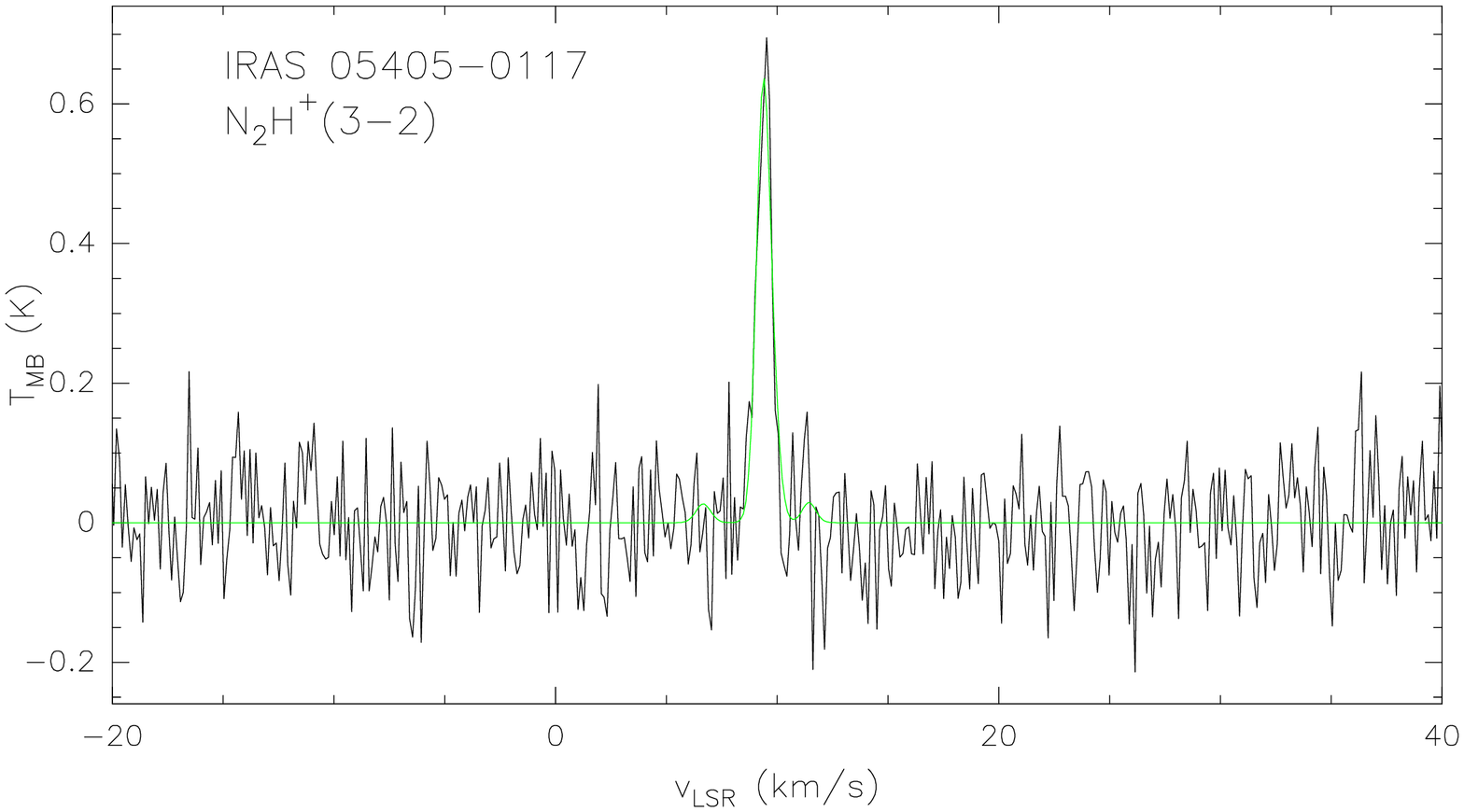}
\includegraphics[width=3.2cm, angle=-90]{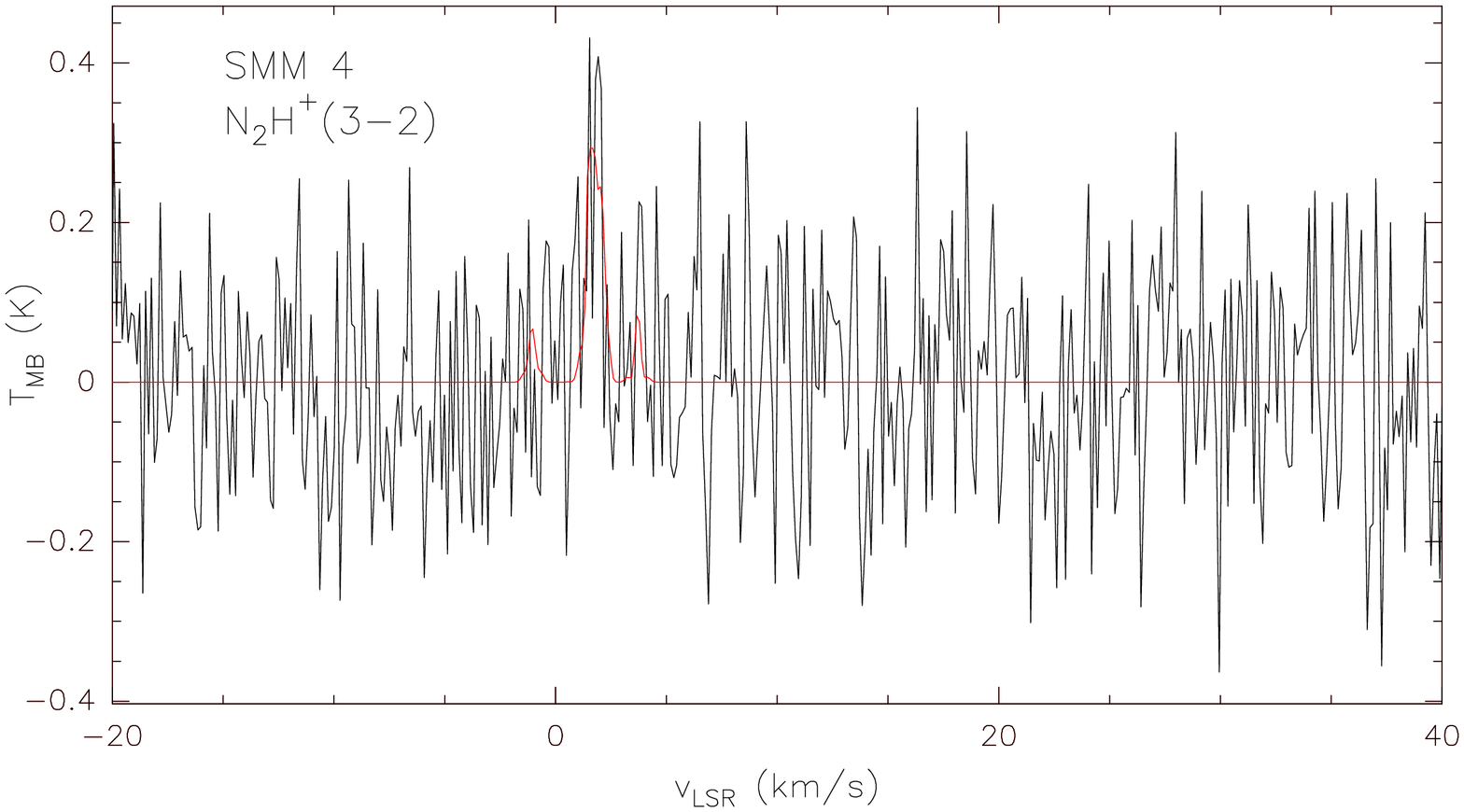}
\includegraphics[width=3.2cm, angle=-90]{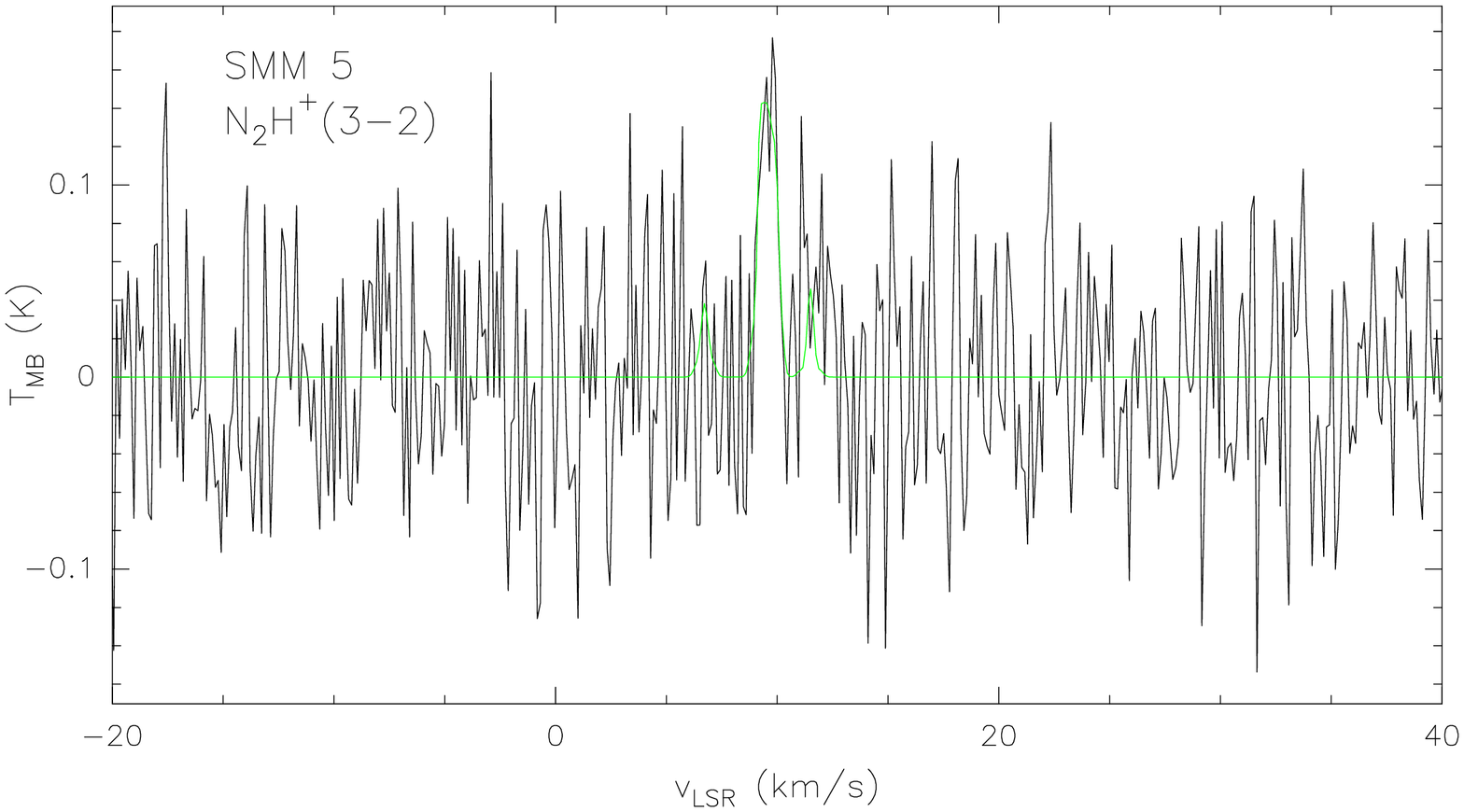}
\includegraphics[width=3.2cm, angle=-90]{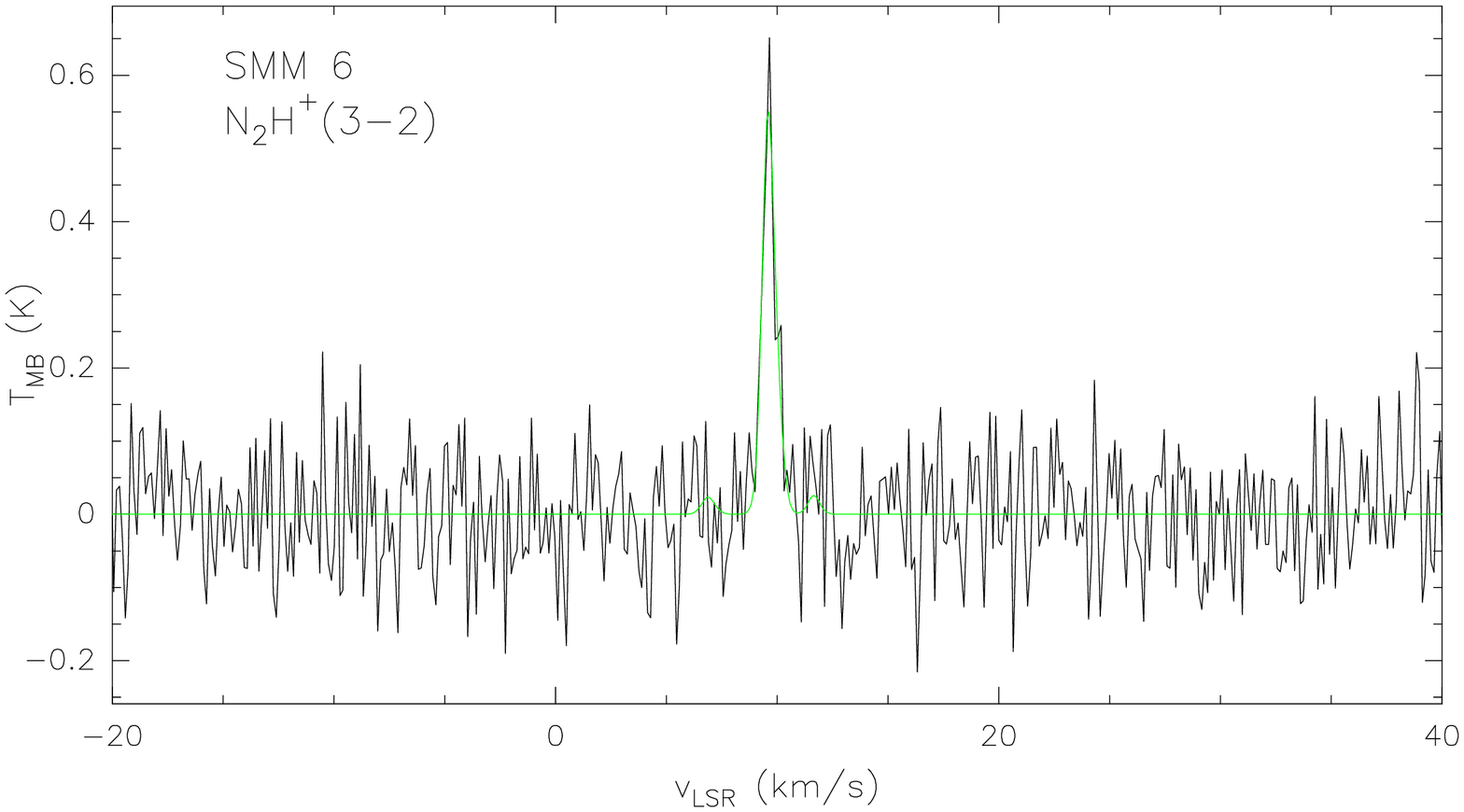}
\includegraphics[width=3.2cm, angle=-90]{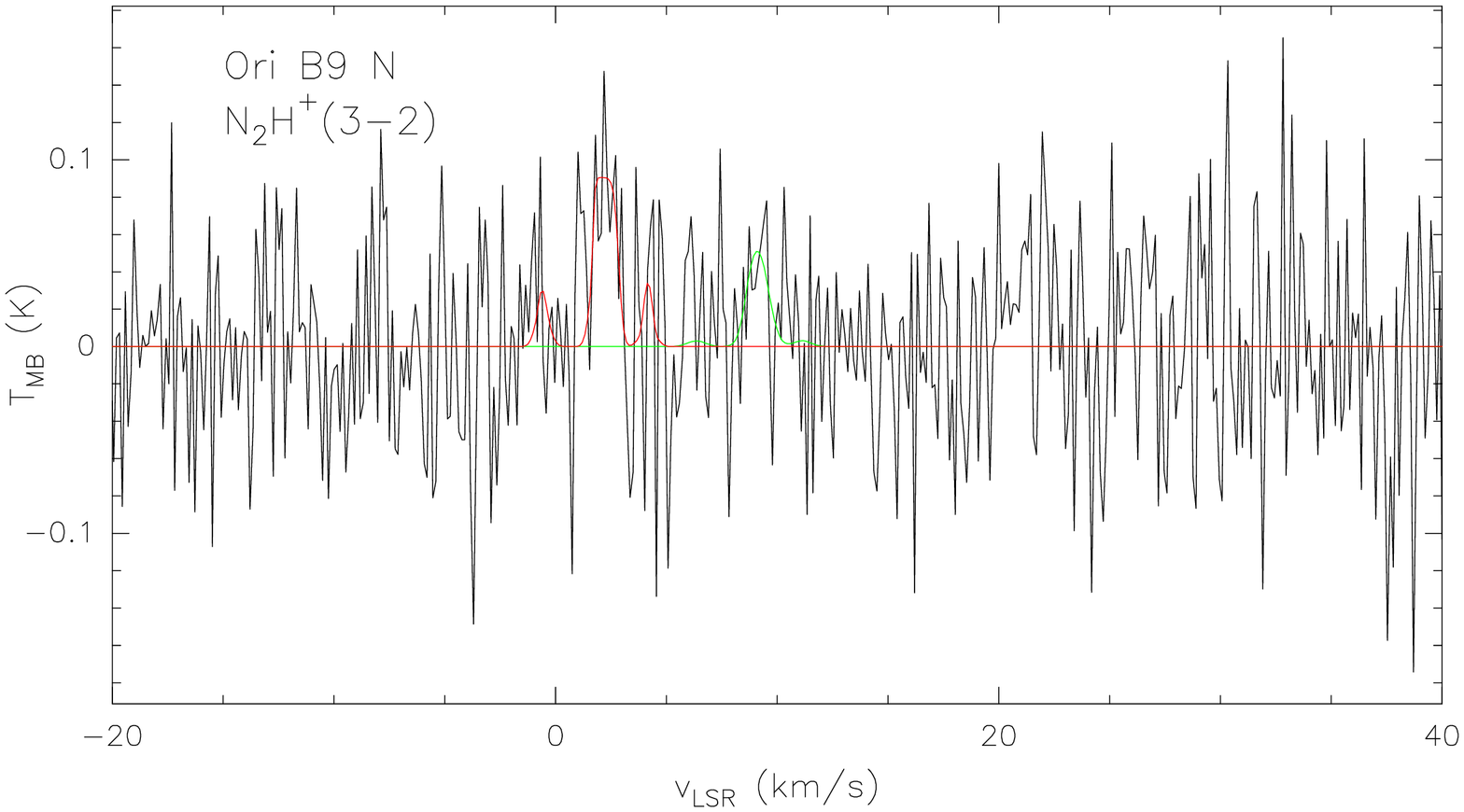}
\includegraphics[width=3.2cm, angle=-90]{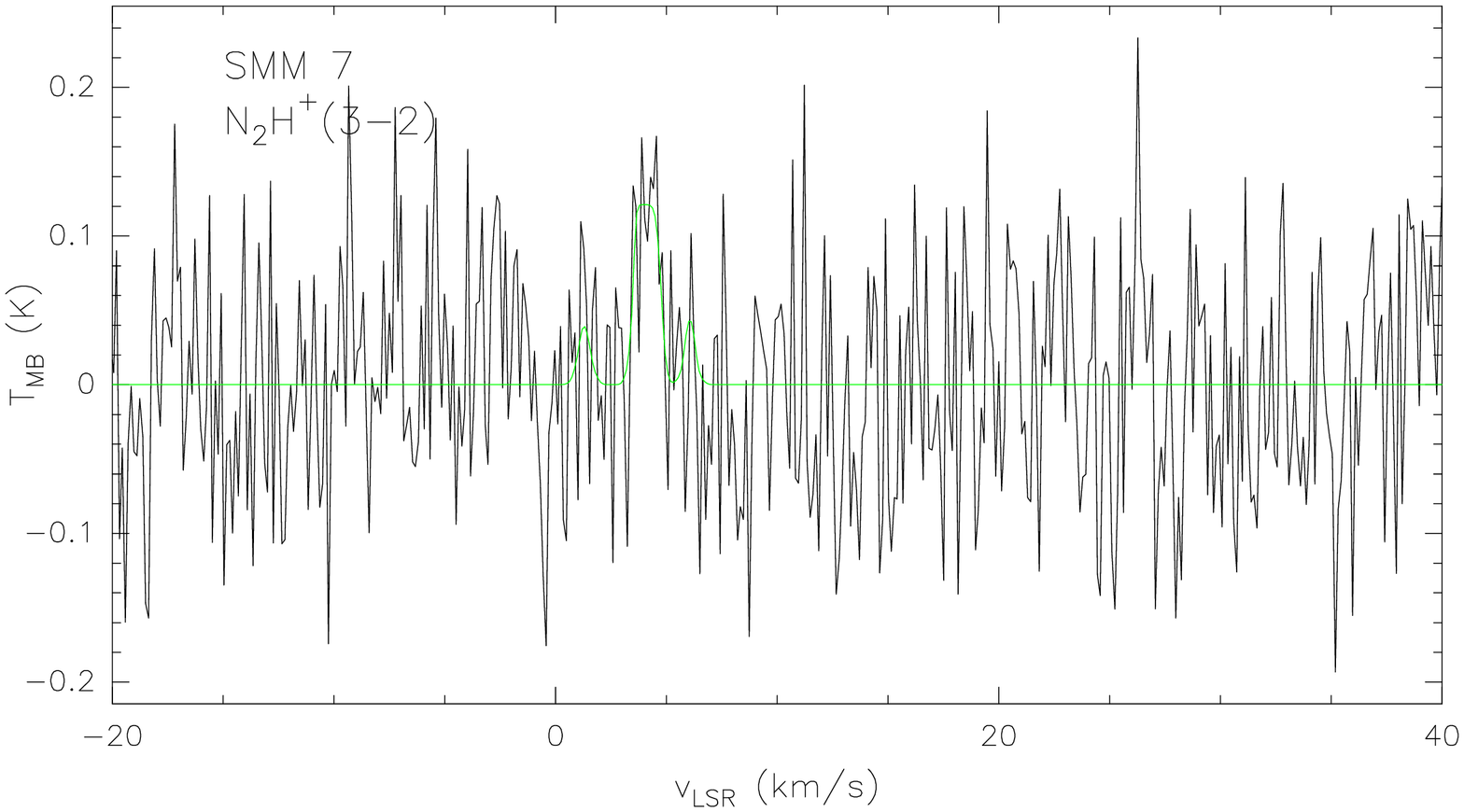}
\caption{N$_2$H$^+(3-2)$ spectra measured from the pre- and 
protostellar cores in Orion B9. Spectra are overlaid with the 38 component 
hyperfine structure fits. Hyperfine fits to the second velocity 
components are indicated by red lines. Note that only the second velocity 
component is detected towards SMM 4. The relative velocity of each 
individual hyperfine component is labelled with a short bar on the spectrum 
towards IRAS 05399-0121 (top left panel). The temperature scale is in 
$T_{\rm MB}$.} 
\label{figure:spectra2}
\end{center}
\end{figure*}

The N$_2$H$^+(3-2)$ spectra are shown in Fig.~\ref{figure:spectra2}.
In Table~\ref{table:diazparameters}, we show the line parameters obtained from 
the fits to the hyperfine structure. Here, the value of $\sigma(T_{\rm MB})$ 
includes only the $1\sigma$ rms noise in the spectrum (Col.~(4) of 
Table~\ref{table:diazparameters}).  

\begin{figure}[!h]
\resizebox{\hsize}{!}{\includegraphics{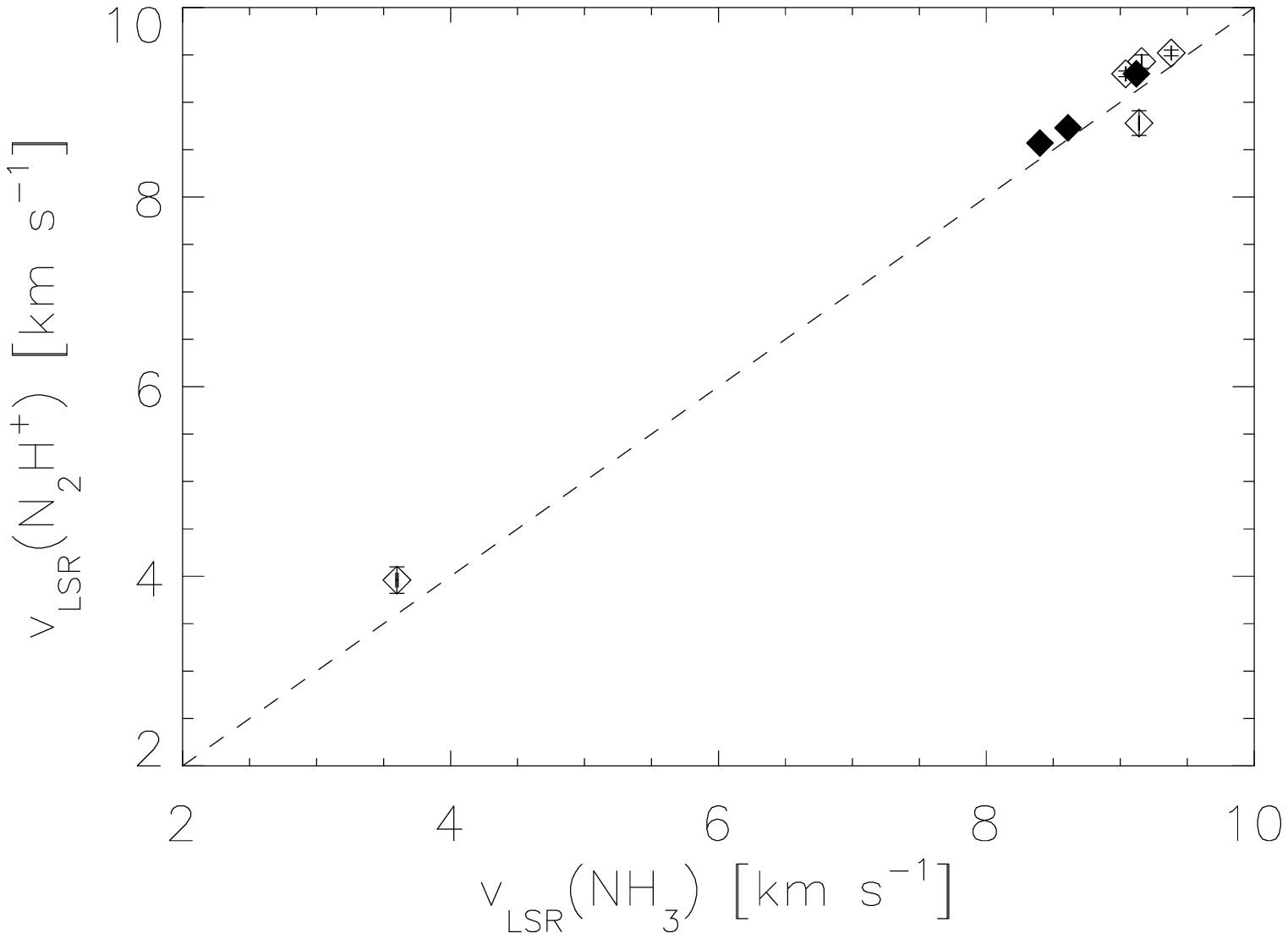}}
\resizebox{\hsize}{!}{\includegraphics{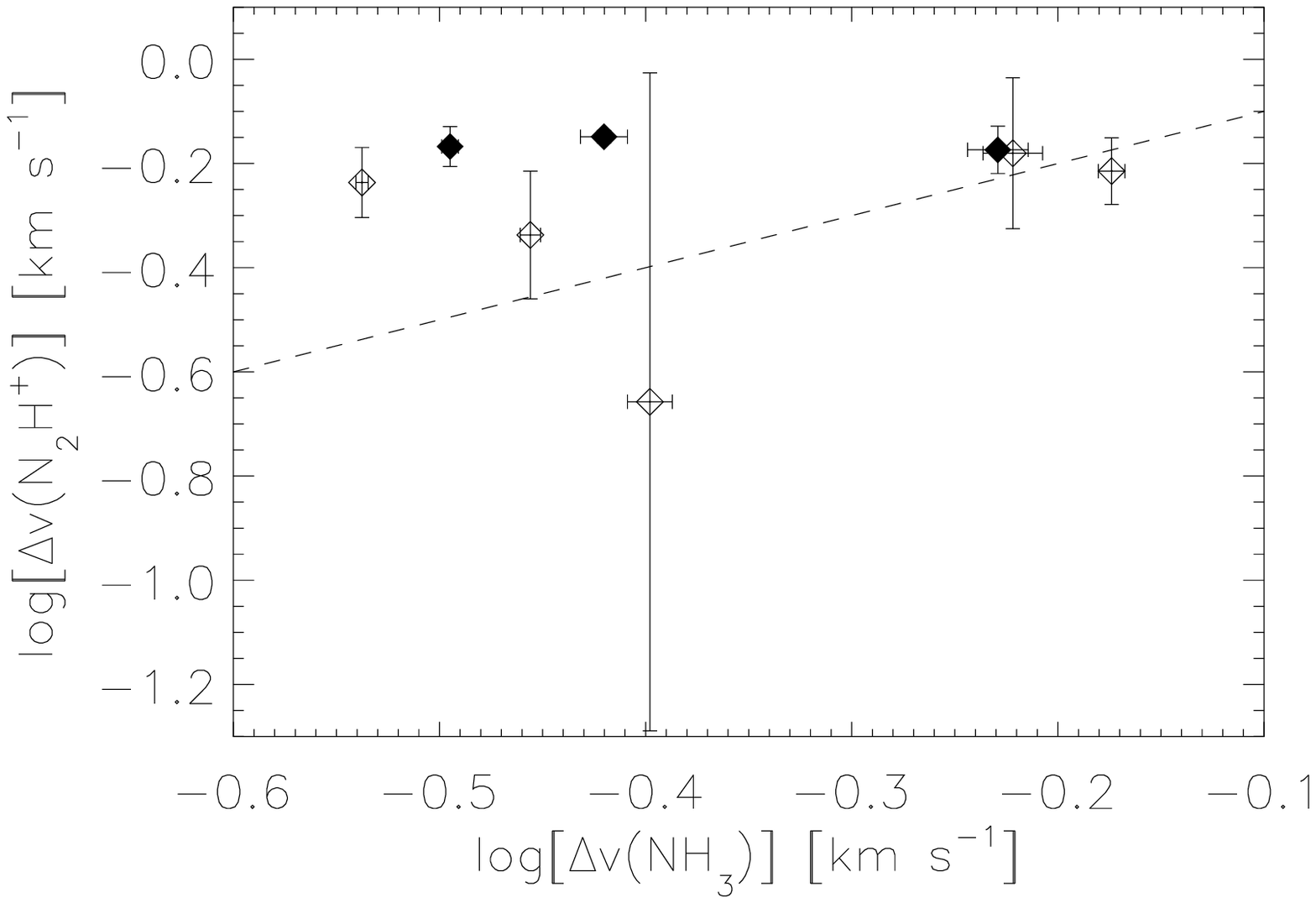}}
\caption{\textbf{Top:} The centroid velocity of the N$_2$H$^+(3-2)$ emission 
versus the velocity of the NH$_3(1,\,1)$ emission. \textbf{Bottom:} Log-log 
plot of the N$_2$H$^+(3-2)$ linewidth versus the NH$_3(1,\,1)$ linewidth.
Prestellar cores are indicated by open diamonds and protostellar cores 
are marked by filled diamonds. In both plots, the dashed line shows the 
equality.}
\label{figure:vcomparison}
\end{figure}

The LSR velocities of the N$_2$H$^+(3-2)$ lines are mostly similar to those 
determined from NH$_3(1,\,1)$, as can be seen in the top panel of 
Fig.~\ref{figure:vcomparison}. On the other hand, there is hardly any 
correlation between the N$_2$H$^+(3-2)$ and NH$_3(1,\,1)$ linewidths 
(bottom panel of Fig.~\ref{figure:vcomparison}). It should be noted,
however, that the hyperfine fitting to the N$_2$H$^+$ spectra is uncertain
due to strongly overlapping components and rather a poor signal-to-noise (S/N) 
ratio.

The N$_2$H$^+(3-2)$ spectra towards SMM 4 and Ori B9 N show the same lower
radial velocity components ($\lesssim2$ km~s$^{-1}$) as seen in NH$_3(1,\,1)$. 
On the other hand, the 'principal component' at $\sim9$ km~s$^{-1}$ is not
detected towards SMM 4 and it is also very weak towards Ori B9 N. 
Similarly, in Paper I we found that the N$_2$H$^+(1-0)$ spectra towards the 
selected position near IRAS 05405-0117/SMM 4 and Ori B9 N have second velocity 
components at $\sim1.3$ and $\sim2.2$ km~s$^{-1}$, respectively. 
The latter position also had additional N$_2$D$^+(2-1)$ line centred at 
$\sim2.3$ km~s$^{-1}$. 

We note that Harju et al. (2006) probably detected N$_2$H$^+(4-3)$
towards IRAS 05405-0117. They were using a double sideband system where
a coincidence with DCO$^+(5-4)$ from image band could not be ruled out.
Using the rest frequencies of N$_2$H$^+$ rotational lines from Pagani et al. 
(2009), the peak velocity of the suggested N$_2$H$^+(4-3)$ line (9.25 
km~s$^{-1}$) is very similar to that of N$_2$H$^+(3-2)$ (9.30 km~s$^{-1}$). 
For DCO$^+(5-4)$ the peak velocity would be about 0.3 km~s$^{-1}$ lower 
(8.95 km~s$^{-1}$). The N$_2$H$^+(4-3)$ linewidth (0.34 km~s$^{-1}$) is similar
to that of NH$_3(1,\,1)$ (0.32 km~s$^{-1}$).

\begin{table*}
\caption{N$_2$H$^+(3-2)$ line parameters.}
\begin{minipage}{2\columnwidth}
\centering
\renewcommand{\footnoterule}{}
\label{table:diazparameters}
\begin{tabular}{c c c c c c}
\hline\hline 
       & ${\rm v}_{\rm LSR}$ & $\Delta {\rm v}$ & $T_{\rm MB}$\footnote{The error in $T_{\rm MB}$ is the $1\sigma$ rms noise error.} & $\tau_{\rm tot}$\footnote{$\tau_{\rm tot}$ is the total optical thickness of the hyperfine multiplet.} & $T_{\rm ex}$\\
Source & [km~s$^{-1}$] & [km~s$^{-1}$] & [K] & & [K]\\
\hline
IRAS 05399-0121 & $8.73\pm0.02$ & $0.67\pm0.07$ & $0.92\pm0.04$ & $3.08\pm1.10$ & $5.1\pm0.3$\\
SMM 1 & $9.30\pm0.03$ & $0.61\pm0.09$ & $0.66\pm0.07$ & $2.77\pm1.48$ & $4.6\pm0.3$\\
SMM 3 & $8.57\pm0.03$ & $0.71\pm0.02$ & $0.65\pm0.08$ & $0.92\pm0.19$ & $5.5\pm0.5$\\
IRAS 05405-0117 & $9.30\pm0.03$ & $0.68\pm0.06$ & $0.66\pm0.07$ & $1.48\pm0.34$\footnote{$\tau_{\rm tot}$ could not be determined through fitting the hyperfine structure. Instead, $\tau_{\rm tot}$ was calculated by assuming $T_{\rm ex}=5.0$ K, and taking into account that the line intensity in Col.~(4) accounts for 92.6\% of the total line strength.} & 5.0\footnote{Assumed value.}\\
SMM 4\footnote{The principal velocity component was not detected.} & - & - & - & - & - \\
SMM 4 (2nd v-comp.) & $1.60\pm0.10$ & $0.44\pm0.25$ & $0.35\pm0.13$ & $0.54\pm0.26^c$ & 5.0$^d$\\
SMM 5 & $9.43\pm0.07$ & $0.46\pm0.13$ & $0.17\pm0.06$ & $0.23\pm0.09^c$ & 5.0$^d$\\
SMM 6 & $9.52\pm0.03$ & $0.58\pm0.09$ & $0.53\pm0.08$ & $0.99\pm0.24^c$ & 5.0$^d$\\
Ori B9 N & $8.78\pm0.13$ & $0.22\pm0.32$\footnote{The associated error is larger than the value.} & $<0.07$ & $<0.10^c$ & 5.0$^d$\\
Ori B9 N (2nd v-comp.) & $2.07\pm0.10$ & $0.43\pm0.22$ & $0.10\pm0.05$ & $0.13\pm0.07^c$ & 5.0$^d$\\
SMM 7 & $3.96\pm0.14$ & $0.66\pm0.22$ & $0.13\pm0.08$ & $0.17\pm0.11^c$ & 5.0$^d$\\
\hline 
\end{tabular} 
\end{minipage}
\end{table*}

\section{Analysis}

\subsection{NH$_3$ analysis}

In this subsection, we derive physical parameters from NH$_3$ data.  
The obtained results are given in Col.~(7) of Table~\ref{table:parameters1} 
($T_{\rm ex}$), and in Table~\ref{table:parameters2}. The analysis follows the 
concept first presented in Ho et al. (1979) and further discussed in Ho \&
Townes (1983).  Here it is assumed that the same excitation
temperature, $T_{\rm ex}$, characterises all the hyperfine transitions
between the $(1,\,1)$ and $(2,\,2)$ inversion doublets.

\subsubsection{Excitation, rotation, and kinetic temperatures}

The $T_{\rm ex}$ was calculated at the $(1,\,1)$ line peak
using the optical thickness, $\tau_{\rm peak}$, at that velocity, from the
equation

\begin{equation}
\label{eq:Tex}
T_{\rm ex}=\frac{T_{11}}{\ln \left[1+\frac{T_{\rm MB}({\rm v})}{T_{11}}\frac{1}{1-{\rm e}^{-\tau({\rm v})}}+F(T_{\rm bg})\right]}\,,
\end{equation}
where $T_{11} = h\nu_{11}/k_{\rm B}$, $h$ is the Planck constant, $\nu_{11}$ 
is the $(1,\,1)$ line frequency, $k_{\rm B}$ 
is the Boltzmann constant, $T_{\rm bg}=2.73$ K is the background temperature, 
and $F(T)=\left({\rm e}^{h\nu/k_{\rm B}T}-1\right)^{-1}$. The uncertainty 
in $T_{\rm ex}$ was calculated by propagating the errors in the peak values 
of $T_{\rm MB}({\rm v})$ and $\tau({\rm v})$. 
The $\tau({\rm v})$ distribution was obtained using the $(1,\,1)$ hyperfine fit
from CLASS which yields the LSR velocity line centroid, the width of an
individual hyperfine component, and the main group optical thickness, 
$\tau_{\rm m}$ (see Sect.~3.1).

The rotational temperature, $T_{\rm rot}$, can be in principle calculated
using the formula

\begin{equation}
\label{eq:Trot2}
T_{\rm rot}=\frac{-41.5}{\ln \left[\frac{3}{5}\frac{N(2,\,2)}{N(1,\,1)}\right]}\, ,
\end{equation}
where $N(1,\,1)$ and $N(2,\,2)$ are the column densities of ammonia molecules
at the $(1,\,1)$ and $(2,\,2)$ rotational levels, respectively. The estimation 
of these column densities is described in Sect.~4.1.2. The ratio 3/5 refers to 
the ratio of the statistical weights of the two states ($g(1,\,1)/g(2,\,2)$).

When the optical thickness, $\tau_{\rm m}$, is known, Eq.~(\ref{eq:Trot2}) can 
be written as 

\begin{equation}
\label{eq:Trot}
T_{\rm rot}=\frac{-41.5}{\ln \left\{\frac{-0.283}{\tau_{\rm m}}\ln \left[1-\frac{T_{\rm MB}(2,\,2)}{T_{\rm MB}(1,\,1)}\left(1-{\rm e}^{-\tau_{\rm m}}\right)\right]\right\}} 
\end{equation}
(\cite{ho1983}; their Eq.~(4)). Here it is assumed that $T_{\rm ex}$ and 
$\Delta {\rm v}$ are the same 
for both the $(1,\,1)$ and $(2,\,2)$ transitions. The error associated with 
$T_{\rm rot}$ was propagated from the uncertainties in $T_{\rm MB}(1,\,1)$, 
$T_{\rm MB}(2,\,2)$, and $\tau_{\rm m}$. 
When $\tau_{\rm m}$ could not be reliably determined (SMM 4, Ori B9 N, 
and their additional velocity components), we estimated the 
NH$_3(1,\,1)$ and NH$_3(2,\,2)$ column densities from the 
integrated intensities assuming $T_{\rm ex} = 5$ K, as described below. 
In these cases, $T_{\rm rot}$ was also calculated using Eq.~(\ref{eq:Trot2}), 
and the uncertainty in $T_{\rm rot}$ was propagated from the uncertainties in 
the column densities. The values of $T_{\rm rot}$ calculated 
using Eq.~(\ref{eq:Trot2}) are mostly similar to those 
resulting from Eq.~(\ref{eq:Trot}) within the errors (see 
Table~\ref{table:parameters2}). The subsequent analysis includes only those 
$T_{\rm rot}$ values which could be derived using Eq.~(\ref{eq:Trot}).

The gas kinetic temperature, $T_{\rm kin}$, was calculated from $T_{\rm rot}$ 
using the relationship given by Tafalla et al. (2004): 

\begin{equation}
\label{eq:Tkin}
T_{\rm kin}=\frac{T_{\rm rot}}{1-\frac{T_{\rm rot}}{42}\ln \left(1+1.1{\rm e}^{-\frac{16}{T_{\rm rot}}}\right)} \,.
\end{equation}
This relationship is recommended for dense cores with gas temperatures 
$\lesssim20$ K. Uncertainty in $T_{\rm kin}$ was propagated from the 
uncertainty in $T_{\rm rot}$.

\subsubsection{NH$_3$ column density calculations}

The column density in the $(J,\,K)=(1,\,1)$ state was calculated using the 
formula (see \cite{harju1993}; Eqs.~(3) and (5) therein)

\begin{eqnarray}
\label{eq:N11}
\lefteqn{N(1,\,1)=\frac{3h\epsilon_0}{2\pi^2\mu^2}\frac{\sqrt{\pi}}{2\sqrt{\ln 2}}\frac{J(J+1)}{K^2}F(T_{\rm ex})\left({\rm e}^{T_{11}/T_{\rm ex}}+1\right)\Delta {\rm v}\tau_{\rm tot}}
\nonumber\\
&=7.83\times10^{12}F(T_{\rm ex})\left({\rm e}^{1.14/T_{\rm ex}}+1\right)\Delta {\rm v}[{\rm km~s^{-1}}]\tau_{\rm tot}\: {\rm cm^{-2}}\,,
\end{eqnarray}
where $\epsilon_0$ is the vacuum permittivity, $\mu$ is the permanent 
electric dipole moment (1.476 D), and $\tau_{\rm tot}=2\times \tau_{\rm m}$.

For SMM 4 and Ori B9 N, and their second velocity components 
(for which $\tau_{\rm m}$ is uncertain in all the cases), $N(1,\,1)$ was 
also calculated 
by assuming optically thin emission ($\tau \ll 1$), and using the 
integrated intensities of clean satellites, i.e., those where the two velocity 
components do not overlap. There are three 
satellite lines in the NH$_3(1,\,1)$ spectra of both SMM 4 and Ori B9 N that 
do not suffer from the contamination of the second velocity component, 
i.e., $F_1=1-0$, $2-1$, and $0-1$ (see Fig.~\ref{figure:spectra}). 
For the second velocity components, the corresponding satellite lines are 
$F_1=1-0$, $1-2$, and $0-1$. Taking into account that these lines comprise 
$R_i=36.1\%$ of the total line strength, and assuming that the filling 
fraction of the emission in the beam is $\eta_{\rm f}=1$, we can derive the 
formula

\begin{eqnarray}
\label{eq:N11thin}
\lefteqn{N(1,\,1)=\frac{3k_{\rm B}\epsilon_0}{2\pi^2\mu^2\nu_{11}}\frac{J(J+1)}{K^2}\frac{{\rm e}^{T_{11}/T_{\rm ex}}+1}{1-\frac{F(T_{\rm bg})}{F(T_{\rm ex})}}\frac{1}{R_i}\int T_{\rm MB}(1,\,1; {\rm s}){\rm dv}}
\nonumber\\
&=6.47\times10^{12}\frac{{\rm e}^{1.14/T_{\rm ex}}+1}{1-\frac{F(T_{\rm bg})}{F(T_{\rm ex})}}\frac{1}{R_i}\int T_{\rm MB}(1,\,1; {\rm s}){\rm dv} [{\rm K~km~s^{-1}}]\: {\rm cm^{-2}}\,,
\end{eqnarray}
where $T_{\rm MB}$ is integrated over the three above mentioned satellite 
lines (the integrated satellite intensities are $2.53\pm0.03$ and 
$0.74\pm0.03$ K~km~s$^{-1}$ for SMM 4 and Ori B9 N, respectively;
for the corresponding second velocity components, the values are 
$2.18\pm0.02$ and $1.18\pm0.03$ K~km~s$^{-1}$, respectively).
For this calculation, the excitation temperature was assumed to be 5 K for 
both SMM 4 and Ori B9 N (and their second velocity components).
This assumption is similar to the values calculated from Eq.~(\ref{eq:Tex})
in all the other cases except the principal velocity 
component towards SMM 4, where $T_{\rm ex}$ might be $>5$ K (see Col.~(7) of 
Table~\ref{table:parameters1}).
Similarly, $N(2,\,2)$ (needed in the calculation of $T_{\rm rot}$ using 
Eq.~(\ref{eq:Trot2})) was calculated by assuming that 
$\tau \ll 1$, and using the integrated intensity of the main $(2,\,2)$ 
hyperfine complex, which accounts for $R_i=79.63\%$ of the total line 
strength. In this case, we can write

\begin{eqnarray}
\label{eq:N22thin}
\lefteqn{N(2,\,2)=\frac{3k_{\rm B}\epsilon_0}{2\pi^2\mu^2\nu_{22}}\frac{J(J+1)}{K^2}\frac{{\rm e}^{T_{22}/T_{\rm ex}}+1}{1-\frac{F(T_{\rm bg})}{F(T_{\rm ex})}}\frac{1}{R_i}\int T_{\rm MB}(2,\,2; {\rm m}){\rm dv}}
\nonumber\\
&=4.85\times10^{12}\frac{{\rm e}^{1.14/T_{\rm ex}}+1}{1-\frac{F(T_{\rm bg})}{F(T_{\rm ex})}}\frac{1}{R_i}\int T_{\rm MB}(2,\,2; {\rm m}){\rm dv}[{\rm K~km~s^{-1}}]\: {\rm cm^{-2}}\,,
\end{eqnarray}
where $\nu_{22}$ is the $(2,\,2)$ line frequency, and 
$T_{22}=h\nu_{22}/k_{\rm B}$. 
The integrated intensities of the $(2,\,2)$ main line for SMM 4 and Ori B 9 N 
are $0.21\pm0.01$ and $0.05\pm0.01$ K~km~s$^{-1}$, respectively ($0.15\pm0.01$ 
and $0.21\pm0.03$ K~km~s$^{-1}$ for the corresponding second velocity 
components).
The uncertainties associated with $N(1,\,1)$ and $N(2,\,2)$ were propagated 
from the formal errors in the integrated intensities.

The total NH$_3$ column density, $N({\rm NH_3})$, was calculated by scaling 
the column density in the $(1,\,1)$ state by the ratio 
$N({\rm NH_3})/N(1,\,1)$ obtained from the partition function $Z$. This yields 
approximately (e.g., \cite{rosolowsky2008}; \cite{busquet2009})

\begin{equation}
\label{eq:Ntot}
N({\rm NH_3})=N(1,\,1)\left(\frac{1}{3}{\rm e}^{\frac{23.4}{T_{\rm rot}}}+1+\frac{5}{3}{\rm e}^{-\frac{41.5}{T_{\rm rot}}}+\frac{14}{3}{\rm e}^{-\frac{101.2}{T_{\rm rot}}}\right)\, .
\end{equation}
The uncertainty in $N({\rm NH_3})$ was 
calculated by propagating the errors associated with 
$T_{\rm ex}$, $\Delta {\rm v}$, $\tau_{\rm m}$, and $T_{\rm rot}$. The values 
of $N({\rm NH_3})$ calculated by substituting $N(1,\,1)$ from 
Eq.~(\ref{eq:N11thin}) into Eq.~(\ref{eq:Ntot}) are mostly in good 
agreement, within the errors, with those resulting from Eqs.~(\ref{eq:N11}) 
and (\ref{eq:Ntot}). In the subsequent analysis, we only use NH$_3$ column 
densities derived for each source from Eqs.~(\ref{eq:N11}) and (\ref{eq:Ntot}) 
to keep the data set homogeneous (see Col.~(4) of 
Table~\ref{table:parameters2}).

\subsubsection{Fractional NH$_3$ abundance}

The fractional NH$_3$ abundance was calculated by dividing the total 
NH$_3$ column density by the H$_2$ column density, i.e., 
$x({\rm NH_3})=N({\rm NH_3})/N({\rm H_2})$. The H$_2$ column densities were 
determined from the submm dust continuum emission (Paper I; Eq.~(3) therein), 
using the kinetic temperatures derived in the present paper and assuming that
$T_{\rm kin}=T_{\rm dust}$ (see also Sect.~4.3). We smoothed the LABOCA 870 
$\mu$m map to correspond the $40\arcsec$ resolution of the Effelsberg NH$_3$ 
observations.

\subsubsection{Non-thermal velocity dispersion and the level of internal 
turbulence}

The measured NH$_3(1,\,1)$ linewidths were used to 
calculate the non-thermal portion of the line-of-sight velocity dispersion 
(averaged over a $\sim40\arcsec$ beam), and the level of internal turbulence. 
The observed velocity dispersion is related to the FWHM linewidth as 
$\sigma_{\rm obs}=\Delta {\rm v}/\sqrt{8\ln 2}$. The non-thermal velocity 
dispersion can then be calculated as follows:

\begin{equation}
\label{eq:sigmaNT}
\sigma_{\rm NT}=\sqrt{\sigma_{\rm obs}^2-\frac{k_{\rm B}T_{\rm kin}}{\mu_{\rm mol}m_{\rm H}}} \, ,
\end{equation}
where $\mu_{\rm mol}$ is the mass of the emitting molecule in units of atomic 
mass number (17 for NH$_3$), and $m_{\rm H}$ is the mass of the hydrogen 
atom. Furthermore, the level of internal turbulence is 
given by $f_{\rm turb}=\sigma_{\rm NT}/c_{\rm s}$, where $c_{\rm s}$ is the 
one-dimensional isothermal sound speed (0.19 km~s$^{-1}$ in a 10 K H$_2$ gas 
with 10\% He). The errors in $\sigma_{\rm NT}$ and $f_{\rm turb}$ were derived
by propagating the errors in $\Delta {\rm v}$ and $T_{\rm kin}$. 

\begin{table*}
\caption{Parameters derived from NH$_3$ data.}
\begin{minipage}{2\columnwidth}
\centering
\renewcommand{\footnoterule}{}
\label{table:parameters2}
\begin{tabular}{c c c c c c c}
\hline\hline 
       & $T_{\rm rot}$ & $T_{\rm kin}$ & $N({\rm NH_3})$ & $x({\rm NH_3})$ & $\sigma_{\rm NT}$ & $f_{\rm turb}$ \\
Source & [K] & [K] & [$10^{14}$ cm$^{-2}$] & [$10^{-8}$] & [km~s$^{-1}$] & \\
\hline
IRAS 05399-0121 & $12.5\pm1.4$ & $13.5\pm1.6$ & $3.2\pm0.6$ & $1.5\pm0.4$ & $0.24\pm0.01$ & $1.1\pm0.1$ \\
SMM 1 & $11.2\pm0.8$ & $11.9\pm0.9$ & $6.2\pm0.9$ & $3.5\pm0.7$ & $0.27\pm0.004$ & $1.3\pm0.1$ \\
SMM 2 & $11.1\pm1.6$ & $11.8\pm1.8$ & $2.6\pm0.7$ & $4.1\pm0.7$ & $0.12\pm0.01$ & $0.6\pm0.1$ \\
SMM 3 & $10.7\pm0.7$ & $11.3\pm0.8$ & $5.6\pm0.8$ & $1.9\pm0.4$ & $0.14\pm0.003$ & $0.7\pm0.04$ \\
IRAS 05405-0117 & $10.7\pm0.6$ & $11.3\pm0.6$ & $10.0\pm1.3$ & $9.8\pm1.6$ & $0.11\pm0.002$ & $0.6\pm0.02$ \\
SMM 4 & $12.8\pm0.6$ & $13.9\pm0.8$ & $3.4\pm0.4$ & $3.8\pm0.6$ & $0.12\pm0.005$ & $0.5\pm0.03$ \\
      & $10.2\pm0.9$\footnote{Calculated by using the Eq.~(\ref{eq:Trot2}).} & $10.7\pm1.1^a$ & $8.8\pm1.4$\footnote{Calculated by using Eqs.~(\ref{eq:N11thin}) and (\ref{eq:Ntot}).} &\\
SMM 4 (2nd v-comp.) & $10.0\pm1.2$ & $10.4\pm1.4$ & $6.8\pm1.6$ & $4.5\pm1.6$ & $0.29\pm0.01$ & $1.5\pm0.1$\\
                    & $9.7\pm1.0^a$ & $10.2\pm1.1^a$ & $8.2\pm1.6^b$ & \\
SMM 5 & $10.7\pm0.6$ & $11.3\pm0.7$ & $5.9\pm0.7$ & $8.0\pm1.3$ & $0.13\pm0.002$ & $0.6\pm0.02$\\
SMM 6 & $10.4\pm0.3$ & $11.0\pm0.4$ & $9.9\pm0.9$ & $7.2\pm0.8$ & $0.10\pm0.001$ & $0.5\pm0.01$ \\
Ori B9 N & $12.4\pm1.1$ & $13.4\pm1.3$ & $2.0\pm0.4$ & $3.8\pm0.9$ & $0.15\pm0.01$ & $0.7\pm0.04$ \\
         & $9.7\pm3.6^a$ & $10.2\pm4.0^a$ & $2.8\pm1.9^b$ &\\
Ori B9 N (2nd v-comp.) & $12.5\pm2.1$ & $13.6\pm2.5$ & $2.5\pm0.8$ & $5.0\pm2.3$ & $0.63\pm0.02$ & $2.9\pm0.3$\\
                       & $12.5\pm1.7^a$ & $13.6\pm1.9^a$ & $3.0\pm0.5^b$ &\\
SMM 7 & $9.0\pm1.0$ & $9.4\pm1.1$ & $8.0\pm2.0$ & $5.3\pm1.9$ & $0.25\pm0.01$ & $1.3\pm0.1$ \\
IRAS 05413-0104 & $12.3\pm1.0$ & $13.4\pm1.2$ & $3.7\pm0.5$ & $3.4\pm0.7$ & $0.18\pm0.004$ & $0.8\pm0.04$ \\
\hline 
\end{tabular} 
\end{minipage}
\end{table*}

\subsection{N$_2$H$^+$ analysis}

In this subsection, we derive physical parameters from  N$_2$H$^+$ data.  
The obtained results are given in Col.~(6) of Table~\ref{table:diazparameters} 
($T_{\rm ex}$), and in Table~\ref{table:diazparameters2}.

\subsubsection{$T_{\rm ex}$, $\sigma_{\rm NT}$, and $f_{\rm turb}$}

The excitation temperature of the N$_2$H$^+(3-2)$ transition was calculated 
as in the case of NH$_3(1,\,1)$ (see Eq.~(\ref{eq:Tex})).
The optical thickness, and thus $T_{\rm ex}$, could be determined only for 
three sources: IRAS 05399-0121, and SMM 1 and 3. For the rest of the sources, 
it was assumed that $T_{\rm ex}=5$ K, and $\tau_{\rm tot}$ was then 
estimated using this value. The value $T_{\rm ex}=5$ K is expected 
to be a reasonable choice because for the above three sources $T_{\rm ex}$ 
is around 5 kelvins.

The values of $\sigma_{\rm NT}$ and $f_{\rm turb}$ were calculated  
as in the case of NH$_3(1,\,1)$ (see Eq.~(\ref{eq:sigmaNT})), by using the 
values of $T_{\rm kin}$ from NH$_3$ measurements. For N$_2$H$^+$, 
$\mu_{\rm mol}$ is 29.

\subsubsection{N$_2$H$^+$ column density and fractional abundance}

The N$_2$H$^+$ column density was calculated using the formula (see, e.g., 
Paper I)


\begin{eqnarray}
\label{eq:Ntot2}
\lefteqn{N({\rm N_2H^+})=\frac{3h\epsilon_0}{2\pi^2\mu^2}\frac{\sqrt{\pi}}{2\sqrt{\ln 2}}\frac{1}{J_{\rm u}}{\rm e}^{E_{\rm u}/k_{\rm B}T_{\rm ex}}F(T_{\rm ex})Z(T_{\rm ex})\Delta {\rm v}\tau_{\rm tot}}
\nonumber\\
&=2.46\times10^{11}{\rm e}^{26.83/T_{\rm ex}}F(T_{\rm ex})Z(T_{\rm ex})\Delta {\rm v}[{\rm km~s^{-1}}]\tau_{\rm tot}\: {\rm cm^{-2}}\,,
\end{eqnarray}
where $\mu$ is 3.4 D, the upper rotational level number $J_{\rm u}=3$, 
$E_{\rm u}=hBJ_{\rm u}(J_{\rm u}+1)$ is the energy 
of the upper transition state, and $B=46\,586.8713$ MHz is the rotational 
constant (\cite{pagani2009}). The rotational partition function was 
approximated by 
$Z(T_{\rm ex})=\left(\frac{k_{\rm B}T_{\rm ex}}{hB}+\frac{1}{3}\right)$.

In those cases where the line was optically thin (i.e., SMM 5, Ori B9 N and 
its second velocity component, and SMM 7), the N$_2$H$^+$ column density was 
also calculated from the integrated intensity. The integrated intensities 
obtained from Gaussian fits are $0.15\pm0.03$, $0.07\pm0.16$ ($0.12\pm0.04$), 
and $0.22\pm0.07$ K~km~s$^{-1}$ for SMM 5, Ori B9 N (and 
its second velocity component), and SMM 7, respectively. Note that the error 
is larger than the value for Ori B9 N. 
By combining Eqs.~(4) and (6) of Paper I, we get the following 
formula for the $N({\rm N_2H^+})$ as a function of integrated intensity:

\begin{eqnarray}
\label{eq:Ntot3}
\lefteqn{N({\rm N_2H^+})=\frac{3\epsilon_0 k_{\rm B}}{2\pi^2\mu^2\nu}\frac{1}{J_{\rm u}}{\rm e}^{E_{\rm u}/k_{\rm B}T_{\rm ex}}Z(T_{\rm ex})\frac{1}{1-\frac{F(T_{\rm bg})}{F(T_{\rm ex})}}\int T_{\rm MB}{\rm dv}}
\nonumber\\
&=1.72\times10^{10}Z(T_{\rm ex})\frac{{\rm e}^{26.83/T_{\rm ex}}}{1-\frac{F(T_{\rm bg})}{F(T_{\rm ex})}}\int T_{\rm MB}{\rm dv}[{\rm K~km~s^{-1}}]\: {\rm cm^{-2}}\, .
\end{eqnarray}
When applying Eq.~(\ref{eq:Ntot3}), it was taken into account that the 
detectable emission feature with 24 hyperfine components contains
92.6\% of the total line strength. The values of $N({\rm N_2H^+})$ calculated 
from Eqs.~(\ref{eq:Ntot2}) and (\ref{eq:Ntot3}) are similar to each other 
within the errors (see Table~\ref{table:diazparameters2}).

The fractional N$_2$H$^+$ abundance was calculated in a similar manner as for 
NH$_3$. In this case, the LABOCA map was smoothed to the resolution
of the N$_2$H$^+$ observations (22\farcs3).

\begin{table*}
\caption{Parameters derived from N$_2$H$^+(3-2)$ data.}
\begin{minipage}{2\columnwidth}
\centering
\renewcommand{\footnoterule}{}
\label{table:diazparameters2}
\begin{tabular}{c c c c c c}
\hline\hline 
       & $N({\rm N_2H^+})$ & $x({\rm N_2H^+})$ & $N({\rm NH_3})/N({\rm N_2H^+})$ & $\sigma_{\rm NT}$ & $f_{\rm turb}$ \\
Source & [$10^{13}$ cm$^{-2}$] & [$10^{-10}$] & & [km~s$^{-1}$] & \\
\hline
IRAS 05399-0121 & $2.0\pm0.7$ & $4.3\pm1.8$ & $16\pm7$ & $0.28\pm0.03$ & $1.3\pm0.2$ \\
SMM 1 & $1.9\pm1.1$ & $5.9\pm3.4$ & $32\pm18$ & $0.25\pm0.04$ & $1.2\pm0.2$ \\
SMM 3 & $0.6\pm0.1$ & $0.8\pm0.2$ & $99\pm25$ & $0.30\pm0.01$ & $1.5\pm0.1$ \\
IRAS 05405-0117 & $1.0\pm0.2$\footnote{In Paper I, we derived the value $N({\rm N_2H^+})\approx0.91\pm0.01\times10^{13}$ cm$^{-2}$ from observations of N$_2$H$^+(1-0)$ emission, which is in good agreement with the present value.} & $5.5\pm1.5$ & $100\pm28$ & $0.28\pm0.03$ & $1.4\pm0.1$ \\
SMM 4\footnote{The principal velocity component was not detected.} & - & - & - & - & -\\
SMM 4 (2nd v-comp.) & $0.2\pm0.2$ & $0.8\pm0.6$ & $288\pm225$ & $0.18\pm0.11$ & $0.9\pm0.6$\\
SMM 5 & $0.1\pm0.1$ & $0.7\pm0.4$ & $562\pm279$ & $0.19\pm0.06$ & $0.9\pm0.3$ \\      & $0.2\pm0.03$\footnote{Calculated by using the optically thin approximation.} & $1.2\pm0.2^c$ & $275\pm54^c$ & \\
SMM 6 & $0.6\pm0.2$ & $2.1\pm0.6$ & $194\pm80$ & $0.24\pm0.04$ & $1.2\pm0.2$ \\
Ori B9 N & $0.02\pm0.03$ & $0.2\pm0.3$ & $916\pm1345$\footnote{The associated error is larger than the value.} & $0.07\pm0.18^d$ & $0.3\pm0.8$ \\
         & $0.1\pm0.2^{c,d}$ & $0.8\pm0.1^c$ & $280\pm190^c$ &\\
Ori B9 N (2nd v-comp.) & $0.1\pm0.04$ & $0.6\pm0.5$ & $250\pm90$ & $0.17\pm0.10$ & $0.8\pm0.5$\\
        & $0.1\pm0.05^c$ & $1.1\pm0.4^c$ & $250\pm154^c$ &\\
SMM 7 & $0.1\pm0.1$ & $0.3\pm0.2$ & $718\pm553$ & $0.28\pm0.10$ & $1.5\pm0.5$ \\
      & $0.3\pm0.1^c$ & $0.7\pm0.3^c$ & $267\pm91^c$ & \\
\hline 
\end{tabular} 
\end{minipage}
\end{table*}

\subsection{Revision of core properties presented in Paper I}

In Paper I, we assumed that the dust temperature is $T_{\rm dust}=10$ K for 
the starless cores, whereas for protostellar cores we used
the temperatures derived from the spectral energy distribution (SED)
fits (Table 6 in Paper I). By using the gas kinetic temperatures determined in 
the present paper, we recalculated several parameters presented in Paper I, 
by assuming that $T_{\rm kin}=T_{\rm dust}$. 
These include the core mass, $M$, H$_2$ column 
density, $N({\rm H_2})$, and the average H$_2$ number density, 
$\langle n({\rm H_2}) \rangle$. In the present paper, we have also amended 
some formula used in the derivation of the physical parameters: in the 
calculation of $\langle n({\rm H_2}) \rangle$, we have used the
effective radius $R_{\rm eff}=\sqrt{A/\pi}$ (instead of the FWHM radius
used in Paper I), and the value $2.8m_{\rm H}$ for the mean molecular weight 
per H$_2$ molecule (instead of $2.33m_{\rm H}$ which is the mean particle 
weight in an H$_2$ + 10\%  He mixture). The former change is motivated by the 
fact that the flux densities used to calculate the core masses refer to
their projected areas, $A$. The results of these calculations are
presented in Table~\ref{table:parameters3}. The revised
values of $M$ and $N({\rm H_2})$, are, on average, about 11\% and 
9\% higher than those reported in Paper I. On the other
hand, the new average densities, $\langle n({\rm H_2}) \rangle$, are only 
about 1/5 of those calculated in Paper I. Note that the total mass of the 
cores is still $\sim50$ M$_{\sun}$, i.e., about 4\% of the total mass in the 
region (Paper I).

In Paper I, we compared the core mass distributions in Orion B9 and Orion B 
North studied by Nutter \& Ward-Thompson (2007) (Sect.~5.2 and Fig.~7 in 
Paper I). The core mass functions (CMFs) were constructed by removing the 
Class I protostellar cores from the samples (i.e., IRAS 05399-0121 was removed 
from the Orion B9 core sample). We performed a two-sample Kolmogorov-Smirnov 
(K-S) test between the two CMFs to examine if they represent the subsamples of 
the same underlying parent distribution. For this purpose, the masses were 
scaled to compensate for the different assumptions about the distance, dust
temperature, and opacity. We found a very high likelihood of $\sim95\%$ for 
the null hypothesis that the two CMFs are drawn from the same parent 
distribution (i.e., the probability that the two samples do not have the same 
parent distribution is $1-0.95=0.05$). We repeated this analysis by using the 
updated masses listed in Col.~(3) of Table~\ref{table:parameters3}. 
In this case, a K-S test yielded a probability of $\sim100\%$, strengthening 
the possibility that the core masses in these two different parts of Orion B 
are drawn from the same distribution. 

\subsection{Virial masses}

In order to study the stability of the cores, we calculated their 
virial masses using the following formula where the effects of external 
pressure and magnetic field are ignored:

\begin{equation}
\label{eq:Mvir}
M_{\rm vir}=\frac{5}{8\ln 2}\frac{R_{\rm eff}\Delta {\rm v_{\rm ave}^2}}{aG} \, ,
\end{equation}
where $G$ is the gravitational constant, and $\Delta {\rm v_{\rm ave}}$ is the 
width of the spectral line emitted by the molecule of mean mass $\mu=2.33$. 
The parameter $a=(1-p/3)/(1-2p/5)$, where $p$ is the power-law index of the 
density profile ($n(r)\propto r^{-p}$), is a correction for deviations from 
constant density. For starless cores we used the value $p=1.0$, whereas for 
protostellar cores we set $p=1.5$ (see Paper I). The value of $M_{\rm vir}$ 
calculated with $p=1.0$ is about 13\% higher compared to that calculated 
with the value $p=1.5$. As a function of the observed
linewidth, $\Delta {\rm v_{\rm obs}}$ (Table~\ref{table:parameters1}, 
Col.~(4)), $\Delta {\rm v_{\rm ave}}$ is given by

\begin{equation}
\label{eq:Dv}
\Delta {\rm v_{\rm ave}^2}=\Delta {\rm v}_{\rm T}^2+\Delta {\rm v}_{\rm NT}^2=\Delta {\rm v_{\rm obs}^2}+8\ln 2 \times \frac{k_{\rm B}T_{\rm kin}}{m_{\rm H}}\left(\frac{1}{\mu}-\frac{1}{\mu_{\rm mol}} \right) \, .
\end{equation}
The virial masses are listed in Col.~(4) of 
Table~\ref{table:parameters3}. The associated error was propagated from 
those of $\Delta {\rm v_{\rm obs}}$ and $T_{\rm kin}$. 

The virial parameters of the cores were calculated following
Bertoldi \& McKee (1992), i.e., $\alpha_{\rm vir}=M_{\rm vir}/M$. 
The uncertainty was derived by propagating the errors in both mass estimates.
The values are given in Col.~(5) of Table~\ref{table:parameters3}. Note 
that $\alpha_{\rm vir}=1$ corresponds to the virial equilibrium, 
$2\langle T \rangle=-\langle U \rangle$, 
where $T$ and $U$ are the internal kinetic and gravitational energies, 
respectively. The value $\alpha_{\rm vir}=2$ corresponds to the 
self-gravitating limit defined by $\langle T \rangle=-\langle U \rangle$. 

\begin{table*}
\renewcommand{\footnoterule}{}
\caption{Revised core parameters presented in Paper I.}
\begin{minipage}{2\columnwidth}
\centering
\label{table:parameters3}
\begin{tabular}{c c c c c c c}
\hline\hline 
       & $R_{\rm eff}$ & $M$\footnote{The masses and H$_2$ column and number densities were calculated using the values of $T_{\rm kin}$ ($=T_{\rm dust}$).} & $M_{\rm vir}$\footnote{The table also lists the core virial masses and virial parameters (Cols.~(4) and (5)).} & $\alpha_{\rm vir}^b$ & $N({\rm H_2})^a$ & $\langle n({\rm H_2}) \rangle^a$\\
Source & [pc] & [M$_{\odot}$] & [M$_{\odot}$] & & [$10^{22}$ cm$^{-2}$] & [$10^4$ cm$^{-3}$]\\
\hline
IRAS 05399-0121 & 0.08 & $6.1\pm1.4$ & $7.8\pm0.5$ & $1.3\pm0.3$ & $4.2\pm0.9$ & $5.5\pm1.3$\\
SMM 1 & 0.09 & $8.4\pm1.5$ & $11.1\pm0.3$ & $1.3\pm0.2$ & $2.8\pm0.4$ & $5.3\pm0.9$\\
SMM 2 & 0.08 & $2.0\pm0.6$ & $4.8\pm0.5$ & $2.4\pm0.8$ & $1.4\pm0.4$ & $1.8\pm0.5$\\
SMM 3 & 0.07 & $7.8\pm1.6$ & $4.0\pm0.2$ & $0.5\pm0.1$ & $8.4\pm1.1$ & $10.5\pm2.1$\\
IRAS 05405-0117 & 0.07 & $2.8\pm0.4$ & $3.5\pm0.1$ & $1.2\pm0.2$ & $1.4\pm0.1$ & $3.8\pm0.5$\\
SMM 4 & 0.07 & $2.8\pm0.3$ & $4.1\pm0.2$ & $1.5\pm0.2$ & $1.4\pm0.1$ & $3.8\pm0.4$\\
SMM 4 (2nd v-comp.) & & & & & $2.4\pm0.7$ &\\
SMM 5 & 0.07 & $1.9\pm0.4$ & $4.2\pm0.2$ & $2.2\pm0.5$ & $1.2\pm0.1$ & $2.5\pm0.5$\\
SMM 6 & 0.12 & $8.2\pm1.1$ & $6.2\pm0.2$ & $0.8\pm0.1$ & $2.0\pm0.1$ & $2.2\pm0.3$\\
Ori B9 N & 0.08 & $2.3\pm0.4$ & $5.8\pm0.4$ & $2.5\pm0.5$ & $0.9\pm0.1$ & $2.1\pm0.4$\\
Ori B9 N (2nd v-comp.) & & & & & $0.9\pm0.1$ &\\
SMM 7 & 0.07 & $3.6\pm1.0$ & $6.9\pm0.4$ & $1.9\pm0.6$ & $3.4\pm0.9$ & $4.8\pm1.3$\\
IRAS 05413-0104 & 0.06 & $2.0\pm0.6$ & $4.4\pm0.2$ & $2.2\pm0.7$ & $3.6\pm0.6$ & $4.3\pm1.3$\\
\hline 
\end{tabular} 
\end{minipage}
\end{table*}

\section{Discussion}

\subsection{The gas kinetic temperature in dense cores in Orion B}

The gas kinetic temperatures within the cores in Orion B9 are in the range 
$\sim9.4-13.9$ K, with a mean value of $12.0\pm0.4$ K (the quoted error is the 
standard deviation of the mean). 
The highest temperatures (13.4-13.9 K) are found towards 
IRAS 05399-0121, SMM 4, Ori B9 N, and IRAS 05413-0104. 
There is, however, no such tendency that the warmest cores should be 
protostellar. For example, the Class 0 sources SMM 3 and IRAS 05405-0117 have 
$T_{\rm kin}$ values at the low end in the sample (11.3 K for both sources). 
This is consistent with the notion that embedded low-mass protostars do not 
heat significantly the parent cores, but the heating is localised in their 
immediate vicinity ($\lesssim100$ AU; \cite{friesen2010} and references 
therein).

Recently, based on the Nobeyama 45-m observations, Ikeda et al. (2009) derived 
the NH$_3$ rotational temperatures towards 144 positions in Orion B. 
The cores No.~55-66 in the Ikeda et al. (2009) sample lie in the Orion B9 
region. For these latter cores, they found the values in the range 
$T_{\rm rot}=10.1-16.1$ K. 
By using the relationship~(\ref{eq:Tkin}), this corresponds to the $T_{\rm kin}$ 
values $10.6-18.5$ K, with the mean value 16.4 K. This is 
clearly higher than the mean $T_{\rm kin}$ of the present study. The difference 
is likely to be caused by the larger beamsize used by Ikeda et al. 
($\sim1\farcm3$). We note that the highest value of 
$T_{\rm kin}=43.9$ K in the Ikeda et al. (2009) sample was found in the 
NGC 2024 \ion{H}{II} region, and the lowest values ($\lesssim11$ K) were 
found in Orion B9 and other regions of relatively low star formation activity, 
i.e., outside the NGC 2023, 2024, 2068, and 2071 regions.

We note that in several previous submm surveys of Orion B dust temperatures 
around 20 K have been derived or assumed for dense cores 
(\cite{johnstone2001}, 2006; \cite{nutter2007}). Together with the results of 
Ikeda et al. (2009) the present temperature determinations demonstrate that 
Orion B contains also several very cold cores resembling cores in nearby
low-mass star forming regions (see, e.g., \cite{rosolowsky2008}; 
\cite{schnee2009}; \cite{friesen2009} and references therein). 

\subsection{Kinematics of the core gas}

The NH$_3$ line profiles show that most cores have subsonic 
non-thermal motions ($f_{\rm turb}<1$).
In two cores (SMM 1 and 7), $f_{\rm turb}$ is higher 
than in the rest of the cores, and non-thermal motions appear to be 
slightly transonic ($f_{\rm turb}=1.3<2$). Recently, Friesen et al. (2009) 
found that the mean $f_{\rm turb}$ value for dense cores in Oph B is 1.5, 
clearly larger than the corresponding value for our cores (a mean $f_{\rm turb}$ 
and its standard deviation is $0.8\pm0.1$).

\begin{figure}[!h]
\resizebox{\hsize}{!}{\includegraphics{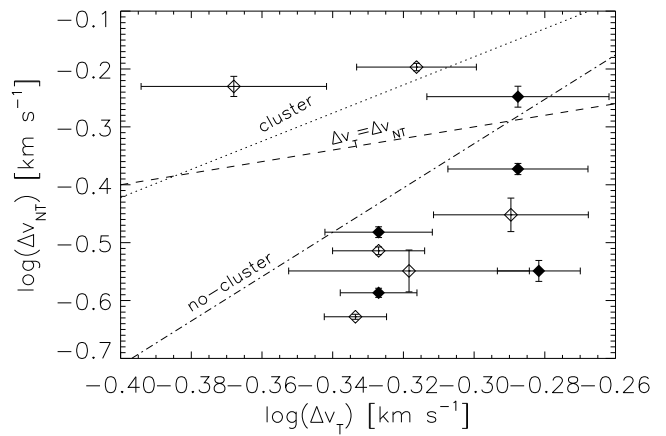}}
\caption{Non-thermal versus thermal (for H$_2$, i.e., the isothermal sound 
speed) linewidths in log-log scales. The dashed line indicates 
$\Delta {\rm v_{NT}}=\Delta {\rm v_{T}}$. The dotted and dash-dotted
lines show relationships found by Jijina et al. (1999) for NH$_3$ cores with 
and without cluster association, respectively. Symbols have the same meaning 
as in Fig.~\ref{figure:vcomparison}.}
\label{figure:widths}
\end{figure}

In Fig.~\ref{figure:widths}, we plot the non-thermal linewidth against 
the thermal linewidth for the cores. Figure~\ref{figure:widths} shows 
that most of the cores in Orion B9 are quiescent, i.e., 
$\Delta {\rm v_{NT}}<\Delta {\rm v_{T}}$. Also shown in this figure are the 
$\Delta {\rm v_{NT}}-\Delta {\rm v_{T}}$ relationships found by Jijina et al. 
(1999) for NH$_3$ cores in clusters and isolated regions. 
The two relationships suggest that non-thermal NH$_3$ linewidths are larger for 
cores in clustered environments than in isolated cores. A majority of Orion B9 
cores lie on the Jijina plot in the region characteristic of isolated cores.

Like in the case of the kinetic temperature, there is no clear
difference in linewidths between starless and protostellar cores.
Protostars are likely to be associated with outflows which could cause the 
linewidths to be larger, but evidence for outflows can only be found in 
IRAS 05399-0121 (HH~92; \cite{bally2002}) and IRAS 05413-0104 (HH~212; e.g., 
\cite{lee2008}). It is possible that the HH~92 outflow from IRAS 05399-0121 
contribute to the relatively broad NH$_3$ lines seen in the adjacent SMM 1 
core which is oriented along the direction of the outflow.

\begin{figure}[!h]
\resizebox{\hsize}{!}{\includegraphics{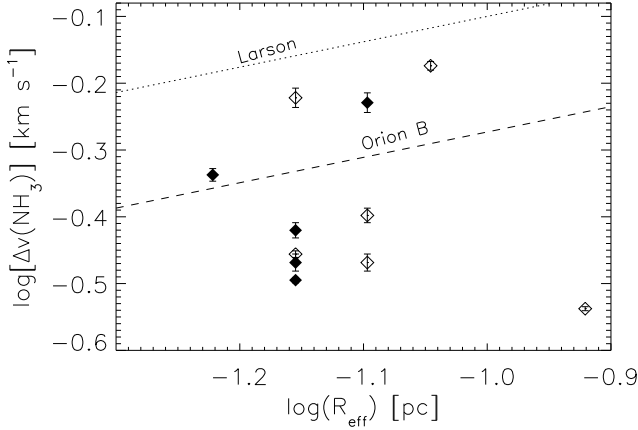}}
\caption{Linewidth versus radius for the cores in Orion B9 in log-log scales. 
The Larson (1981) relation is shown by the dotted line. The dashed line 
indicates the relation derived by Ikeda et al. (2009) for the H$^{13}$CO$^+$ 
cores in Orion B. Symbols have the same meaning as in 
Fig.~\ref{figure:vcomparison}.}
\label{figure:linewidthsize}
\end{figure}

In Fig.~\ref{figure:linewidthsize}, we plot the NH$_3(1,\,1)$ 
linewidth as a function of core effective radius. Also shown are the 
linewidth-size relation derived by Larson (1981), and the relationship recently 
found by Ikeda et al. (2009) for the H$^{13}$CO$^+$ cores in Orion B. 
The Larson relation, thought to arise from the interstellar turbulence, 
was originally presented for the 3D velocity dispersion, $\sigma_{\rm 3D}$, 
and the maximum linear size of the source, $L$. In terms of our definitions, 
it can be written as  
$\Delta {\rm v}[{\rm km~s^{-1}}]=1.9\left(R_{\rm eff}[{\rm pc}]\right)^{0.38}$, 
where we have assumed that $L=2R_{\rm eff}$.
The linewidths of Orion B9 cores mostly deviate downward from the Larson 
relation, and the cores lie around the Ikeda et al. (2009) relationship, 
namely $\Delta {\rm v}[{\rm km~s^{-1}}]=1.28\left(R_{\rm core}[{\rm pc}]\right)^{0.38}$ (we remind the reader that the Ikeda et al. sample contains Orion B9 
members). On the other hand, it is possible that the 
more tenuous gas in Orion B9 follows the Larson relation: the mean linewidth 
($\sim1.3$ km~s$^{-1}$) and half-maximum radius ($\sim0.6$ pc) measured from 
the $^{13}$CO$(1-0)$ emission towards the Orion B9 region by Caselli \& Myers 
(1995) are quite similar to those expected from the Larson relation.
As pointed out by Maruta et al. (2010), this could be explained if the cores 
were formed in regions where supersonic turbulence was dissipated. 
Dissipation of turbulence could be due to shocks in the converging turbulent 
flows, where the formation of density enhancements is expected to take place 
(see Sect.~5.9). The timescale for the dissipation of turbulence in dense 
cores is comparable to the free-fall timescale (e.g., \cite{maclow2004}). 
Thus, because the observed linewidth-size relation appears to be quite flat, 
it is possible that the turbulent gas motions at small scales 
($\lesssim0.1$ pc) are not yet settled into equilibrium state (i.e., relaxed). 
In this case, the core formation should be a rapid process 
with the corresponding timescale being comparable to the free-fall time. 

\subsection{Internal pressure support}

\begin{figure*}
\begin{center}
\includegraphics[width=0.45\textwidth]{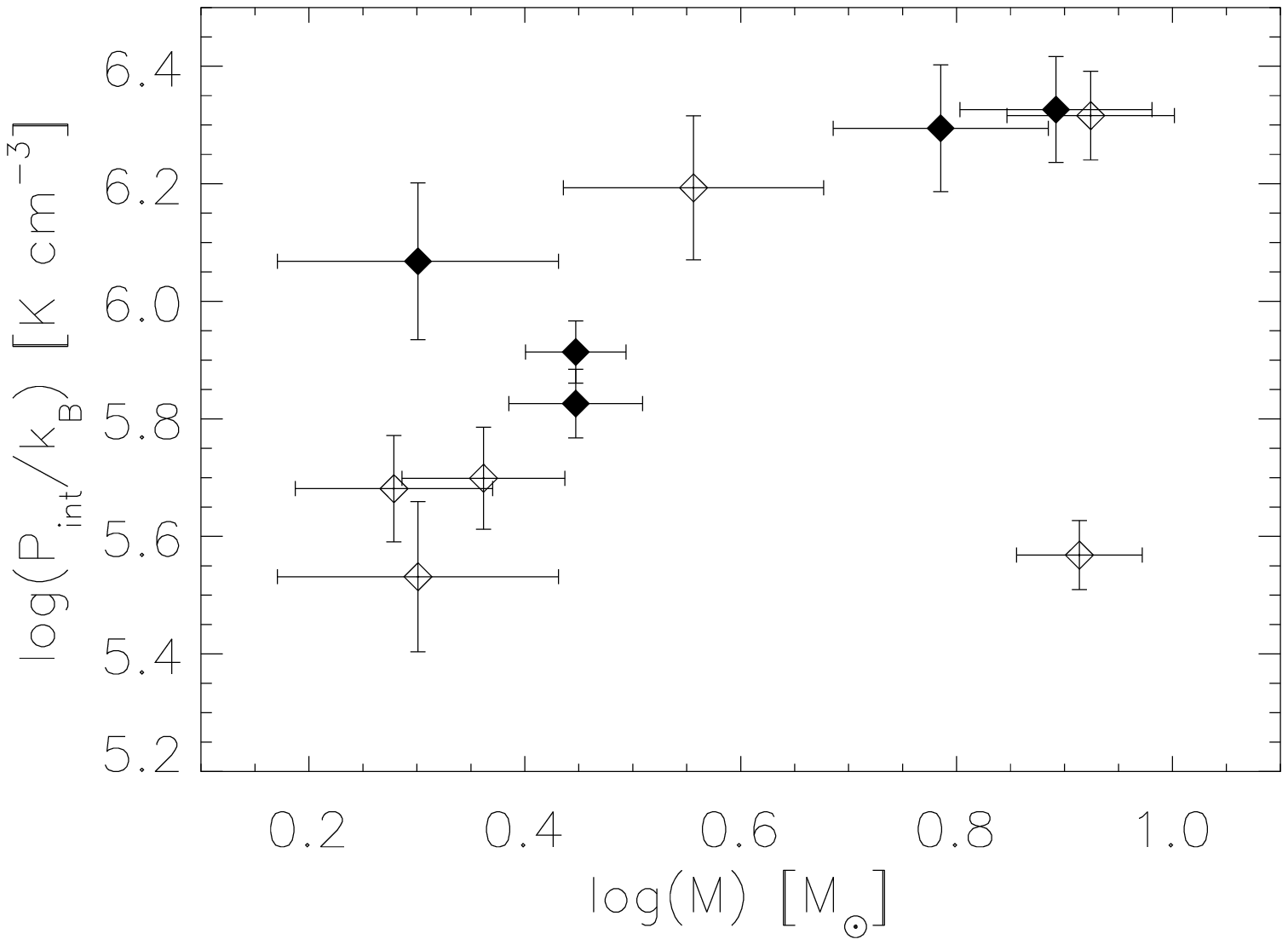}
\includegraphics[width=0.45\textwidth]{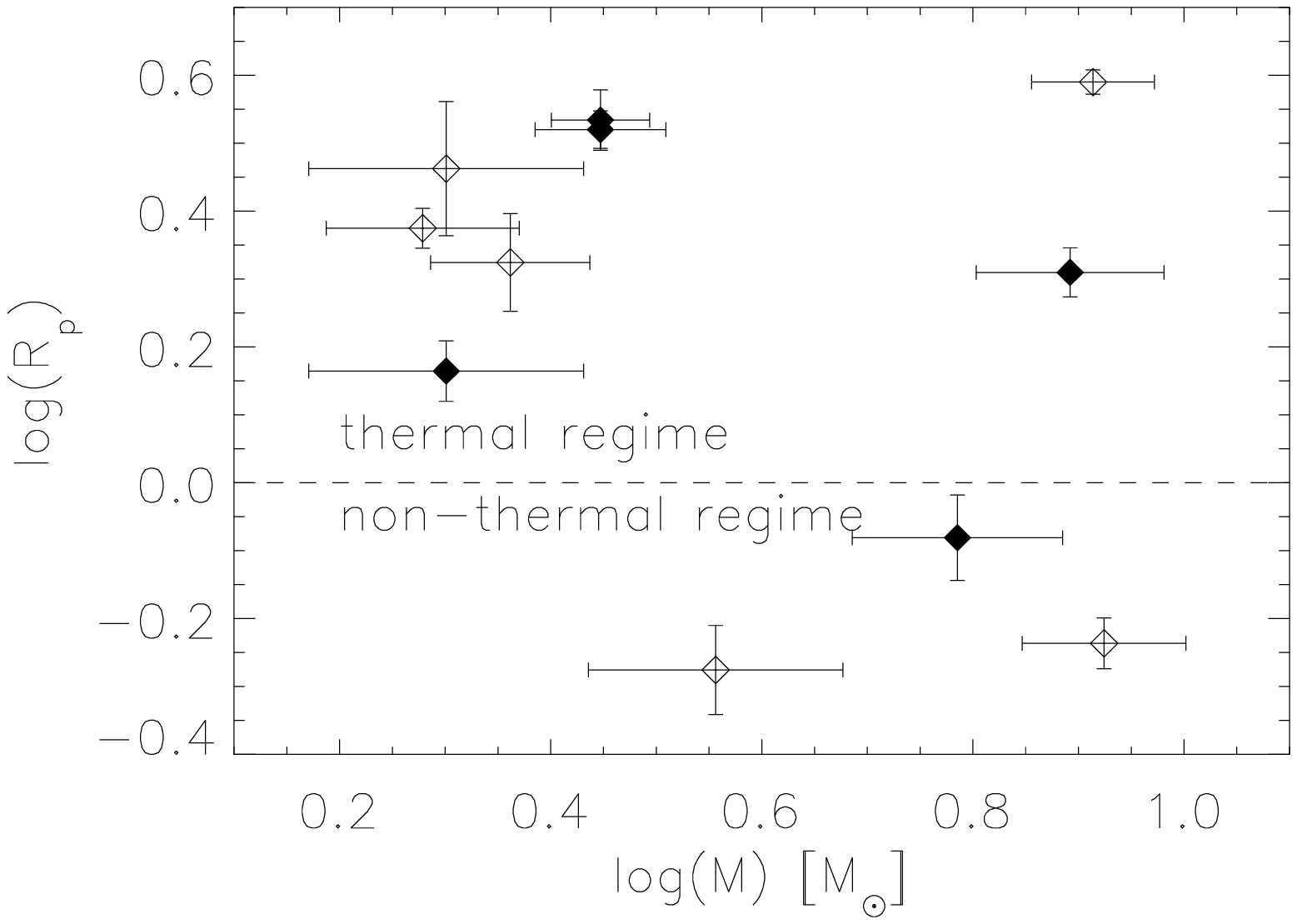}
\caption{\textbf{Left:} Total internal kinetic gas pressure versus core mass. 
\textbf{Right:} The ratio of thermal to non-thermal
pressure as a function of mass. The dashed horizontal line marks the point at
which thermal and non-thermal pressure support are equal. Both plots are in 
log-log scales, and the symbols have the same meaning as in 
Fig.~\ref{figure:vcomparison}.} 
\label{figure:pressure}
\end{center}
\end{figure*}

The internal kinetic gas pressure, $P_{\rm int}$, within the core consists of a
thermal pressure, $p_{\rm T}=nk_{\rm B}T_{\rm kin}$, and a non-thermal 
pressure, $p_{\rm NT}=\mu m_{\rm H}n\sigma_{\rm NT}^2$, where $n$ is 
the number density, and $\mu=2.33$ ($P_{\rm int}=p_{\rm T}+p_{\rm NT}$). 
In the left panel of Fig.~\ref{figure:pressure}, we plot 
$P_{\rm int}/k_{\rm B}$ versus core mass. Excluding the one outlier in the
plot (SMM 6), there is a trend of increasing internal pressure with increasing 
core mass. This indicates that the cores are not in pressure equilibrium 
with an external pressure due to, e.g., the weight of the cloud in which the 
cores are embedded (\cite{lada2008}). The mean internal pressure of 
the Orion B9 core population is $\sim1.1\times10^6$ K~cm$^{-3}$. 
This is in excellent agreement with the value $\sim10^6$ K~cm$^{-3}$ estimated 
by Johnstone et al. (2001) for the cores in the northern part of Orion B. 
For comparison, the overall pressure of the ISM in the Galactic midplane, 
which consists of thermal and turbulent kinetic pressures, and the pressures 
of magnetic fields and cosmic rays, has been estimated to be 
$P_{\rm ISM}/k_{\rm B}\approx2.8\times10^4$ K~cm$^{-3}$ (\cite{boulares1990}).
The high pressures inside cores are likely to result from 
compression by gravity (see Sect.~5.4). 

To examine the relative role of turbulence in the core internal pressure, 
we calculated the ratio of thermal to non-thermal pressure, which is given 
by $R_{\rm p}=1/f_{\rm turb}^2$ (see Sect.~4.1.4). The values of $R_{\rm p}$ 
are plotted as a function of mass in the right panel of 
Fig.~\ref{figure:pressure}. 
As can be seen in this figure, thermal pressure is clearly the dominant source 
of internal gas pressure for 8 of 11 cores. For the rest of the cores, 
thermal pressure support is still significant as they have $R_{\rm p}>0.5$. 
Also in this regard, the Orion B9 cores appear to be similar to low-mass dense
cores in nearby molecular clouds, which are commonly found to be thermally 
dominated (e.g., \cite{myers1983}; \cite{kirk2007}; \cite{lada2008}).

\subsection{Dynamical state and gravitational boundedness of the cores}

\begin{figure*}
\begin{center}
\includegraphics[width=0.45\textwidth]{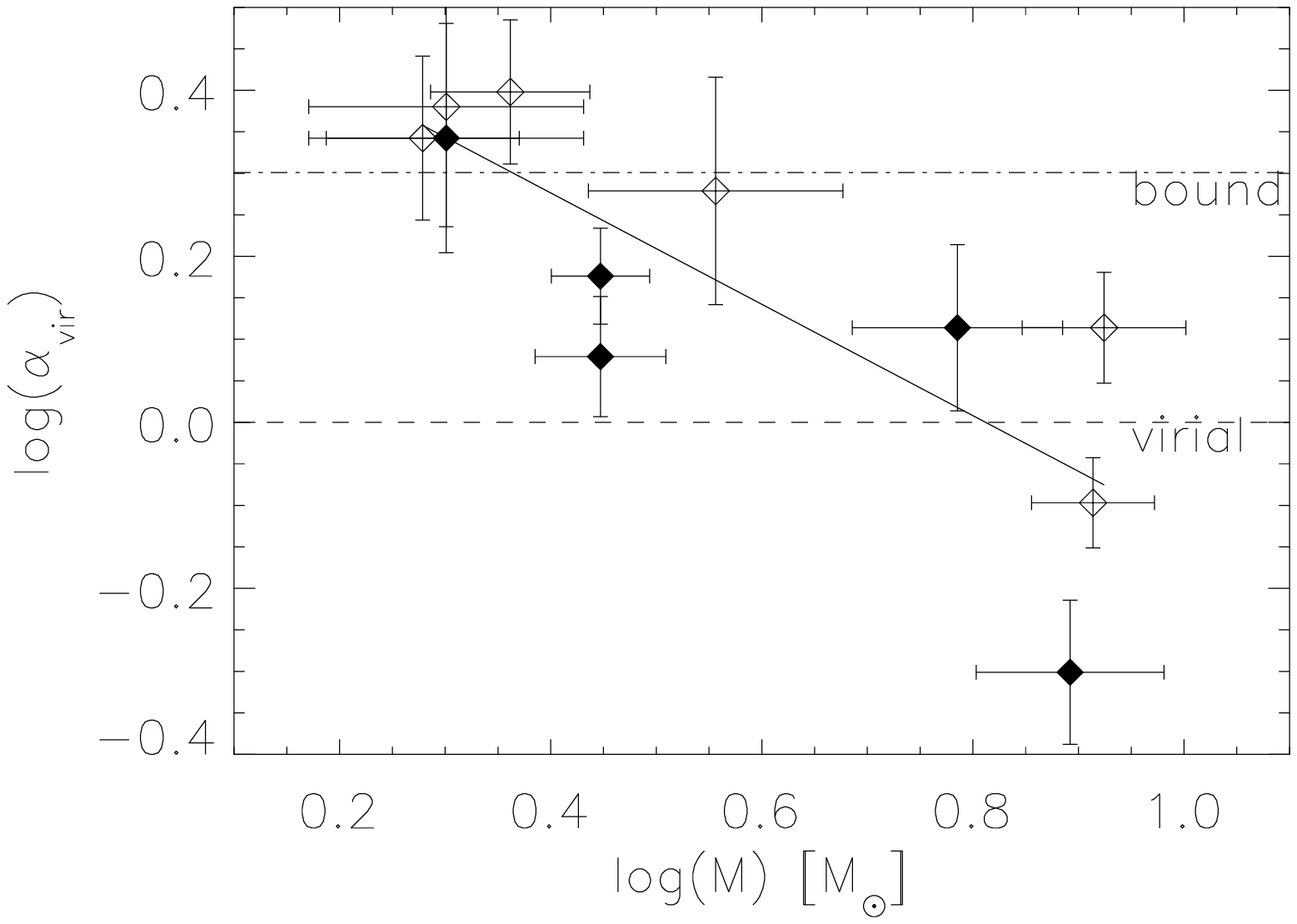}
\includegraphics[width=0.45\textwidth]{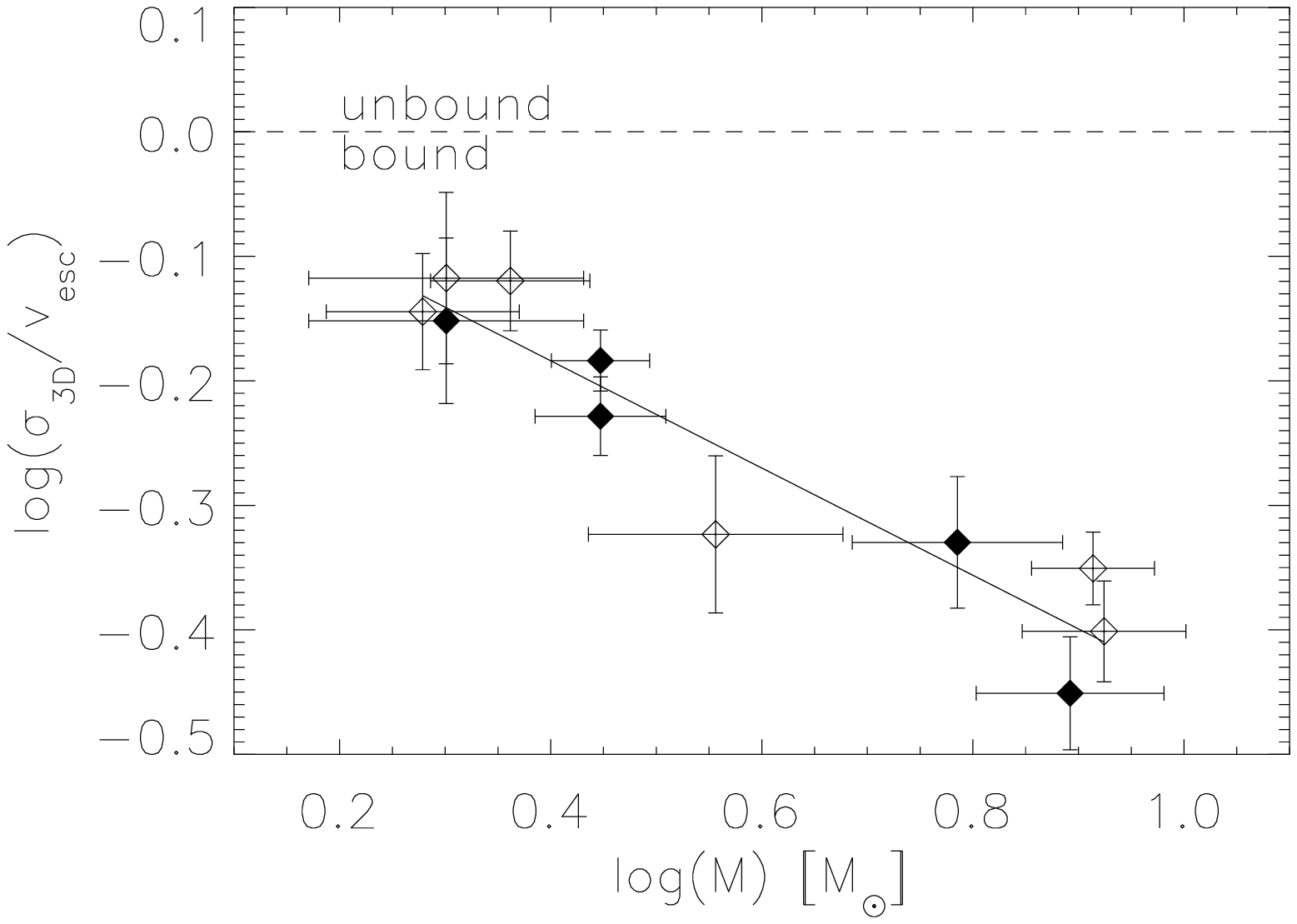}
\caption{\textbf{Left:} Virial parameter versus core mass. 
The dashed line indicates the virial equilibrium limit ($\alpha_{\rm vir}=1$), 
and the dash-dotted line shows the 
limit of gravitational boundedness ($\alpha_{\rm vir}=2$). \textbf{Right:} 
Ratio of 3D velocity dispersion to escape velocity against core mass. 
The dashed line indicates $\sigma_{\rm 3D}/{\rm v_{\rm esc}}=1$. 
All cores are likely to be gravitationally bound. The solid line in both 
plots indicates the least-squares fit to the data. Both plots are in log-log 
scales, and the symbol key is identical to that of 
Fig.~\ref{figure:vcomparison}.} 
\label{figure:state}
\end{center}
\end{figure*}

To further examine the dynamical state of the cores, we inspect 
the virial parameters derived in Sect.~4.4.
Figure~\ref{figure:state} (left panel) shows the distribution of the virial 
parameter for the Orion B9 cores as a function of mass. 
Within the errors, all eleven cores are self-gravitating
($\alpha_{\rm vir}<2$), and five of them seem to be close to the virial
equilibrium or collapsing ($\alpha_{\rm vir}\leq1$). One should note that the
virial calculation presented above did not include any external
pressure or magnetic fields. An external presssure of $P_{\rm ext}/k_{\rm B}
\sim 2-3 \times10^5$ K~cm$^{-3}$ would bring all starless cores except SMM 7
to virial equilibrium or make them collapse (neglecting the possible
magnetic support). Assuming that the gas surrounding the cores is
characterised by $\sigma_{\rm NT}\sim 0.6$ km~s$^{-1}$ and 
$n({\rm H_2})=2\times10^3$ cm$^{-3}$, corresponding to the typical width of
$^{13}$CO$(1-0)$ lines observed by Caselli \& Myers (1995) in this region
and the critical density of the transition in question (which is lower
than the critical density of the NH$_3$ inversion lines), its turbulent
ram pressure would roughly equal to the required confining pressure
quoted above. As discussed by Lada et al. (2008) in the case of Pipe
Nebula, the supersonic intercore turbulence can be a manifestation of
the self-gravity of the surrounding massive cloud.

It seems likely that most if not all starless cores in Orion B9 detected
in this survey are prestellar, i.e. they will eventually collapse to
stars. The situation resembles that observed in Perseus by Foster et
al. (2009): most cores are gravitationally bound, and $\sim1/3$ are in
virial equilibrium.

The virial parameter, $\alpha_{\rm vir}$ seems to decrease as a
function of core mass (see Fig.~\ref{figure:state}, left). The slope of a 
least-squares fit, $\log(\alpha_{\rm vir})=(0.54\pm0.10)-(0.67\pm0.16)\log(M)$, 
with the linear correlation coefficient $r=-0.81$, is very close to that found 
by Lada et al. (2008) for cores in the Pipe Nebula which are believed to be 
predominantly confined by external pressure. Our cores with masses in the range
$2-8$ M$_{\sun}$ correspond to the most massive Pipe cores. The slope 
($\sim -2/3$) is consistent with the theoretical prediction of
Bertoldi \& McKee (1992) for pressure confined cores, and basically
results from the fact that there is not much core-to-core variation in
the average densities and velocity dispersions. It should be noted,
however, that the overall level of $\alpha_{\rm vir}$, characterised by 
the constant of the fit, is lower than derived in the Pipe, probably 
reflecting the different environments and evolutionary stages of the two
regions.

Like Lada et al. (2008) we have also plotted in the right panel
of Fig.~\ref{figure:state} the correlation diagram between the ratio 
$\sigma_{\rm 3D}/{\rm v_{\rm esc}}$ and the core mass, where the three-dimensional 
velocity dispersion, $\sigma_{\rm 3D}$, is calculated from 
$\sigma_{\rm 3D}=\sqrt{3c_{\rm s}^2+3\sigma_{\rm NT}^2}$, and the escape velocity, 
${\rm v_{\rm esc}}$ is given by ${\rm v_{\rm esc}}=\sqrt{2GM/R_{\rm eff}}$.
For all cores, $\sigma_{\rm 3D}$ is smaller 
than ${\rm v_{\rm esc}}$, supporting the above conclusion that the cores are 
gravitationally bound. Morever, the $\sigma_{\rm 3D}/{\rm v_{\rm esc}}$ 
ratio appears to decrease as a function of core mass. A least-squares 
fit to the whole sample gives 
$\log\left(\sigma_{\rm 3D}/{\rm v_{\rm esc}}\right)=(-0.01\pm0.03)-(0.43\pm0.05)\log(M)$, with $r=-0.94$. This relationship implies that the transition from 
unbound to bound core occurs at about 0.8-0.9 M$_{\sun}$. 
This is very close to the peak of the CMF for Orion B North, i.e., 
$\sim1$ M$_{\sun}$ (\cite{nutter2007}). 
This is expected because the CMFs for Orion B9 and Orion B North seem to 
represent the subsamples of the same parent distribution (Sect.~4.3). 
A similar correspondence was found by Lada et al. (2008) for the dense cores 
in the Pipe Nebula.

\subsection{Correlating the properties of protostellar cores}

In Paper I, we derived the SEDs for the 
protostellar cores in Orion B9. For all sources, the observed flux densities 
were fitted by a two-temperature (warm$+$cold) composite model. The bolometric
dust temperatures and bolometric luminosities of the sources derived from the
SEDs are given in Table~6 of Paper I. 

\begin{figure*}
\begin{center}
\includegraphics[width=0.45\textwidth]{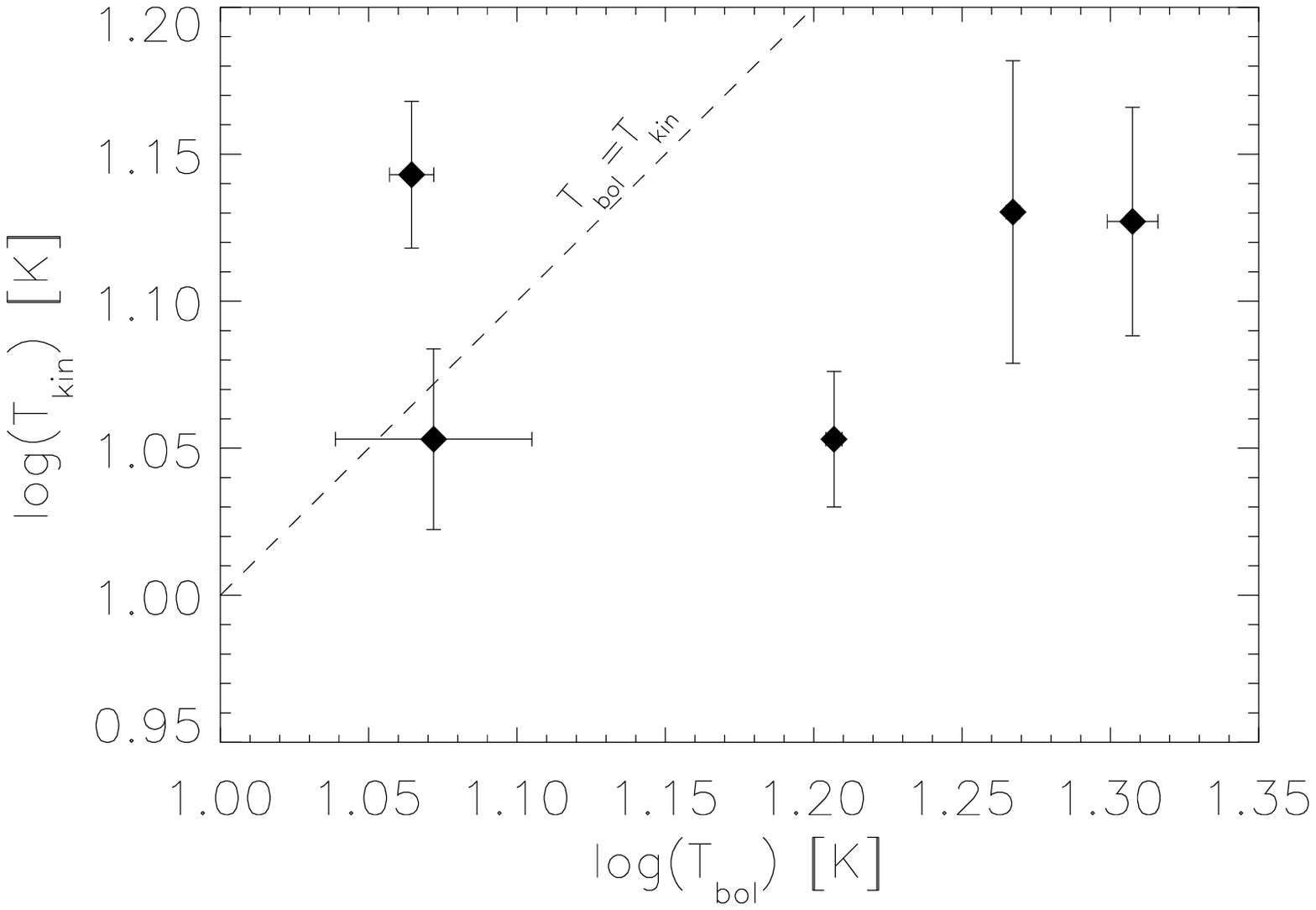}
\includegraphics[width=0.45\textwidth]{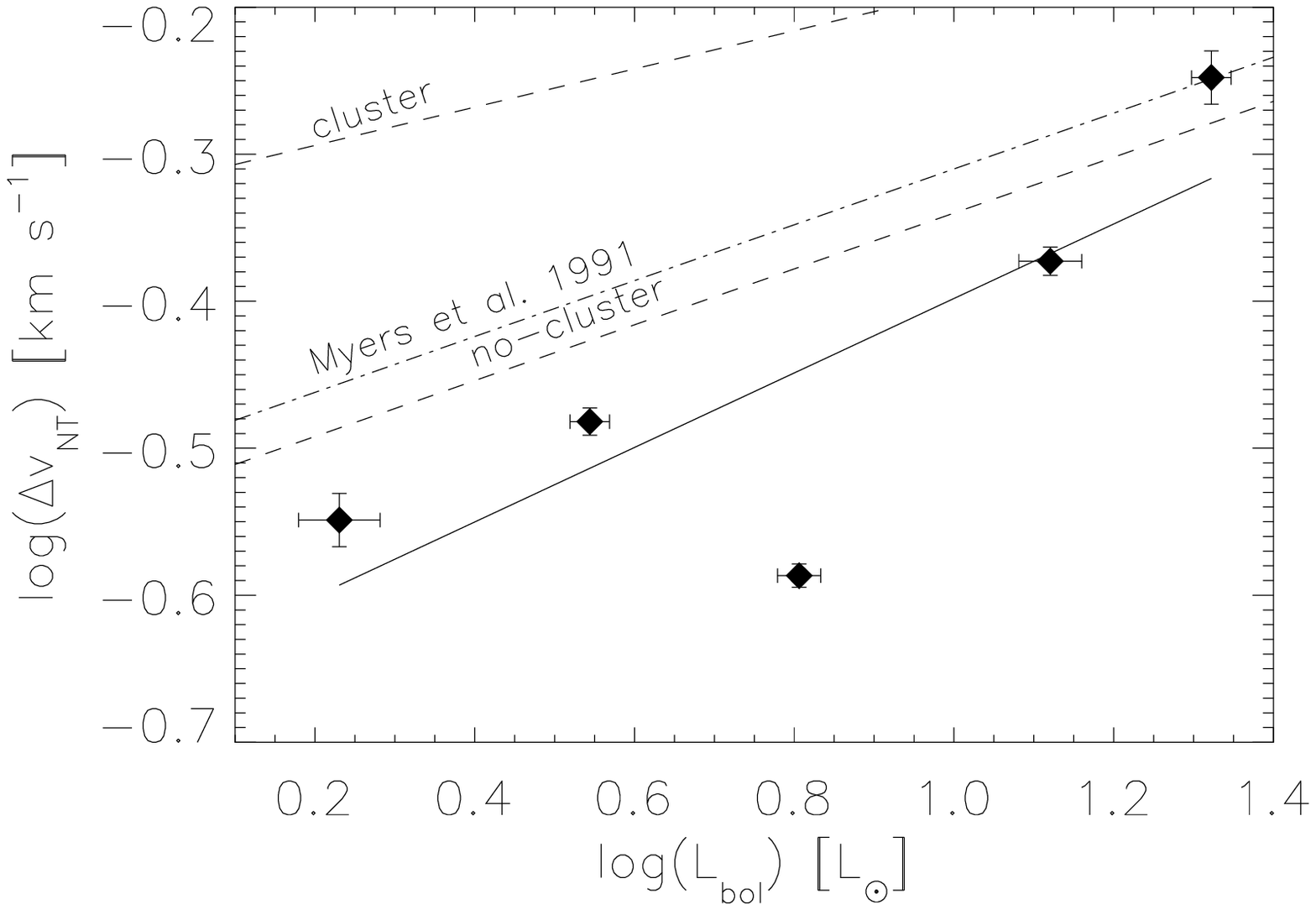}
\caption{\textbf{Left:} $T_{\rm kin}$ versus $T_{\rm bol}$. The dashed line 
indicates $T_{\rm kin}=T_{\rm bol}$. \textbf{Right:} $\Delta {\rm v_{\rm NT}}$ 
versus $L_{\rm bol}$. The solid line indicates the least-squares fit to the 
data. The upper and lower dashed lines represent relationships
found by Jijina et al. (1999) for NH$_3$ cores with and without cluster 
association, respectively. The dash-dotted line shows the relationships found 
by Myers et al. (1991) for IRAS point sources. Both plots are in log-log 
scales.} 
\label{figure:protostars}
\end{center}
\end{figure*}

In the left panel of Fig.~\ref{figure:protostars}, we show the correlation 
plot between $T_{\rm kin}$ and the bolometric temperature, $T_{\rm bol}$. 
Only for SMM 3, the two values are similar within the errors, and in the case 
of SMM 4, $T_{\rm kin}$ and $T_{\rm bol}$ agree within a few kelvins. 
For the other sources (i.e., the three IRAS sources), however, the 
temperatures estimated from the SEDs are clearly higher than those
obtained from NH$_3$ measurements. The SEDs of SMM 3 and 4 were
constructed by using only three flux density values at 24, 70, and 870 $\mu$m
(see Paper I; Fig.~6 therein). In the case of other sources, 
also the flux densities at all four IRAS bands (12, 25, 60, and 100 $\mu$m) 
were included in the SED fits. This may have caused the temperature of the 
cold part of the spectrum to be overestimated. Shetty et al. (2009) recently 
concluded that if using flux density values at widely separated wavelengths,
and including short-wavelength flux densities in the SED fit, 
the obtained temperatures may be too high.

The right panel of Fig.~\ref{figure:protostars} shows a plot of 
$\Delta {\rm v_{\rm NT}}$ as a function of $L_{\rm bol}$. 
There appears to be a positive correlation between the two quantities, and 
the least-squares linear fit to the data gives
$\log(\Delta {\rm v_{\rm NT}})=(-0.65\pm0.10)+(0.25\pm0.11)\log(L_{\rm bol})$, 
with $r=0.80$. We also show the relationships found by Jijina et al. (1999) for 
NH$_3$ cores in clusters and cores without cluster association (see their 
Table~B7), and the Myers et al. (1991) relation for a sample of 61 IRAS point 
sources (their Eq.~(3)). Our result is similar to the relationship 
found in the latter two studies, 
$\Delta {\rm v_{\rm NT}}\propto L_{\rm bol}^{0.19}$. The fact that 
$\Delta {\rm v_{\rm NT}}$ increases as a function of $L_{\rm bol}$ can be 
explained by the effects of winds driven by the embedded protostar$+$disk 
system, and by the fact that more turbulent initial conditions give rise to 
more massive, i.e., more luminous stars (\cite{myers1991}). 
As in the case of $\Delta {\rm v_{\rm NT}}-\Delta {\rm v_{\rm T}}$
relation (see Fig.~\ref{figure:widths}), the Orion B9 cores more 
closely follow the relationship for isolated cores. 

\subsection{Column densities and fractional abundances of NH$_3$ and 
N$_2$H$^+$}

The mean NH$_3$ fractional abundance in our sample is $4.8\pm0.7\times10^{-8}$, 
where the $\pm$-error represents the standard deviation of the mean. 
The value of $\langle x({\rm NH_3})\rangle$ in Orion B9 is 
similar to those recently found by Foster et al. (2009) and Friesen et al. 
(2009) for the dense cores in Perseus and Ophiuchus, respectively 
(i.e., a few to several $\times10^{-8}$). For comparison, the NH$_3$ 
abundances in dense cores in nearby regions of isolated star formation, 
such as Taurus, are commonly found to be $\sim0.4-5\times10^{-8}$ 
(e.g., \cite{hotzel2001}; \cite{tafalla2002}, 2006). 
The average N$_2$H$^+$ abundance in Orion B9, 
$\langle x({\rm N_2H^+})\rangle=2.5\pm0.8\times10^{-10}$, is very similar to 
those in isolated low-mass cores (e.g., 
$\langle x({\rm N_2H^+})\rangle=2\pm1\times10^{-10}$ and $3\pm2\times10^{-10}$ 
for the starless and protostellar core samples of Caselli et al. (2002)). 
On the other hand, the value of $\langle x({\rm N_2H^+})\rangle$ is about 2--6 
times lower than those determined by Friesen et al. (2010) in the Oph B1 and 
B2 cores which lie in the region of clustered star formation. 

\begin{figure*}
\begin{center}
\includegraphics[width=0.33\textwidth]{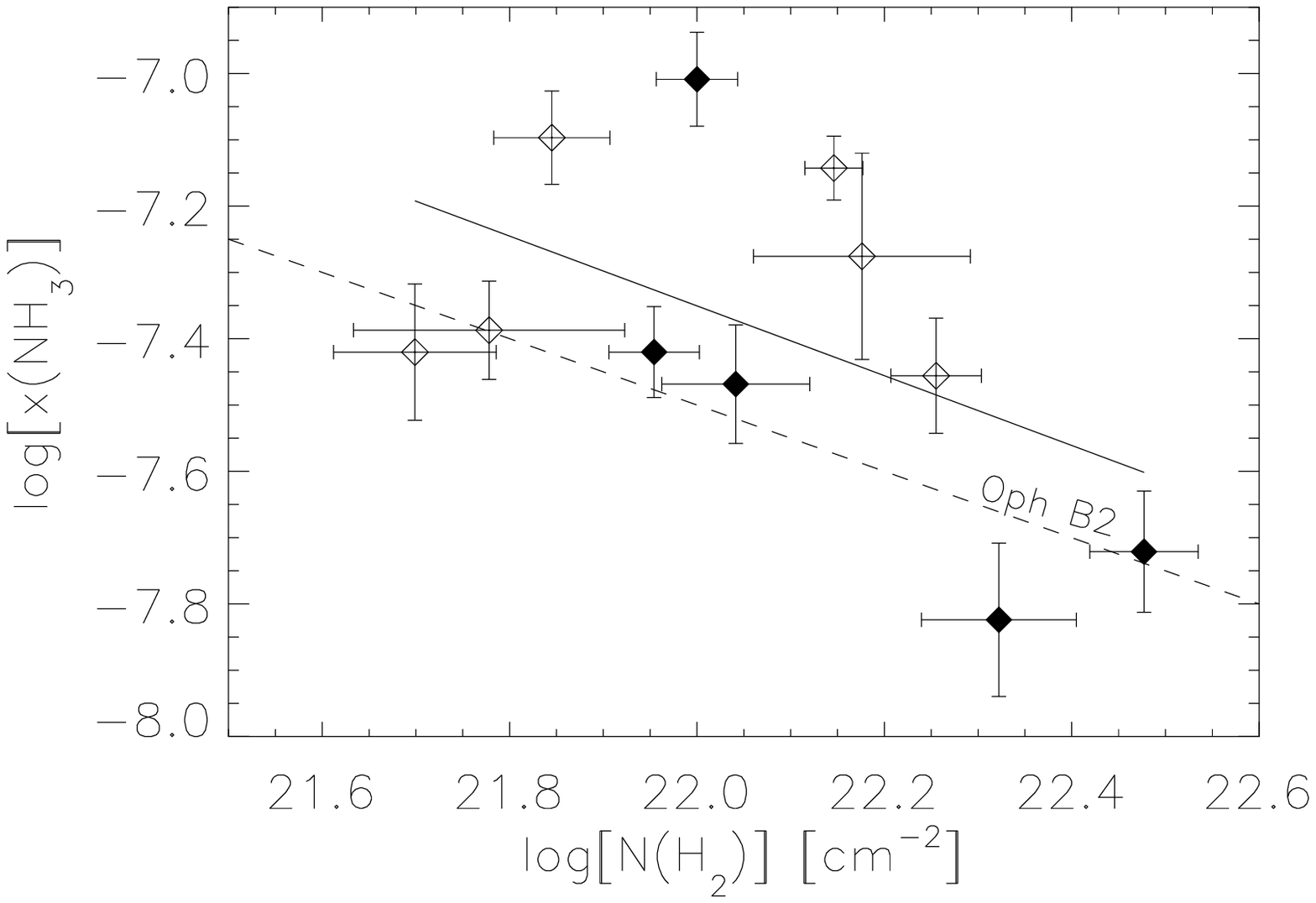}
\includegraphics[width=0.33\textwidth]{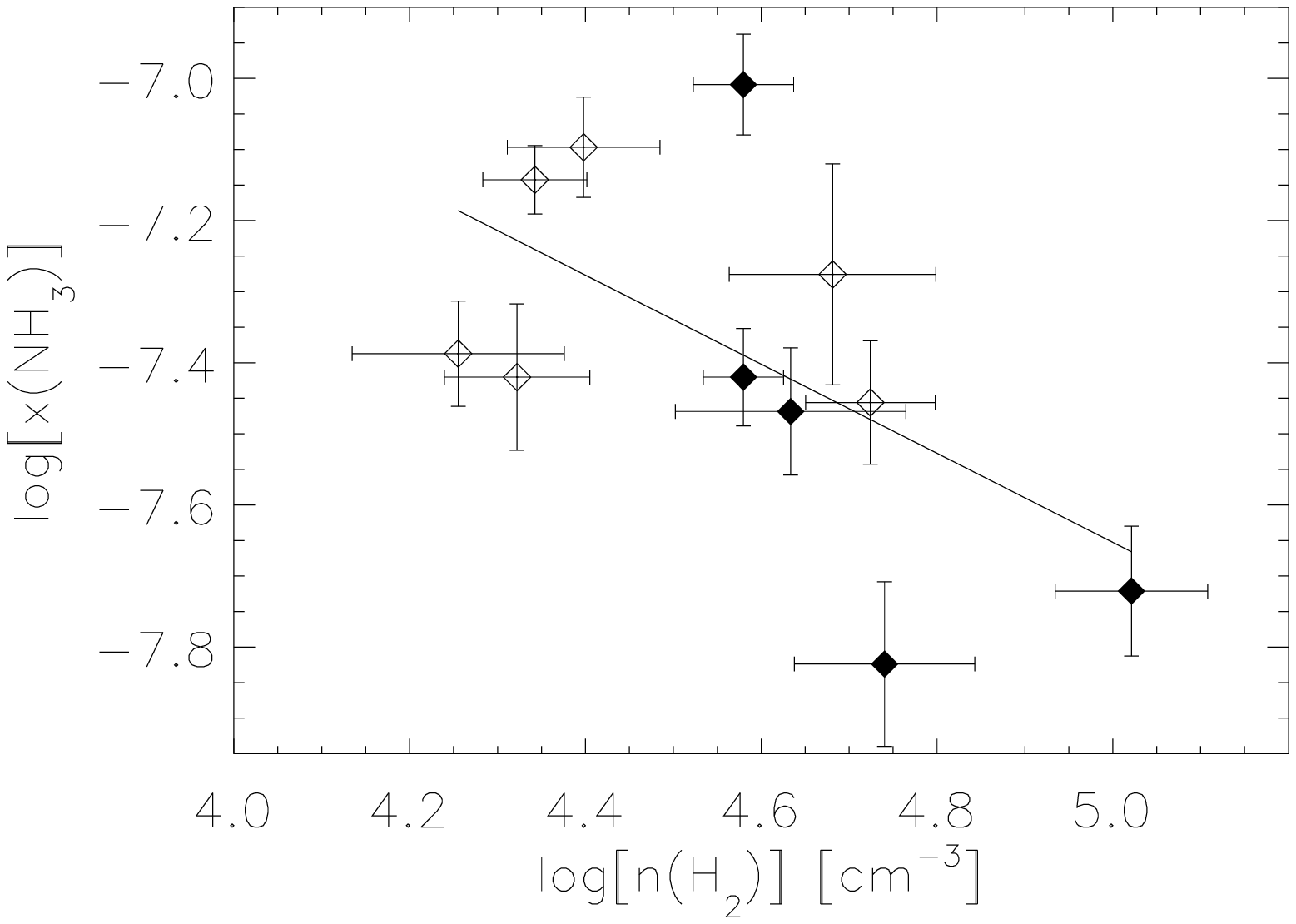}
\includegraphics[width=0.33\textwidth]{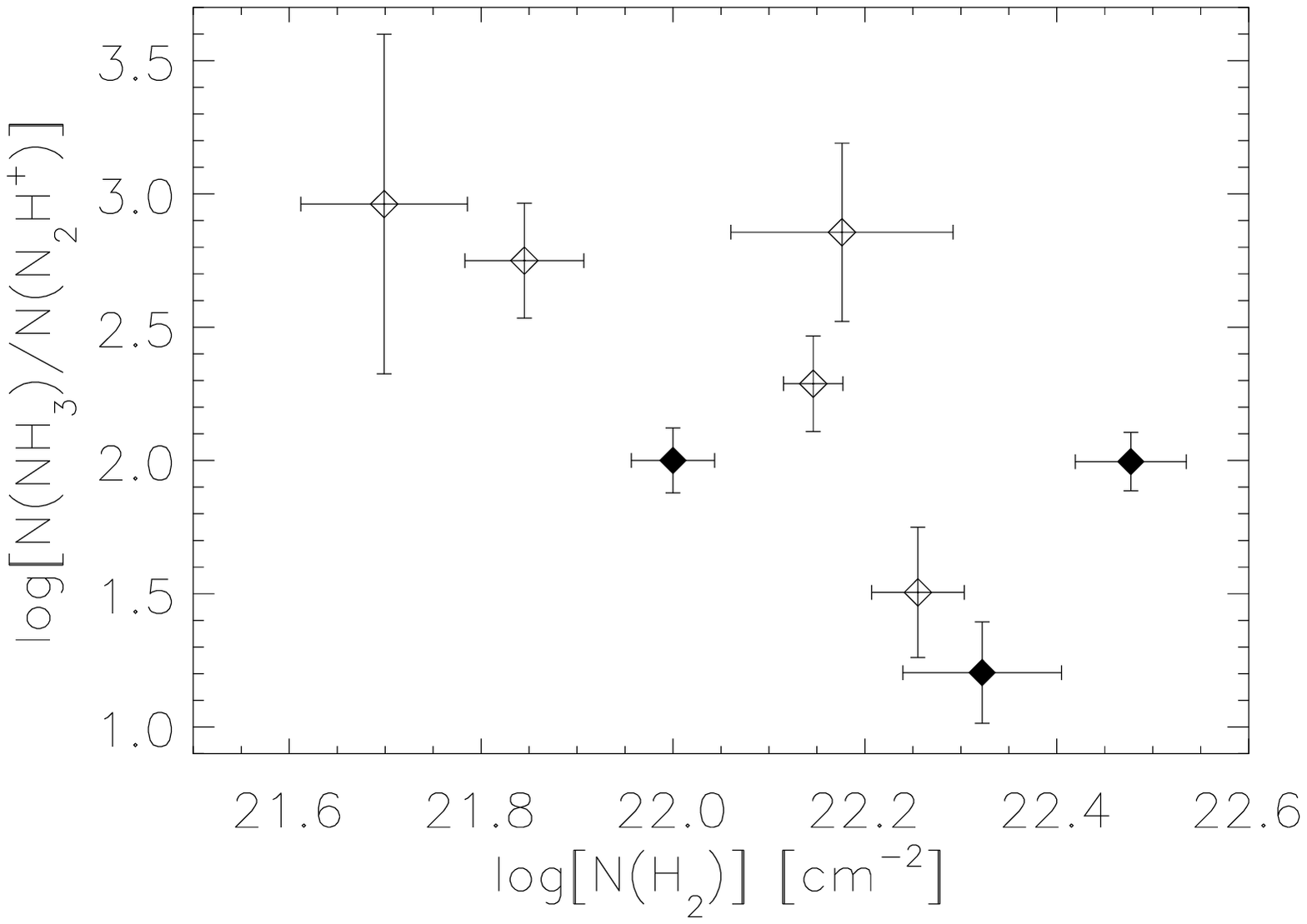}
\caption{Comparison of chemical abundances as a function of H$_2$ column 
and number density. All plots are in log-log scales, and the symbol key is 
identical to that of Fig.~\ref{figure:vcomparison}. The solid lines indicate 
the least-squares fits to the data. \textbf{Left:} $x({\rm NH_3})$ versus 
$N({\rm H_2})$. The dashed line indicates the relationship found by Friesen 
et al. (2009, 2010) for Oph B2. \textbf{Middle:} $x({\rm NH_3})$ versus 
$n({\rm H_2})$. \textbf{Right:} NH$_3$/N$_2$H$^+$ column density ratio 
versus $N({\rm H_2})$.} 
\label{figure:abundances}
\end{center}
\end{figure*}

As shown in the left panel of Fig.~\ref{figure:abundances}, 
the fractional NH$_3$ abundance appears to decrease with increasing H$_2$ 
column density (as traced by 870 $\mu$m continuum emission smoothed to 
$40\arcsec$ resolution). A linear-least-squares fit to the data gives
$\log x({\rm NH_3})=(4.2\pm6.6)-(0.5\pm0.3)\log N({\rm H_2})$, with 
$r=-0.51$. The slope of this relationship, 
$x({\rm NH_3}) \propto 1/\sqrt{N({\rm H_2})}$, is very
similar to that found by Friesen et al. (2009, 2010) in the Oph B2
core. In the middle panel of Fig.~\ref{figure:abundances}, we plot 
$x({\rm NH_3})$ as a function of $n({\rm H_2})$. The trend is similar as 
above, and the least-squares linear fit to the data indicates 
$\log x({\rm NH_3})=(-4.5\pm1.4)-(0.6\pm0.3)\log n({\rm H_2})$, with $r=-0.57$.
The tendencies described above are likely to manifest the fact that NH$_3$, 
like species containing carbon or oxygen, accrete onto grain surfaces at high
densities. According to chemistry models of Aikawa et al. (2005) and
Flower et al. (2006), the NH$_3$ abundance becomes heavily depleted at
densities $n_{\rm H}\gtrsim$ a few $\times10^6$ cm$^{-3}$ (due to freeze-out of 
the parent species N$_2$). The $x({\rm NH_3})-n({\rm H_2})$ relationship shown 
in Fig.~\ref{figure:abundances} agrees reasonably well with the predictions for 
the early stages of collapse presented in Flower et al. (2006; see, e.g., 
their Fig.~4). In the Flower et al. model, the protostellar collapse occurs
on the free-fall timescale which is about $4\times10^5$ yr at 
$n_{\rm H}=10^4$ cm$^{-3}$. This is in agreement with statistical estimates of 
the prestellar core lifetime (see Sect.~5.9).

Unlike in Friesen et al. (2010), no correlation was found
between $x({\rm N_2H^+})$ and H$_2$ column or number densities. 
In the right panel of Fig..~\ref{figure:abundances}, we plot the 
NH$_3$/N$_2$H$^+$ column density ratio as a function of 
$N({\rm H_2})$ (determined using $40\arcsec$ resolution). A possible negative 
correlation can be seen between the two quantities, resembling the result 
found by Friesen et al. (2010) in Oph B1, and by Johnstone et al. (2010) 
in Perseus. As can be seen from the figure and Col.~(4) of 
Table~\ref{table:diazparameters2}, prestellar cores generally have 
higher values of $N({\rm NH_3})/N({\rm N_2H^+})$ ratio than in protostellar 
cores. The tendency has been observed also in other star-forming regions 
(Paper I and references therein; Friesen et al. 2010). The reason for this 
differentiation is not quite clear. The abundances of both molecules build up 
slowly, and they benefit from increasing density and the freezing of CO and 
other heavier species (e.g., \cite{aikawa2005}). For both molecules the
enhanced formation at high densities should be counteracted by
accretion onto dust grains. In the time evolution of the column
densities these contradictory tendencies are probably reflected as NH$_3$
and N$_2$H$^+$ peaks coming one after the other. Because of the different
radial distributions of these molecules (and probably also because of
the different critical densities of their transitions) the derived
column density ratios depend on the spatial resolution. The fact that
Friesen et al. (2010) could see the effect of N$_2$H$^+$ depletion towards
the highest densities can probably be explained by a higher resolution
and better data quality than available here.

Our data contains, however, evidence that N$_2$H$^+$ is frozen out in the
core SMM 4, as suggested in Paper I. The core was not detected in
N$_2$H$^+(3-2)$, and the absence of this molecule is most probably not
caused by efficient desorption of CO, which is one of the principal
destructors of N$_2$H$^+$ in the gas phase. The gas kinetic temperature
derived from ammonia ($\sim14$ K) is clearly lower than the sublimation
temperature of CO ($\sim20$ K; \cite{aikawa2008}). Depletion of N$_2$H$^+$ 
could also be the reason for the very weak N$_2$H$^+(3-2)$ emission seen 
towards Ori B9 N. An other indication of depletion in the core Ori B9 N is 
that the H$_2$D$^+(1_{10}-1_{11})$ line at $\sim9$ km~s$^{-1}$ is detected in 
this source, about 38\arcsec southeast from the dust peak position 
(\cite{harju2006}). This molecule is known to resist depletion ``to last'' 
and is observed in highly depleted cold cloud cores (\cite{caselli2008}).

\subsection{Low-velocity gas emission seen towards Orion B9}

The LSR velocities of SMM 7 (3.6 km s$^{-1}$) and IRAS 05413-0104 
(1.5 km~s$^{-1}$) are about 6--8 km s$^{-1}$ less than the systemic velocity of 
Orion B (see Sect.~3.1). We note that the systemic velocity of IRAS 05413-0104 
is known to be low before the present study (e.g., from NH$_3$ measurements by 
Wiseman et al. (2001)). Given that the sources north-east from the 
central part of Orion B9 appear to have lower radial velocities, the same is 
likely to be true for IRAS 05412-0105. Indeed, IRAS 05412-0105 and 05413-0104 
could be embedded in a common 1.2 mm clump (\cite{kauffmann2008}).
No NH$_3(1,\,1)$ line was detected towards IRAS 05412-0105 (Sect.~2.1), and we
could not found any other spectral line observations for it from the 
literature (providing kinematic information). Also, the second velocity 
components seen in the NH$_3$ and N$_2$H$^+(3-2)$ spectra towards SMM 4 and 
Ori B9 N have centroid velocities $\sim7-8$ km s$^{-1}$ lower than the 
systemic velocity. In Paper I, the $\sim2$ km~s$^{-1}$ velocity components 
were detected in N$_2$H$^+(1-0)$ and/or N$_2$D$^+(2-1)$ towards a few target 
positions near IRAS 05405-0117 (called Ori B9 E in Paper I) and Ori B9 N. 
Similarly, Aoyama et al. (2001) found that the C$^{18}$O$(1-0)$ and 
H$^{13}$CO$^+(1-0)$ clumps associated with IRAS 05405-0117 have centroid LSR 
velocities of about 2 km~s$^{-1}$ (at the resolutions 2\farcm7 and 
3\farcm8, respectively). This raises the question whether the 
sources and/or high-density gas showing the lower velocity emission are 
associated with the same cloud complex as the ``regular'' 9 km~s$^{-1}$ 
component ?

The CO$(1-0)$ integrated intensity maps in the velocity ranges $-1.0-2.3$ and 
$2.3-5.5$ km~s$^{-1}$ by Wilson et al. (2005; their Fig.~3) show relatively 
strong emission in the direction of Orion B9. As discussed by Wilson et al. 
(2005), the stellar winds from the Ori OB 1b subgroup have likely interacted 
with the low-longitude and -latitude part of Orion B, resulting in an 
accelerated motion towards the Sun. 
This fraction of the gas is likely to be located a few tens of parsecs closer 
to the Sun than the higher velocity gas, and it is estimated to have a mass of 
about 15\% of the total mass of Orion B, i.e., $\sim10^4$ M$_{\sun}$ 
(\cite{wilson2005}). Our pointed molecular line observations in high density 
tracers, including deuterated species, show that the low-velocity gas in 
Orion B also contains dense cores, possibly originating from the fragmentation 
of the cloud region compressed by the feedback from the massive stars of the 
Ori OB 1b group.

\subsection{Core spatial distribution revisited}

In Paper I, we studied the spatial distribution of cores in Orion B9 
in order to examine the fragmentation length-scale. We determined the 
core-separation distribution and the number distribution of the projected 
separation distance between nearest neighbours. These were compared with the 
corresponding random distributions. We concluded that the observed 
distributions are random-like, and thus the origin of cores is possibly caused 
by turbulent fragmentation; random distribution is expected if the cloud 
fragmentation is driven by a stochastic turbulent process.
As discussed in the previous subsection, some of the cores are probably 
located at somewhat different distance than the ``9 km~s$^{-1}$ members 
of Orion B9''. For this reason, we re-examined the core spatial distribution 
in the region. 

\begin{figure*}
\begin{center}
\includegraphics[width=0.45\textwidth]{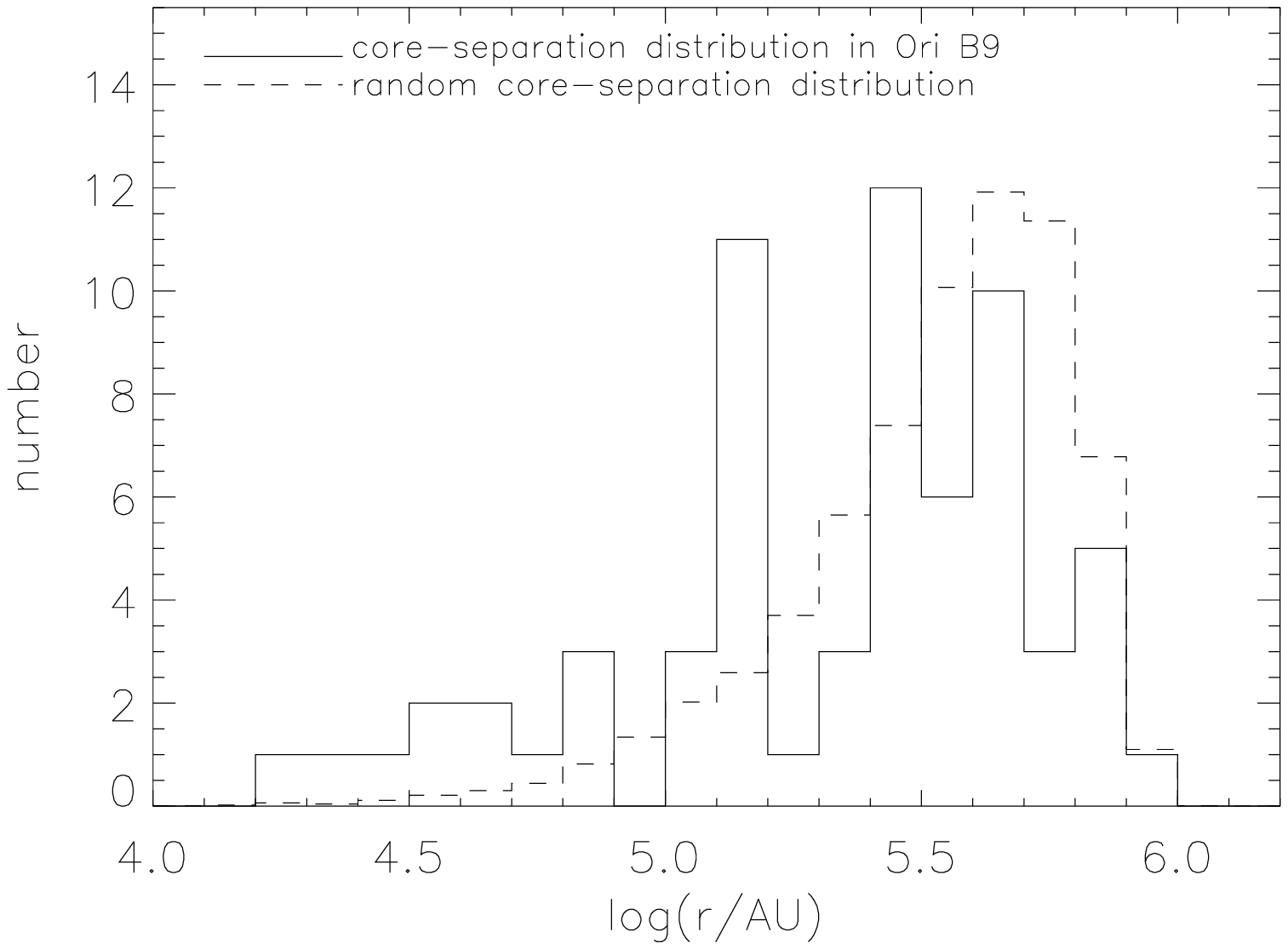}
\includegraphics[width=0.45\textwidth]{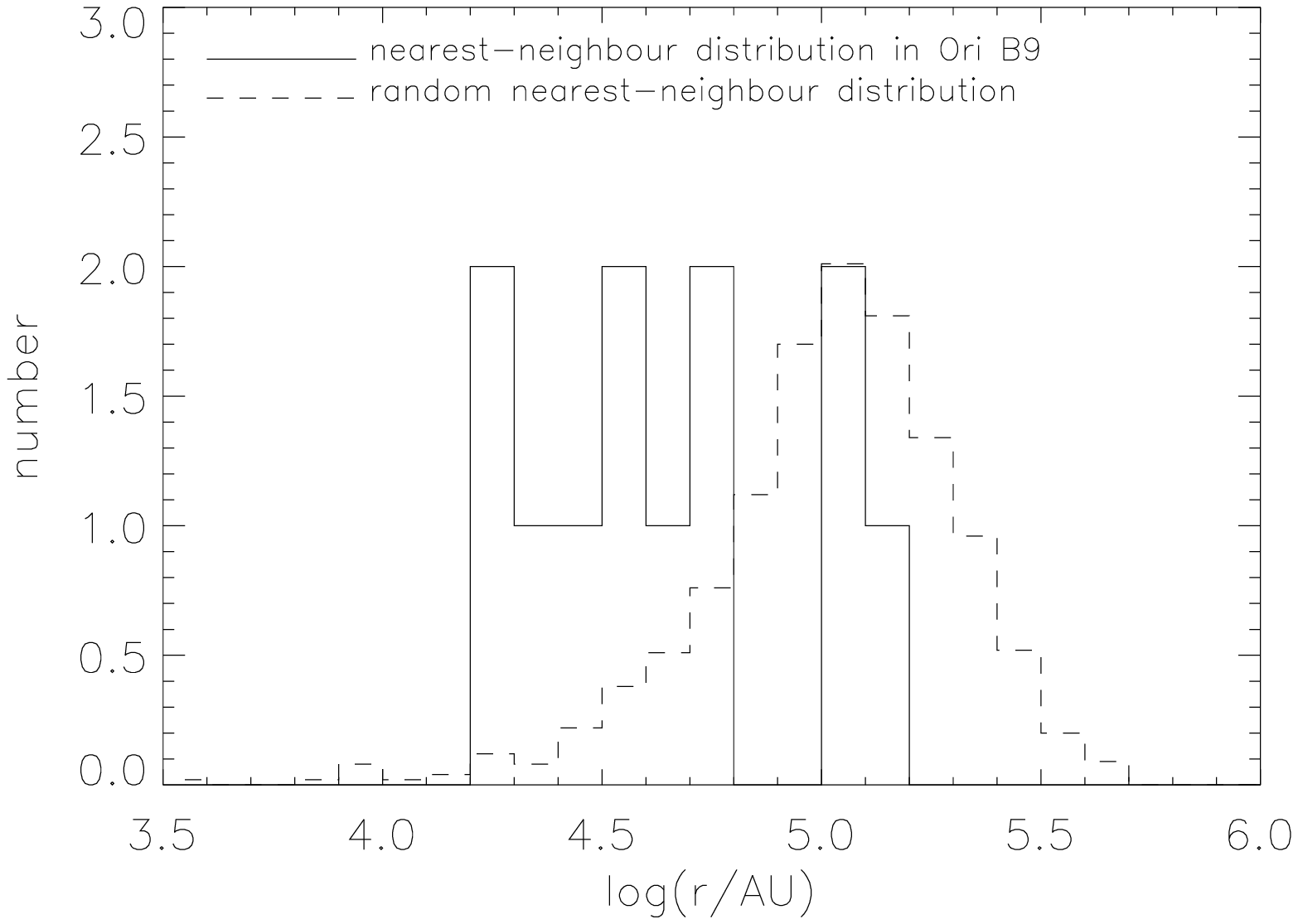}
\includegraphics[width=0.45\textwidth]{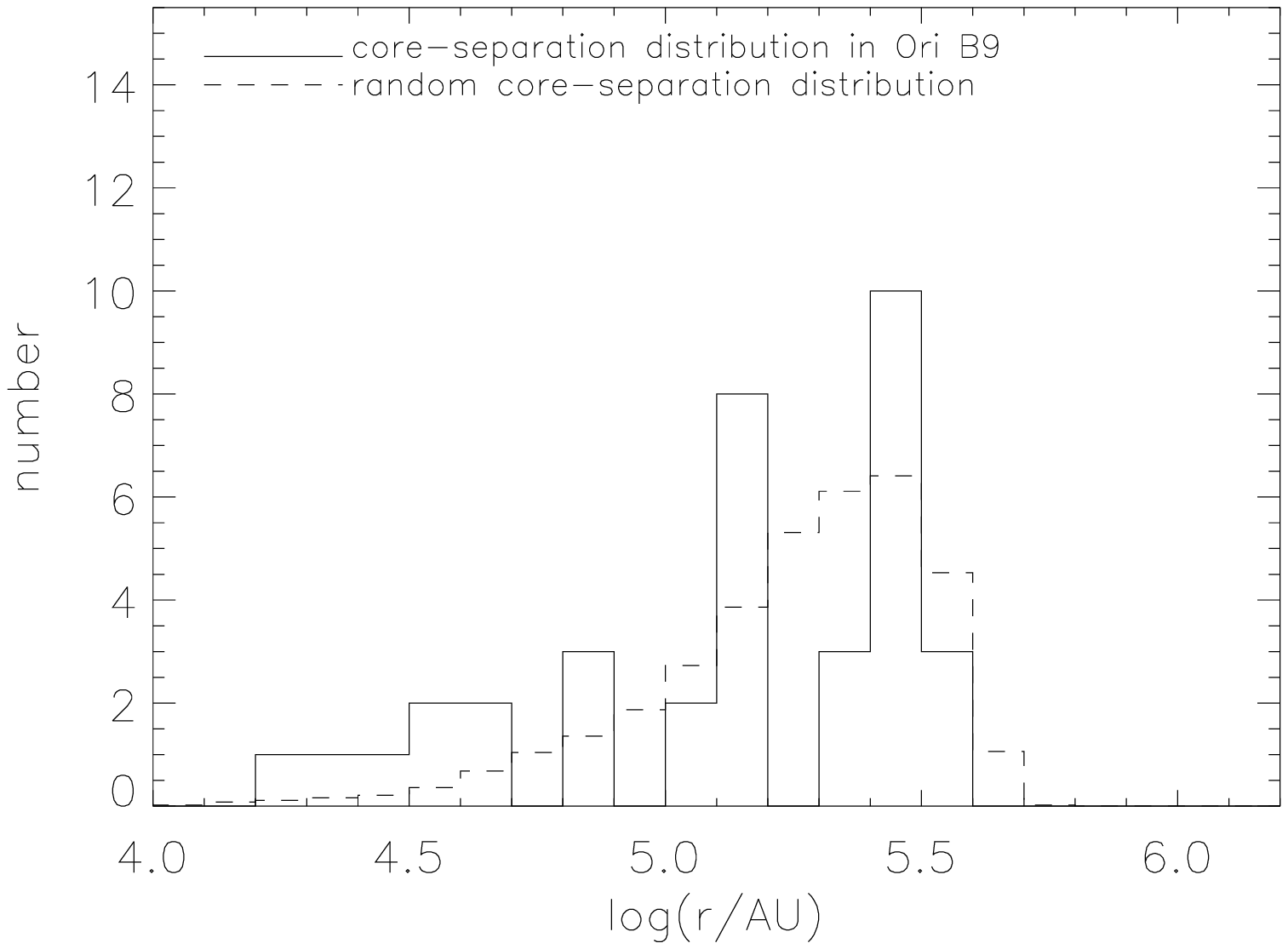}
\includegraphics[width=0.45\textwidth]{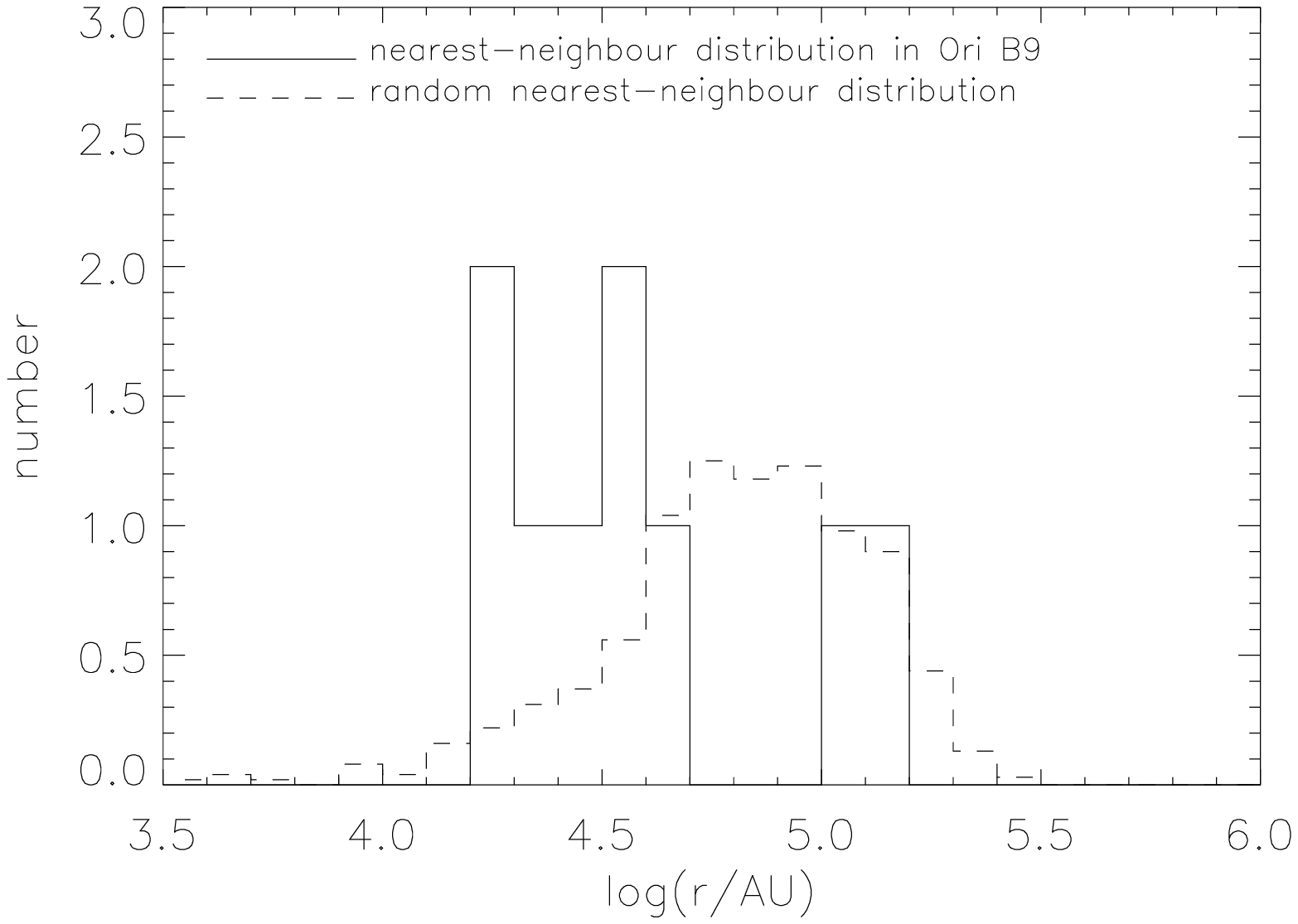}
\caption{\textbf{Top left:} Observed core-separation distribution 
(solid line) compared with the expected distribution for random distribution of
the same number of cores as the observed sample over an identical area 
(dashed line). \textbf{Top right:} Same as in the top left panel but for 
the nearest-neighbour distribution. \textbf{Bottom panels:} Same as in the top 
panels, but SMM 7, IRAS 05412-0105, and IRAS 05413-0104 have been excluded 
from the samples.} 
\label{figure:separations}
\end{center}
\end{figure*}

We note furthermore that there was a flaw in our IDL procedure used to generate
the histograms shown in Fig.~8 of Paper I. The number of separations 
(66 for the core separations, and 12 for the nearest neighbours) 
were not correct, and the numbers in the observed and model (random) 
distributions were not equal. Here we reproduce the histograms and show the 
new plots for the core separations and nearest neighbours in the top 
panels of Fig.~\ref{figure:separations}. In the bottom panels of 
Fig.~\ref{figure:separations}, we show the same distributions but
without the sources SMM 7, IRAS 05412-0105, and IRAS 05413-0104. 
In the latter figure, the size of the region used to generate the random 
distributions was decreased from 0.22 $\Box \degr$ to 0.06 $\Box \degr$.
The obtained statistics of the distributions are presented in 
Table~\ref{table:separations}. In this table, we give the 
mean (and its standard deviation) and median of the observed core spatial
distribution (Cols.~(2) and (3)), those of the corresponding random 
distribution (Cols.~(4) and (5)), ratios between the 
observed and random mean and median separations (Cols.~(6) and (7)), and 
probability given by the two-sample K-S test that the observed and random 
distributions are drawn from the same underlying distribution, see below 
(Col.~(8)). The random distributions were generated a 
hundred times and the averaged histograms were compared with those derived
from observations. The numerical values for the random distributions presented 
in Table~\ref{table:separations} are the averages and the standard deviations 
of these 100 runs. Only a few values differ from those reported in Paper 
I (see the footnotes in the table). Also, for the core 
positions in Orion B North (\cite{nutter2007}), the values are the same 
as those reported in Paper I. We note that for the core-separation 
distributions
$\langle r \rangle_{\rm OriB9}/\langle r \rangle_{\rm OriBN}=0.63\pm0.01$ 
and $\tilde{r}_{\rm OriB9}/\tilde{r}_{\rm OriBN}=0.62$, and for the 
nearest-neighbour distribution 
$\langle r \rangle_{\rm OriB9}/\langle r \rangle_{\rm OriBN}=1.95\pm0.27$ 
and $\tilde{r}_{\rm OriB9}/\tilde{r}_{\rm OriBN}=1.87$.

In Paper I, we suggested that the comparable mean and median values
between the observed and random distributions indicate that the cores are 
likely to be randomly distributed within the region. 
To examine this in more detail, we carried out K-S tests between the observed 
and model distributions. As shown in the last column of 
Table~\ref{table:separations}, the probability that the observed 
distribution and the generated random distribution represent the same 
underlying distribution is very small in all cases.
Such low probabilities call into question the similarity between the observed 
and random distributions. Thus, even the observed and model mean and median 
separation-distances are comparable, the above K-S test probabilities suggest 
that the distributions as a whole are not similar. The K-S probability 
is expected to be a more robust measure of the similarity between the two 
distributions than the comparison of the mean and median values because 
the latter two can be the same for two different distributions.

The observed mean and median distances between the nearest-neighbours are in 
the range $\sim3\times10^4-6\times10^4$ AU. Assuming that the parental cloud 
region is characterised by the average gas kinetic temperature and density 
values of 15 K and $\sim2\times10^3-10^4$ cm$^{-3}$, respectively (Sect.~5.4), 
the thermal Jeans length, $\lambda_{\rm J}=\sqrt{\pi c_{\rm s}^2/G\rho}$, is about 
$5\times10^4-10^5$ AU. These are comparable to the observed core separations 
indicated above. This suggests that the fragmentation of the region into 
cores is caused by gravitational instability. The parental cloud region 
in which the cores have formed could have been initially compressed by the 
winds from the nearby massive stars as discussed in Sect.~5.7.

\begin{table*}
\caption{Statistics of the core spatial distributions in Orion B9.}
\begin{minipage}{2\columnwidth}
\centering
\renewcommand{\footnoterule}{}
\label{table:separations}
\begin{tabular}{c c c c c c c c}
\hline\hline 
  & $\langle r \rangle_{\rm obs}$ & $\tilde{r}_{\rm obs}$ & $\langle r \rangle_{\rm ran}$ & $\tilde{r}_{\rm ran}$ & $\langle r \rangle_{\rm obs}/\langle r \rangle_{\rm ran}$ & $\tilde{r}_{\rm obs}/\tilde{r}_{\rm ran}$ & Prob.\\ 
  & [$\log$ AU] & [$\log$ AU] & [$\log$ AU] & [$\log$ AU] & & & [\%]\\  
\hline
\textbf{Core separation-distribution} &\\
Original sample & $5.467\pm0.037$ & 5.420 & $5.525\pm0.051$\footnote{In Paper I, 
this was reported to be $5.59\pm0.05$.} & $5.584\pm0.054$ & $0.87\pm0.03$ & $0.69\pm0.09$ & $1.7\pm0.6$\\
Reduced sample\footnote{SMM 7, IRAS 05412-0105, and IRAS 05413-0104 have been 
excluded from the sample.} & $5.246\pm0.043$ & 5.195 & $5.245\pm0.068$ & $5.305\pm0.075$ & $1.00\pm0.06$ & $0.78\pm0.13$ & $4.7\pm0.7$\\
\textbf{Nearest neighbour-distribution} &\\
Original sample & $4.750\pm0.090$ & 4.619 & $5.018\pm0.047$\footnote{$5.08\pm0.08$ in Paper I.} & $5.035\pm0.045$\footnote{$5.07\pm0.11$ in Paper I.} & $0.54\pm0.05$ & $0.38\pm0.04$ & $5.7\pm1.2$\\
Reduced sample$^b$ & $4.699\pm0.123$ & 4.511 & $4.806\pm0.130$ & $4.822\pm0.144$ & $0.78\pm0.01$ & $0.49\pm0.16$ & $23.8\pm2.6$\\
\hline 
\end{tabular} 
\end{minipage}
\end{table*}

\subsection{The lifetime and origin of dense cores in Orion B9}

According to the scenario of turbulence-regulated star formation, the 
prestellar core evolution is dynamic and the corresponding lifetime is
only a few times the free-fall time (\cite{maclow2004}; \cite{vazquez2005}).
Several results of our study conform to this scenario: as was discussed in 
Sect.~5.2, the Orion B9 cores could have been formed 
in turbulent shocks where the dissipation of the kinetic energy of turbulent 
motions took place. It was also concluded that the timescale for core 
formation should be comparable to the free-fall time. On the other hand, 
the abundances of the N-bearing species studied 
in the present paper are consistent with those predicted by chemical models 
based on dynamical prestellar core evolution (Sect.~5.6). 
Moreover, in Paper I we deduced that there are equal numbers of pre- and 
protostellar cores in the Orion B9 region, which suggest that the 
corresponding lifetimes are comparable. Based on similar statistical results, 
i.e., $N_{\rm pre}/N_{\rm proto}\approx1$, recent studies have shown that the 
duration of the prestellar phase of core evolution at densities 
$n({\rm H_2})\gtrsim10^4$ cm$^{-3}$ is only a few times the free-fall time 
(see references in Paper I; \cite{evans2009}). It is possible that some of the 
cores do not belong to the same volume of the cloud as those with centroid 
velocities around 9 km~s$^{-1}$. Even if the cores with lower 
centroid velocities are excluded (i.e., SMM 7, and IRAS 05413-0104 and 
05412-0105), the above ratio is $5:4=1.25$, which is still consistent with 
the dynamic core evolution scenario. 

In the present study (and in Paper I), we found that the CMFs in Orion B9 and 
Orion B North very likely represent the subsamples of the 
same parent distribution, and resemble the stellar IMF (\cite{nutter2007}). 
Numerical simulations of turbulent fragmentation have been able to reproduce 
the general shape of the IMF (e.g., \cite{padoan2002}; \cite{ballesteros2006}).
On the other hand, recent models concerning fragmentation initiated by the 
ambipolar diffusion (AD) have likewise succeeded in explaining the connection 
between the CMF and the IMF (\cite{kunz2009}).

In Paper I, we used the spatial distribution of cores, which seemed to
mimic a random distribution, as an argument for the dominance of
turbulent fragmentation in core formation. Moreover, we stated that
because region represents clustered star formation, the turbulence is
most likely driven on large scales (e.g., \cite{klessen2000}; 
\cite{klessen2001}). However, the present molecular line data and a more careful
analysis of the spatial distribution have called both arguments into
question: According to the K-S test, the core distribution is unlikely
to be random, and the core properties resemble closely those in the
regions of isolated star formation. 

To summarise, the current data suggest that the evolution of the Orion
B9 cores is dynamic. As discussed in Sect.~5.7, the Orion B9 region is likely 
to be influenced by the massive stars of the Ori OB 1b group. This interaction 
process could have led to the driving of the large-scale turbulence and/or 
to the compression of the cloud material into dense sheets and filaments 
as observed in molecular clouds. These structures could have then been 
fragmented into dense cores via gravitational instability as suggested by the 
Jeans-length analysis in Sect.~5.8. Further investigations of the chemistry, 
deuteration, and the degree of ionisation can help to fix the evolutionary 
timescales, and thereby estimate the roles of turbulence, thermo-gravitational 
instability, and ambipolar diffusion in the core formation.

\section{Summary and conclusions}

We observed the NH$_3(1,\,1)$ and $(2,\,2)$ inversion transitions, and 
the N$_2$H$^+(3-2)$ lines towards the 
submm cores in Orion B9 star-forming region with the Effelsberg 100-m and 
APEX telescopes, respectively. These line observations were combined with our 
previous 870 $\mu$m submm dust continuum data of the region. The data were 
used to determine the physical characteristics of dense cores in this region. 
We mainly investigated the temperatures and kinematics of the cores, and 
recalculated the temperature-dependent quantities presented in Paper I 
(\cite{miettinen2009}). Our main results and conclusions are as follows:

1. The gas kinetic temperatures of the cores are in the range $\sim9.4-13.9$ K,
with an average value of about 12 K. No significant difference in 
$T_{\rm kin}$ was found between starless and protostellar cores. 
The temperature values are similar to those in nearby low-mass 
star-forming regions. 

2. The cores are characterised by subsonic, or at most transonic, non-thermal 
motions. Thus, the cores are kinematically similar to those in nearby 
low-mass star-forming regions.

3. In the case of protostellar cores, we found a positive correlation between 
the non-thermal NH$_3$ linewidth and bolometric luminosity. This can be
understood so that on one hand, the embedded central protostar enhances the 
level of turbulent motions within the parent core, and on the other hand, 
more turbulent initial conditions lead to the formation of more massive stars 
with higher luminosity.

4. The core masses are, on average, about 11\% higher than those reported 
in Paper I when calculated by assuming that $T_{\rm dust}=T_{\rm kin}$
(in Paper I, we assumed that $T_{\rm dust}=10$ K for starless cores). 
These masses are very likely drawn from the same parent distribution 
as the core masses in the northern part of Orion B (\cite{nutter2007}).  

5. Almost all cores were found to be close to virial equilibrium
when the internal kinetic pressure and self-gravity are taken into
account. In addition, they are likely to be exposed to a substantial
turbulent ram pressure from the intercore gas (as can be evaluated
from previous $^{13}$CO data), which is sufficient to bring them to
equilibrium or make them collapse. In particular, it seems probable that
most if not all starless cores detected in this survey are prestellar.

6. The fractional NH$_3$ and N$_2$H$^+$ abundances in the cores are 
$\sim1.5-9.8\times10^{-8}$ and $\sim0.2-5.9\times10^{-10}$, 
respectively. The NH$_3$/N$_2$H$^+$ column density ratio is higher in 
prestellar cores than protostellar cores. Similar trend has also been observed 
in other star-forming regions. The NH$_3$ abundance appears to decrease with 
increasing gas density (as traced by 870 $\mu$m continuum emission). 
This tendency is likely to be caused by the accretion of NH$_3$ onto grains at 
very high densities, and conforms to the recent results of Friesen et al. 
(2009) in Ophiuchus.

7. A few cores have a much lower radial velocity compared to the 
systemic velocity of the region ($\sim9$ km~s$^{-1}$), and do not necessarily 
belong to the same volume of the cloud as the rest of the cores. Instead, 
they are likely to be members of the ``low-velocity'' part of Orion B.
This increases the relative number of pre- and protostellar cores in Orion B9 
and consequently, the statistical lifetime of the prestellar phase of core 
evolution (by 25\% in both cases).

8. Many of the properties of dense cores in Orion B9 suggest that their 
evolution is likely to be dynamic, i.e., comparable to the free-fall 
timescale. On the other hand, the new data and an improved analysis of spatial 
distribution have made the role of turbulent fragmentation in the core 
formation less evident than it appeared in Paper I. According to the new 
results, the cores are unlikely to be randomly positioned within the region, 
and they are not clustered in the way expected if the fragmentation is driven 
on large scale turbulence. Futhermore, the core properties are similar to 
those found in the regions of isolated star formation.

9. The Orion B9 region is likely to be influenced by the massive stars of the 
nearby Ori OB 1b group. This explains the origin of the low-velocity material 
in the southern parts of Orion B (\cite{wilson2005}). Moreover, this 
interaction process could have led to formation of dense filaments, from 
which the dense cores were fragmented out by the action of 
gravitational instability.

The present paper demonstrates the importance of a kinematic and temperature 
information for the studies of dense cores and star formation. A lack of 
radial velocity information, for instance, may cause errors when examining 
the relative numbers of starless and protostellar cores. This, in turn, 
affects the lifetime estimates based on statistical arguments. The core 
lifetime is an important discriminator between different theories of core 
and star formation, particularly between the turbulence and 
ambipolar diffusion driven star formation scenarios. 

\begin{acknowledgements}

We thank the referee for helpful comments, and the staff at the Effelsberg for 
their hospitality and support during our observations. We also thank the staff 
at the APEX telescope for performing the observations in the service mode.
The authors acknowledge support from the Academy of Finland through grants 
117206 and 132291.

\end{acknowledgements}

\end{document}